\documentclass[a4paper,11pt]{article}
\pdfoutput=1

\usepackage{jheparxiv}
\usepackage[T1]{fontenc}
\usepackage[normalem]{ulem}
\usepackage{tensor}
\usepackage{amsmath}
\usepackage{amssymb}
\usepackage{mathrsfs}
\usepackage{graphicx}
\usepackage{stmaryrd}
\usepackage{twistor}
\usepackage{tikz}
\usepackage{commath}
\usepackage{xcolor}
\usepackage{mathtools}
\usepackage{enumitem}

\newcommand{\dd}{\mathrm{d}}
\newcommand{\ii}{\mathrm{i}}
\newcommand{\calA}{\mathcal{A}}
\newcommand{\calM}{\mathcal{M}}

\newcommand{\calI}{\mathcal{I}}
\newcommand{\calO}{\mathcal{O}}
\newcommand{\CC}{\mathbb{C}}
\newcommand{\RR}{\mathbb{R}}

\title{{\Large Topics in Celestial holography: A bottom-up perspective}}

\author[a]{Bin Zhu}

\affiliation[a]{School of Physics, Nankai University, Weijin Road 94, Tianjin 300071, P.R. China}

\emailAdd{bzhu@nankai.edu.cn}

\abstract{We review some selected topics in celestial holography on the search for a celestial dual to quantum gravity in flat spacetimes. We focus on the
bottom-up approach, emphasizing symmetries, key elements in celestial CFT, interplay with twistor theory, and connection to AdS/CFT.}

\dedicated{Invited review for Physics Reports}

\begin{document}
\maketitle

\flushbottom

\makeatletter
\@starttoc{toc}
\makeatother
\thispagestyle{plain}

\section{Introduction}

Holography has been one of the most successful approaches to quantum gravity, originating in the holographic principle of 't Hooft and Susskind \cite{tHooft:1993dmi,Susskind:1994vu}. In anti-de Sitter spacetime, the AdS/CFT correspondence gives a non-perturbative definition of quantum gravity in terms of a conformal field theory living on the timelike boundary \cite{Maldacena:1997re,Gubser:1998bc,Witten:1998qj}.  This framework has transformed our understanding of black holes, strongly coupled quantum field theory, and the emergence of bulk geometry.  It also provides the clearest evidence that gravitational dynamics in a spacetime region can be encoded in a lower-dimensional quantum system.

It is natural to wonder if holography still works in the scenario that is closer to our real world.  The world in which scattering experiments are naturally formulated is not asymptotically AdS.  At laboratory and astrophysical scales, isolated processes are more naturally idealized by asymptotically flat spacetime, where the incoming and outgoing states are defined at past and future null infinity \cite{Bondi:1962px,Sachs:1962wk,Sachs:1962zza}.  If holography is a general principle of quantum gravity rather than a special property of AdS boundary conditions, one would like to understand what replaces the AdS boundary CFT in flat space.  This is the broad motivation behind flat-space holography: to find a boundary description of gravitational physics in a spacetime whose natural observables are scattering amplitudes, radiation data, and asymptotic charges rather than correlation functions on a timelike conformal boundary.

Celestial holography is one of the most concrete approaches to this problem.  Its starting point is conservative: it does not postulate a new microscopic theory at the outset, but rewrites the flat-space S-matrix in a basis adapted to the Lorentz group.  Massless particles reaching null infinity are labeled by a point $(z,\bar z)$ on the celestial sphere and by a light-cone energy $\omega$.  Mellin transforming scattering amplitudes with respect to the external energies trades the energy variables for conformal dimensions.  The resulting objects transform covariantly under the four-dimensional Lorentz group, which acts as the global conformal group $SL(2,\mathbb C)$ on the celestial sphere.  In this representation, flat-space scattering amplitudes take the form of correlation functions of putative operators in a two-dimensional celestial conformal field theory \cite{Pasterski:2016qvg,Pasterski:2017kqt,Pasterski:2017ylz,Pasterski:2021rjz,Raclariu:2021zjz}.

This change of basis is powerful because it brings together structures that had previously appeared in different languages.  Soft theorems become Ward identities for conformally soft operators \cite{Weinberg:1964ew,Weinberg:1965nx,Strominger:2017zoo}.  Asymptotic symmetries, such as large gauge transformations and BMS transformations, become current-algebra constraints on celestial correlators \cite{Bondi:1962px,Sachs:1962wk,Sachs:1962zza,Donnay:2023mrd}.  Collinear factorization becomes a celestial operator product expansion \cite{Pate:2019lpp,Fan:2019emx,Fotopoulos:2019vac}.  The infrared triangle relating soft theorems, memory effects, and asymptotic charges is therefore reorganized as a set of two-dimensional symmetry statements.  These observations are central to the bottom-up approach adopted in this review.

The present review is written from this point of view.  We treat celestial amplitudes as an organizing representation of scattering data, while keeping track of the additional structures needed for a genuine holographic interpretation.  This requires moving between several related boundary languages.  The celestial basis diagonalizes boosts and makes Lorentz covariance look like two-dimensional conformal covariance.  The Carrollian basis keeps retarded time at null infinity and makes radiation, memory and supertranslations more direct.  Twistor and ambitwistor methods expose chiral and worldsheet structures behind celestial OPEs.  Liouville-type descriptions and non-distributional amplitudes provide laboratories in which celestial correlators become closer to ordinary two-dimensional functions.  Flat limits of AdS/CFT and extrapolate dictionaries ask how much of the AdS holographic toolkit survives when the cosmological constant is removed.

Several reviews and lecture notes provide complementary entry points to this subject.  Broad introductions to celestial amplitudes, conformal primary wavefunctions and the celestial transform can be found in \cite{Pasterski:2021rjz,Raclariu:2021zjz}, while the Snowmass overview emphasizes celestial holography as a program connecting scattering amplitudes, asymptotic symmetries, and two-dimensional conformal field theory methods \cite{Pasterski:2021raf}.  The infrared structure of gauge theory and gravity, which supplies much of the symmetry input for celestial holography, is reviewed in \cite{Strominger:2017zoo}, and the SAGEX review chapter gives a complementary perspective on soft theorems and celestial amplitudes \cite{McLoughlin:2022ljp}.  The asymptotic-symmetry approach to celestial holography is reviewed in \cite{Donnay:2023mrd}.  Reviews and lecture notes on Carrollian physics and holography explain the geometry of null boundaries and the Carrollian formulation of flat-space holography \cite{Ruzziconi:2026bix,Bagchi:2025vri,Nguyen:2025zhg}.  Our emphasis in this review is different.  We focus on selected developments that illuminate how the celestial description might become a genuine holographic framework: analytic structures of celestial amplitudes, non-distributional correlators, celestial Liouville theory \cite{Stieberger:2022zyk,Taylor:2023bzj,Stieberger:2023fju}, loop and UV/IR issues, $w_{1+\infty}$ algebra \cite{Guevara:2021abz,Strominger:2021mtt}, twistor methods, and connections to AdS/CFT.

The scope is therefore deliberately selective and not intended to be exhaustive.  We do not attempt a complete review of all recent developments in celestial holography.  Instead, we ask how a bottom-up holographic dictionary is assembled from complementary perspectives and how far these perspectives can be made mutually compatible. This organization is intended to give researchers entering the field a coherent route through the main ideas without obscuring the fact that a complete holographic dual of flat space is still not known.

The organization is as follows.  Section~\ref{sec:celestial-amplitudes} introduces the celestial transform, the massless scattering dictionary and the relation between celestial and Carrollian bases.  Section~\ref{sec:symmetry-ope} reviews asymptotic symmetries, conformally soft operators and celestial OPEs.  Section~\ref{sec:blocks} discusses conformal blocks, distributional support and non-distributional celestial amplitudes.  Section~\ref{sec:celestial-liouville} reviews the appearance of Liouville theory and related analytic structures.  Section~\ref{sec:uv-ir} explains the ultraviolet and infrared behavior of celestial amplitudes, including loop effects, shockwave correlators, and string-theoretic improvements.  Section~\ref{sec:w-infinity} studies celestial $w_{1+\infty}$ algebra and its deformations.  Section~\ref{sec:twistor-ambitwistor} reviews twistor and worldsheet approaches.  Section~\ref{sec:ads-flat-limits} discusses connections to AdS/CFT, Carrollian limits and alternative extrapolate dictionaries.  We end in Section~\ref{sec:outlook} with open questions and future directions.

\section{Celestial amplitudes and transform dictionaries}
\label{sec:celestial-amplitudes}

For simplicity, we focus on scattering amplitudes of massless external particles. Momentum-space amplitudes of massless particles are functions of four-dimensional momenta and helicities.  Celestial amplitudes reorganize the same data in a basis adapted to the Lorentz group, so that the direction of each external massless particle is a point on the celestial sphere while its energy is traded for a conformal dimension.  This is why the subject naturally combines scattering amplitudes, asymptotic symmetries, and two-dimensional conformal field theory: the Lorentz group $SL(2,\CC)$ acts on the celestial sphere as the global conformal group \cite{Pasterski:2016qvg,Pasterski:2017kqt,Pasterski:2017ylz,Pasterski:2021rjz,Raclariu:2021zjz,Pasterski:2021raf}. See also \cite{Kapec:2014opa,Kapec:2016jld} on how the global conformal symmetry is enhanced to the Virasoro symmetry by the subleading soft graviton theorem.

This section fixes the transform dictionary used throughout the review.  The basic distinction is simple but important.  The celestial transform diagonalizes boosts and makes Lorentz covariance manifest.  The Carrollian transform keeps the retarded time at null infinity and is closer to the radiative phase space.  These are not competing descriptions of different physics.  They are different integral transforms of the same scattering data, and they emphasize different parts of the flat-space holographic problem.

\subsection{Massless kinematics}
\label{subsec:massless-kinematics}

For a massless external particle in four-dimensional Minkowski space we write
\be
p^\mu=\epsilon\,\omega\,q^\mu(z,\bar z),\qquad
\epsilon=+1\ {\rm(out)},\quad \epsilon=-1\ {\rm(in)},\quad \omega>0,
\label{eq:massless-momentum-dictionary}
\ee
where
\be
q^\mu(z,\bar z)=
\left(1+z\bar z,\,z+\bar z,\,-\ii(z-\bar z),\,1-z\bar z\right).
\label{eq:null-vector-dictionary}
\ee
With the standard Minkowski convention this vector is null, $q^2=0$, and labels a point on the projective light cone.  Eqs.(\ref{eq:massless-momentum-dictionary}) and (\ref{eq:null-vector-dictionary}) are the standard parametrization used in the celestial-amplitude literature \cite{Pasterski:2016qvg,Pasterski:2017kqt,Pasterski:2021rjz}.  The variable $\omega$ is the light-cone energy, while $(z,\bar z)$ specifies the direction of the null ray.  In Lorentzian signature $\bar z$ is the complex conjugate of $z$ on the physical sphere.  In split-signature or analytically continued discussions the two variables are often treated independently.

The Lorentz group acts on the celestial coordinate by a Mobius transformation,
\be
z\longmapsto z'=\frac{a z+b}{c z+d},\qquad ad-bc=1,
\label{eq:mobius-transform}
\ee
and the null vector transforms as
\be
q^\mu(z',\bar z')=
|c z+d|^{-2}\,\Lambda^\mu{}_\nu\,q^\nu(z,\bar z).
\label{eq:q-lorentz-covariance}
\ee
The compensating conformal factor in Eq.(\ref{eq:q-lorentz-covariance}) is the kinematic origin of the two-dimensional conformal covariance of celestial amplitudes.  A Lorentz transformation changes a null direction by a conformal transformation of the sphere and rescales the energy.  The Mellin transform below precisely diagonalizes this rescaling.

For helicity amplitudes it is often useful to write the same dictionary in spinor-helicity variables,
\be
\lambda_\alpha=\sqrt{\omega}\begin{pmatrix}1\\ z\end{pmatrix},\qquad
\tilde\lambda_{\dot\alpha}=\epsilon\sqrt{\omega}\begin{pmatrix}1\\ \bar z\end{pmatrix},\qquad
p_{\alpha\dot\alpha}=\lambda_\alpha\tilde\lambda_{\dot\alpha}.
\label{eq:spinor-celestial-map}
\ee
This form makes the relation to collinear limits transparent: two massless momenta become collinear when $z_{12}=z_1-z_2$ and $\bar z_{12}=\bar z_1-\bar z_2$ approach zero. We shall see later how the collinear splittings of scattering amplitudes lead to the operator product expansion (OPE) of celestial operators.

\subsection{Celestial basis and conformal primaries}
\label{subsec:mellin-transform}

Momentum eigenstates diagonalize translations.  The celestial basis instead diagonalizes boosts along a fixed null direction.  For a scalar external leg, the corresponding conformal primary wavefunction is obtained by Mellin transforming a plane wave in the energy variable \cite{Pasterski:2016qvg,Pasterski:2017kqt},
\be
\Phi^\epsilon_\Delta(X;z,\bar z)
=\int_0^\infty \dd\omega\,\omega^{\Delta-1}
e^{\ii\epsilon\omega q(z,\bar z)\cdot X-\varepsilon_{\rm reg}\omega}
=\frac{\Gamma(\Delta)}
{\big[\varepsilon_{\rm reg}-\ii\epsilon q(z,\bar z)\cdot X\big]^\Delta}.
\label{eq:scalar-conformal-primary-wavefunction}
\ee
The sign $\epsilon=\pm1$ is the same incoming/outgoing sign as in Eq.(\ref{eq:massless-momentum-dictionary}), and the small positive $\varepsilon_{\rm reg}$ acts as a convergence regulator.  The second expression shows explicitly that the wavefunction transforms as a conformal primary on the celestial sphere.  For spin $J$, the scalar wavefunction is dressed by polarization data, and the associated two-dimensional weights are
\be
h=\frac{\Delta+J}{2},\qquad
\bar h=\frac{\Delta-J}{2}.
\label{eq:h-hbar-definition}
\ee
The principal series $\Delta=1+\ii\lambda$, $\lambda\in\RR$, gives the delta-function normalizable basis for massless one-particle scattering states \cite{Pasterski:2017kqt,Lam:2017ofc}.  The conformal primary wavefunctions of gauge bosons and gravitons are described in \cite{Pasterski:2017kqt}, while the half-integer spins in four dimensions can be found in \cite{Fotopoulos:2020bqj}. See also \cite{Iacobacci:2024laa} for a detailed discussion on all half-integer spins in any dimension and the massive case.

The $n$-point celestial amplitude is the Mellin transform of the momentum-space amplitude with respect to all external energies,
\be
\widetilde{\calA}_n(\Delta_i,J_i;z_i,\bar z_i)
=\prod_{i=1}^n
\left[\int_0^\infty \dd\omega_i\,\omega_i^{\Delta_i-1}\right]
\calA_n(\epsilon_i\omega_i q_i,J_i),
\label{eq:celestial-amplitude-definition}
\ee
where $q_i=q(z_i,\bar z_i)$ and the momentum-conserving delta function is included in $\calA_n$ unless stated otherwise.  In Section \ref{sec:blocks} we show some examples of celestial amplitudes.

The central statement of celestial holography is that the same object may be
interpreted as a correlation function of operators on the celestial sphere,
\be
\left\langle
\calO^{\epsilon_1}_{\Delta_1,J_1}(z_1,\bar z_1)\cdots
\calO^{\epsilon_n}_{\Delta_n,J_n}(z_n,\bar z_n)
\right\rangle_{\rm CCFT}
=
\prod_{i=1}^n
\left[\int_0^\infty \dd\omega_i\,\omega_i^{\Delta_i-1}\right]
\calA_n(\epsilon_i\omega_i q_i,J_i) ,
\label{eq:celestial-holography-main-statement}
\ee
where the two-dimensional spin \(J_i=h_i-\bar h_i\) is identified with the
four-dimensional helicity of the corresponding external particle.  In this
form, celestial holography asserts that flat-space scattering data, after a
change of basis from energy eigenstates to boost eigenstates, can be organized
as correlators of a putative celestial conformal field theory \cite{Pasterski:2016qvg,Pasterski:2017kqt,Pasterski:2017ylz}.

A more conservative point of view is to interpret Eq.(\ref{eq:celestial-amplitude-definition}) as a change of basis applied leg by leg to the same $S$-matrix. As we have seen, it has been useful for exposing some holographic properties of the $S$-matrix.

In the principal series, the inverse transform of Eq.(\ref{eq:celestial-amplitude-definition}) is
\be
\calA_n(\epsilon_i\omega_iq_i,J_i)
=\prod_{i=1}^n
\left[\int_{-\infty}^{\infty}\frac{\dd\lambda_i}{2\pi}\,
\omega_i^{-\ii\lambda_i}\right]
\widetilde{\calA}_n(1+\ii\lambda_i,J_i;z_i,\bar z_i).
\label{eq:inverse-celestial-transform}
\ee
In Section~\ref{sec:celestial-liouville}, we shall see an example of using the inverse Mellin transform to extract four-dimensional physics from two-dimensional correlators.

\subsection{Covariance and Translation Invariance}
\label{subsec:what-transform-makes-manifest}

The most immediate consequence of the Mellin transform is conformal covariance.  Under the Lorentz transformation Eq.(\ref{eq:mobius-transform}), a celestial amplitude of external helicities $J_i$ transforms as a two-dimensional correlator,
\be
\widetilde{\calA}_n(\Delta_i,J_i;z_i',\bar z_i')
=\prod_{i=1}^n
(c z_i+d)^{2h_i}(\bar c \bar z_i+\bar d)^{2\bar h_i}
\widetilde{\calA}_n(\Delta_i,J_i;z_i,\bar z_i).
\label{eq:celestial-covariance}
\ee
In other words, celestial amplitudes are covariant under the global conformal group $SL(2,\CC)$ of the sphere \cite{Pasterski:2017kqt,Pasterski:2017ylz}. This identification of four-dimensional Lorentz symmetry with $SL(2,\CC)$ covariance is made explicit in \cite{Stieberger:2018onx}.

The same integral transform makes translations less manifest.  Multiplication by an energy in momentum space becomes a shift in conformal dimension, so the total momentum-conservation Ward identity becomes
\be
\sum_{i=1}^n \epsilon_i\,q_i^\mu\,e^{\partial_{\Delta_i}}\,
\widetilde{\calA}_n(\Delta_i,J_i;z_i,\bar z_i)=0.
\label{eq:translation-shift-ward}
\ee
Eq.(\ref{eq:translation-shift-ward}) was derived in this form in \cite{Stieberger:2018onx} and is one of the simplest examples of a celestial difference equation \cite{Pasterski:2017kqt,Lam:2017ofc}.  It expresses exact four-dimensional translation invariance, but in the boost basis translations shift the conformal dimensions rather than acting as local differential operators on the celestial sphere.  This observation foreshadows several later themes: differential and difference equations in Section~\ref{sec:blocks}, infrared logarithms in Section~\ref{sec:uv-ir}, and the comparison with Carrollian variables below.

Momentum conservation also explains why low-point celestial amplitudes often have distributional support.  A four-point massless amplitude in real Lorentzian kinematics depends on a single real cross ratio, and the momentum-conserving delta function localizes the celestial correlator on the physical locus $z=\bar z$ with an ordering appropriate to the scattering channel \cite{Pasterski:2017ylz,Schreiber:2017jsr,Lam:2017ofc}.  Thus celestial amplitudes transform like two-dimensional conformal correlators, but they are not automatically ordinary Euclidean CFT correlators.  This tension is central to Section~\ref{sec:blocks}, where we shall see various approaches to this problem.

\subsection{Carrollian basis}
\label{subsec:null-infinity-carroll}

The celestial basis is not the only natural way to organize flat-space scattering.  Radiation reaches future null infinity $\calI^+$ with coordinates $(u,z,\bar z)$, where $u$ is retarded time.  A basis that keeps $u$ explicit is closer to the geometry of radiative data and to the BMS action on null infinity.  This is the language of Carrollian holography, in which the boundary is a null hypersurface rather than a timelike AdS boundary \cite{Donnay:2022aba,Ciambelli:2018wre,Laddha:2020kvp,Figueroa-OFarrill:2021sxz,Donnay:2022wvx,Mason:2023mti,Donnay:2023mrd,Ruzziconi:2026bix}.
 Null-boundary phase-space analyses make the relation between slicings, news and memory especially explicit \cite{Adami:2021nnf}.

For a massless mode, the half-line Fourier transform
\be
\Psi_J^\epsilon(u,z,\bar z)
=\int_0^\infty \dd\omega\,
e^{-\ii\epsilon\omega u}\,
a_J(\epsilon\omega q(z,\bar z))
\label{eq:carrollian-fourier-mode}
\ee
keeps the retarded time conjugate to the energy.  Here $a_J$ denotes the momentum-space annihilation or creation data, with the sign $\epsilon$ distinguishing outgoing and incoming conventions.  The precise normalization is convention-dependent, but the structural point is universal: Carrollian variables diagonalize time translations at null infinity, while celestial variables diagonalize boosts. In the Carrollian holography literature, the modified Mellin transform \cite{Banerjee:2018gce,Banerjee:2019prz} is commonly used as well. See \cite{Bagchi:2025vri} for details.

The BMS group acts naturally on $(u,z,\bar z)$.  Supertranslations shift $u$ by a function on the sphere, while Lorentz transformations act by conformal transformations of $(z,\bar z)$ accompanied by a conformal rescaling of $u$.  This makes the Carrollian representation especially transparent for memory effects, fluxes and asymptotic charges.  The link between BMS supertranslations, gravitational memory and the leading soft graviton theorem was established in \cite{Strominger:2014pwa}, while spin and center-of-mass memory effects were identified in \cite{Pasterski:2015tva}.  

Beyond scalar singlets, local Carrollian primaries can belong to
reducible but indecomposable multiplets under Carrollian boosts.
Chain and more general net representations, together with the
constraints imposed on their two- and three-point functions by
Carrollian conformal Ward identities, were analyzed in
Ref.~\cite{Chen:2021xkw}. Complementary intrinsic and embedding-space
constructions clarify how Carrollian conformal fields at null infinity
create massless scattering states and how their two- and three-point
correlators are constrained by Poincar\'e symmetry
\cite{Salzer:2023jqv,Nguyen:2023vfz,Nguyen:2023miw}. Explicit
dynamical realizations include conformal Carroll scalar actions with
local boost and Weyl invariance \cite{Baiguera:2022lsw}, as well as
electric and magnetic scalar and electromagnetic theories obtained by
null reduction of Bargmann-invariant parent theories
\cite{Chen:2023pqf}.

\subsection{Celestial and Carrollian connections}
\label{subsec:transform-relation}

The relation between the two bases is an integral transform.  Formally, a celestial operator is obtained from a Carrollian operator by Mellin transforming the energy, or equivalently by integrating the retarded-time operator against a power-law kernel.  Schematically,
\be
\calO_{\Delta,J}^{\epsilon}(z,\bar z)
=\int_0^\infty \dd\omega\,\omega^{\Delta-1}
a_J(\epsilon\omega q)
\sim
\int_{-\infty}^{\infty}\dd u\,
(u-\ii\epsilon 0)^{-\Delta}\,
\Psi_J^\epsilon(u,z,\bar z),
\label{eq:celestial-carrollian-kernel}
\ee
where the $i0$ prescription is tied to the incoming/outgoing sign $\epsilon$, while the overall normalization depends on Fourier-transform conventions.  Eq.(\ref{eq:celestial-carrollian-kernel}) is the practical bridge between the celestial and Carrollian descriptions \cite{Donnay:2022aba,Donnay:2022wvx,Mason:2023mti}.  It explains why the two frameworks often see the same physics in different forms.  A soft theorem that is local in $u$ may become a pole in $\Delta$; a time translation at null infinity becomes a dimension-shift operator in the celestial basis.

The celestial basis is natural for conformal covariance and the OPE language, as we show in the next section. The Carrollian basis is natural for radiative phase space, fluxes and asymptotic symmetry charges.  Recent work on the definition of Carrollian amplitudes in general dimensions sharpens this distinction by formulating the boundary observable directly at null infinity and relating it to momentum-space scattering through a Fourier transform in retarded time \cite{Liu:2024llk,Kulkarni:2025qcx}.

The rest of the review uses both languages.  Sections~\ref{sec:symmetry-ope} and \ref{sec:w-infinity} emphasize the celestial operator algebra obtained from soft and collinear limits.  Sections~\ref{sec:uv-ir} and \ref{sec:ads-flat-limits} repeatedly return to the Carrollian viewpoint because infrared physics, retarded-time dependence and flat limits of AdS/CFT are more naturally formulated at null infinity.  The dictionary developed in this section should therefore be considered as a change of basis rather than a choice between mutually exclusive holographic proposals.

\section{Symmetries and OPEs}
\label{sec:symmetry-ope}

The main theme of the bottom-up approaches in celestial holography is to study the universal features of celestial amplitudes that are well defined. These come from universal limits of scattering amplitudes.  Soft limits reveal asymptotic symmetries, while collinear limits reveal local products of operators on the celestial sphere.  In momentum space, these limits are factorization statements of the $S$-matrix.  In the celestial basis they become Ward identities, current algebras, and OPEs.  This section reviews these structures in the form needed later for conformal blocks, non-distributional amplitudes and $w_{1+\infty}$ symmetries.

\subsection{Soft theorems and conformally soft operators}
\label{subsec:soft-theorems}

The leading soft photon and graviton theorems state that an amplitude with one additional low-energy massless particle factorizes universally \cite{Weinberg:1964ew,Weinberg:1965nx}. 
For a soft photon with momentum \(p_s^\mu=\omega_s q_s^\mu\) and polarization
\(\varepsilon_\mu\), Weinberg's theorem gives
\be
\lim_{\omega_s\rightarrow0}\mathcal A_{n+1}(\omega_s q_s;\{p_i\})
=
\frac{e}{\omega_s}
\sum_{i=1}^n Q_i\,
\frac{p_i\cdot \varepsilon_s}{p_i\cdot q_s}\,
\mathcal A_n(\{p_i\})
+O(\omega_s^0),
\label{eq:soft-photon-theorem}
\ee
while the leading soft graviton theorem is
\be
\lim_{\omega_s\rightarrow0}\mathcal M_{n+1}(\omega_s q_s;\{p_i\})
=
\frac{\kappa}{2\omega_s}
\sum_{i=1}^n
\frac{(p_i\cdot \varepsilon_s)^2}{p_i\cdot q_s}\,
\mathcal M_n(\{p_i\})
+O(\omega_s^0).
\label{eq:soft-graviton-theorem}
\ee
Here $Q_i$ is the charge of the $i$th hard particle, $\kappa^2=32\pi G_N$, and the sign of each external momentum is absorbed into the incoming/outgoing convention.  These equations are standard soft-factorization results \cite{Weinberg:1964ew,Weinberg:1965nx,Strominger:2017zoo,McLoughlin:2022ljp}.  Their importance for celestial holography is that the soft energy dependence becomes a pole in conformal dimension.

Indeed, inserting the Mellin transform Eq.(\ref{eq:celestial-amplitude-definition}), the soft limit $\omega_s\to0$ is controlled by the endpoint of the Mellin integral.  A leading soft behavior $\calA_{n+1}\sim \omega_s^{-1}\calA_n$ produces a pole at $\Delta_s=1$.  The conformally soft operator is therefore defined by the residue at this pole.  For a positive-helicity gluon one writes schematically
\be
J^a(z)=\lim_{\Delta\to1}(\Delta-1)\,\calO^{a,+}_\Delta(z,\bar z),
\label{eq:gluon-current-definition}
\ee
and analogous conformally soft photon and graviton operators are obtained from the corresponding soft poles \cite{Donnay:2018neh,Pate:2019mfs,Adamo:2019ipt,Puhm:2019zbl}.  As we shall see, Eq.(\ref{eq:gluon-current-definition}) explains why soft particles become currents in the celestial CFT.

The subleading soft graviton theorem is similarly related to the celestial stress tensor.  In a conformal basis, the conformally soft graviton mode at $\Delta=0$ generates the two-dimensional stress-tensor Ward identity, while the leading graviton soft mode is associated with supertranslations \cite{Donnay:2018neh,Adamo:2019ipt,Puhm:2019zbl,Donnay:2023mrd}.  This is one of the places where celestial and Carrollian interpretations meet: the same physical theorem can be read as a Ward identity on the sphere or as an asymptotic charge acting at null infinity.

\subsection{Soft currents}
\label{subsec:current-algebra}

The leading soft-gluon theorem implies the celestial current OPE
\be
J^a(z)\calO_{\Delta,J}^{b}(w,\bar w)
\sim \frac{(T^a\calO_{\Delta,J})^b(w,\bar w)}{z-w},
\label{eq:kac-moody-current-ope}
\ee
where $T^a$ acts in the color representation of the hard operator \cite{Pate:2019mfs,Nande:2017dba,Pasterski:2021rjz,Donnay:2023mrd}.  This is the celestial version of a Kac-Moody Ward identity. The current is not introduced as an independent two-dimensional field in a microscopic CFT. It is extracted from a universal feature of the four-dimensional scattering amplitudes.  Its OPE with hard operators is therefore guaranteed by gauge invariance and soft factorization.  The same infrared-symmetry logic was extended to Low's subleading soft photon theorem in QED in \cite{Lysov:2014csa}, providing an early example of how subleading soft factors can also be organized as asymptotic symmetry Ward identities.

For gravity, the subleading soft theorem leads to an operator that acts as a two-dimensional stress tensor on celestial primaries \cite{Kapec:2016jld,Fotopoulos:2019tpe}.  In the corresponding Ward identity one obtains the standard form
\be
T(z)\calO_{h,\bar h}(w,\bar w)
\sim \frac{h}{(z-w)^2}\calO_{h,\bar h}(w,\bar w)
+\frac{1}{z-w}\partial_w\calO_{h,\bar h}(w,\bar w),
\label{eq:celestial-stress-ope}
\ee
as reviewed from the asymptotic-symmetry perspective in \cite{Raclariu:2021zjz,Donnay:2023mrd}.  The equation states that celestial operators transform as Virasoro primaries under the symmetry generated by the subleading soft graviton.  At tree level this provides an elegant explanation of how the Lorentz group can be enhanced to local conformal transformations on the celestial sphere.  

The current-algebra viewpoint is one of the clearest successes of the celestial program.  It turns universal theorems about adding a zero-energy particle to an amplitude into operator statements intrinsic to the celestial sphere.  At the same time, it exposes a major challenge: the currents are often conformally soft limits of scattering states rather than ordinary normalizable states.  A complete celestial theory must explain how these symmetry generators act on the physical Hilbert space and how their Ward identities coexist with infrared divergences.

The extended-BMS analysis of Ref.~\cite{Fotopoulos:2019vac} sharpened this picture by deriving the $TT$ OPE in celestial CFT. In that normalization, the holomorphic stress tensor is the shadow transform of the negative-helicity conformally soft graviton,
\be
   T(z)=\widetilde{\cal O}_{\Delta\to0,-2}(z,\bar z)
   =\frac{3!}{2\pi}\int d^2z'\,\frac{1}{(z'-z)^4}\,
   {\cal O}_{0,-2}(z',\bar z') ,
   \label{eq:fstz-shadow-stress-tensor}
\ee
and the supertranslation current is obtained from the positive-helicity soft graviton by
\be
   P(z)=\partial_{\bar z}{\cal O}_{\Delta\to1,+2}(z,\bar z) .
   \label{eq:fstz-supertranslation-current}
\ee
Historically, this stress-tensor interpretation grew out of the semiclassical Virasoro symmetry of the gravitational S-matrix \cite{Kapec:2014opa} and the construction of a two-dimensional stress tensor for four-dimensional gravity \cite{Kapec:2016jld}.  Sugawara-type constructions on the celestial sphere provide another way to relate current algebras and stress-tensor-like operators \cite{Fan:2020xjj}.  These formulae were derived by inserting the corresponding soft gravitons into celestial amplitudes and using the leading and subleading soft-graviton theorems \cite{Fotopoulos:2019vac}.

The same work assembled all supertranslations into a primary operator on the celestial sphere.  If $\phi^{h,\bar h}$ denotes a celestial primary with $\Delta=h+\bar h$, the supertranslation modes act as
\be
   [P_{n-\frac12,m-\frac12},\phi^{h,\bar h}(z,\bar z)]
   =z^n\bar z^m\,\phi^{h+\frac12,\bar h+\frac12}(z,\bar z),
   \label{eq:fstz-supertranslation-mode-action}
\ee
or, equivalently,
\be
   {\cal P}(w,\bar w)\phi^{h,\bar h}(z,\bar z)
   \sim \frac{1}{(w-z)(\bar w-\bar z)}
   \phi^{h+\frac12,\bar h+\frac12}(z,\bar z).
   \label{eq:fstz-supertranslation-primary-ope}
\ee
Both equations are taken from the extended-BMS celestial algebra of \cite{Fotopoulos:2019vac}.  Their physical meaning is simple but important: a translation multiplies a momentum-space state by the null momentum $q^\mu\omega$, and in the Mellin basis multiplication by $\omega$ shifts the conformal dimension by one.  
With the Virasoro modes $L_n,\bar L_n$ generated by $T,\bar T$, these operators obey
\be
\begin{gathered}
   [P_{k,l},P_{k',l'}]=0,\\
   [L_n,P_{k,l}]=\left(\frac{n}{2}-k\right)P_{n+k,l},\qquad
   [\bar L_n,P_{k,l}]=\left(\frac{n}{2}-l\right)P_{k,n+l}.
\end{gathered}
\label{eq:fstz-extended-bms-algebra}
\ee
This is a concrete celestial realization of the extended $\mathfrak{bms}_4$ algebra: local conformal transformations act on the angular dependence, while supertranslations shift the boost weight.  It also clarifies why the BMS algebra is more than the ordinary global Poincare algebra written in a conformal basis.

The supersymmetric extension gives an instructive further test of the same logic.  In ${\cal N}=1$ Einstein-Yang-Mills theory the conformal primary wave functions can be organized into supermultiplets, and the soft gravitino theorem produces fermionic celestial currents \cite{Fotopoulos:2020bqj}.  The holomorphic and antiholomorphic supercurrents are defined by the conformally soft gravitino limits
\begin{align}
   S(z)&=\lim_{\Delta\to\frac12}\frac{\Delta-\frac12}{\pi}
   \int d^2z'\,\frac{1}{(z-z')^3}{\cal O}_{\Delta,-\frac32}(z',\bar z'),
   \qquad \nonumber\\
   \bar S(\bar z)&=\lim_{\Delta\to\frac12}\frac{\Delta-\frac12}{\pi}
   \int d^2z'\,\frac{1}{(\bar z-\bar z')^3}{\cal O}_{\Delta,+\frac32}(z',\bar z') .
   \label{eq:fstz-supercurrents}
\end{align}
Here the pole at $\Delta=1/2$ is the Mellin-space image of the leading soft-gravitino factor.  The resulting Ward identities can be summarized by the OPEs
\be
   S(z){\cal O}_{\Delta,\ell^c}(w,\bar w)
   \sim \frac{1}{z-w}{\cal O}_{\Delta+\frac12,\ell}(w,\bar w),
   \qquad
   \bar S(\bar z){\cal O}_{\Delta,\ell}(w,\bar w)
   \sim \frac{1}{\bar z-\bar w}{\cal O}_{\Delta+\frac12,\ell^c}(w,\bar w),
   \label{eq:fstz-supercurrent-hard-ope}
\ee
where $\ell$ and $\ell^c$ label the two members of the relevant supermultiplet \cite{Fotopoulos:2020bqj}.  The supercurrent therefore shifts the conformal dimension by one half and changes the helicity within the multiplet, exactly as a four-dimensional supersymmetry charge should.

Expanding $S$ and $\bar S$ in modes,
\be
   S(z)=\sum_{n\in\mathbb Z+\frac12}\frac{G_n}{z^{n+\frac32}},
   \qquad
   \bar S(\bar z)=\sum_{n\in\mathbb Z+\frac12}\frac{\bar G_n}{\bar z^{n+\frac32}},
   \label{eq:fstz-supercurrent-modes}
\ee
one obtains the super-BMS relation
\be
   \{G_n,\bar G_m\}=P_{n,m},\qquad
   \{G_n,G_m\}=\{\bar G_n,\bar G_m\}=0,
   \qquad
   [L_m,G_n]=\left(\frac{m}{2}-n\right)G_{m+n},
   \label{eq:fstz-super-bms-algebra}
\ee
together with the antiholomorphic counterpart for $\bar G_n$ \cite{Fotopoulos:2020bqj}.  This is the celestial version of the statement that the anticommutator of supersymmetry charges gives translations.  It strengthens the interpretation of the soft sector as an operator algebra on the celestial sphere: gluon, graviton and gravitino soft theorems become, respectively, Kac-Moody, Virasoro/BMS and super-BMS Ward identities. Unlike in superstring theory, however, the two-dimensional language here does
not refer to a worldsheet supersymmetry; it refers to the organization of
four-dimensional asymptotic symmetry data on the celestial sphere.

The same super-BMS structure can also be approached intrinsically from
Carrollian symmetry rather than inferred from soft limits.  An
infinite-dimensional super-BMS algebra obtained from Carrollian
superconformal symmetry was developed in Ref.~\cite{Bagchi:2022owq},
while explicit electric and magnetic Carrollian fermions and
off-shell ${\cal N}=1$ super-Carrollian multiplets were constructed in
Ref.~\cite{Koutrolikos:2023evq}.  A systematic algebraic classification
subsequently identified singlet and multiplet chiral super-BMS$_4$
extensions \cite{Zheng:2025cuw}.  This analysis was refined into
electric and magnetic families with free-field realizations, for which
the supercharge anticommutators close on supertranslations in the
electric case but can involve superrotations in the magnetic case
\cite{Zheng:2025rfe}.  These results complement the celestial-current
construction above by distinguishing the universal asymptotic algebra
from the additional representation-theoretic and dynamical input
required to realize it in a Carrollian field theory.

\subsection{Collinear limits and celestial OPEs}
\label{subsec:collinear-ope}

The second universal limit is the collinear limit.  In four-dimensional amplitudes, when two massless particles become collinear the amplitude factorizes into a splitting function times a lower-point amplitude.  Since collinearity is the short-distance limit $z_{12},\bar z_{12}\to0$ on the celestial sphere, this factorization becomes an OPE of celestial CFT.

In the holomorphic collinear limit \(z_{12}\to0\) with \(\bar z_1,\bar z_2\) held fixed, adjacent gluons in a color-ordered partial amplitude obey the standard factorization formula \cite{Pate:2019lpp}
\be
\begin{split}
  A_{s_1\cdots s_n}(p_1,\ldots,p_i,p_j,\ldots,p_n)
  \ \longrightarrow\
  \sum_{s=\pm1}
  {\rm Split}^{s}_{s_i s_j}(p_i,p_j)\,
  A_{s_1\cdots s\cdots s_n}(p_1,\ldots,P,\ldots,p_n),
\end{split}
\label{eq:gluon-collinear-splitting}
\ee
where \(P^\mu=p_i^\mu+p_j^\mu\) and \(\omega_P=\omega_i+\omega_j\) at leading order in the collinear expansion.  The nonzero holomorphic splitting functions relevant for the leading celestial OPEs are
\be
  {\rm Split}^{+}_{++}(p_i,p_j)
  =
  \frac{1}{z_{ij}}\frac{\omega_P}{\omega_i\omega_j},
  \qquad
  {\rm Split}^{-}_{+-}(p_i,p_j)
  =
  \frac{1}{z_{ij}}\frac{\omega_j}{\omega_i\omega_P}.
\label{eq:gluon-splitting-functions}
\ee
The first of these is the momentum-space origin of the \(O^+O^+\) celestial OPE below.  After setting \(\omega_i=t\omega_P\), \(\omega_j=(1-t)\omega_P\), the energy-fraction integral produces the Euler beta function.

For two positive-helicity gluons one obtains the leading celestial OPE
\be
\calO^{a,+}_{\Delta_1}(z_1,\bar z_1)\,
\calO^{b,+}_{\Delta_2}(z_2,\bar z_2)
\sim
\frac{f^{ab}{}_{c}}{z_{12}}\,
B(\Delta_1-1,\Delta_2-1)\,
\calO^{c,+}_{\Delta_1+\Delta_2-1}(z_2,\bar z_2),
\label{eq:gluon-celestial-ope}
\ee
up to the overall phase convention used for color-ordered amplitudes.  Here $B(x,y)=\Gamma(x)\Gamma(y)/\Gamma(x+y)$ is the Euler beta function.  It arises from Mellin transforming the energy fraction in the collinear splitting process.  The shift $\Delta_1+\Delta_2-1$ reflects the energy scaling of the Yang-Mills three-point coupling.  This OPE was derived in the early celestial OPE literature \cite{Fan:2019emx,Pate:2019lpp,Fotopoulos:2019vac}.

The gravitational collinear limit is similar in structure but differs in its energy scaling and angular dependence.  For an \(n\)-graviton tree amplitude one has \cite{Pate:2019lpp}
\be
\begin{split}
  \mathcal A_{s_1\cdots s_n}(p_1,\ldots,p_n)
  \ \longrightarrow\
  \sum_{s=\pm2}
  {\rm Split}^{s}_{s_i s_j}(p_i,p_j)\,
  \mathcal A_{s_1\cdots s\cdots s_n}
  (p_1,\ldots,P,\ldots,p_n),
\end{split}
\label{eq:graviton-collinear-splitting}
\ee
with
\be
  {\rm Split}^{+2}_{+2,+2}(p_i,p_j)
  =
  -\frac{\kappa}{2}\,
  \frac{\bar z_{ij}}{z_{ij}}\,
  \frac{\omega_P^2}{\omega_i\omega_j},
  \qquad
  {\rm Split}^{-2}_{+2,-2}(p_i,p_j)
  =
  -\frac{\kappa}{2}\,
  \frac{\bar z_{ij}}{z_{ij}}\,
  \frac{\omega_j^3}{\omega_i\omega_P^2}.
\label{eq:graviton-splitting-functions}
\ee
The factor \(\bar z_{ij}/z_{ij}\) is the characteristic holomorphic collinear singularity of gravity in these variables.

The same analysis applies in gravity.  For two positive-helicity gravitons the leading OPE takes the form
\be
G^+_{\Delta_1}(z_1,\bar z_1)\,
G^+_{\Delta_2}(z_2,\bar z_2)
\sim
-\frac{\kappa}{2}\,
\frac{\bar z_{12}}{z_{12}}\,
B(\Delta_1-1,\Delta_2-1)\,
G^+_{\Delta_1+\Delta_2}(z_2,\bar z_2).
\label{eq:graviton-celestial-ope}
\ee
This formula, first obtained as part of the celestial OPE analysis of gluons and gravitons \cite{Pate:2019lpp}, is the starting point for later derivations of the celestial $w_{1+\infty}$ algebra \cite{Guevara:2021abz,Himwich:2021dau}.  The factor $\bar z_{12}/z_{12}$ is the angular form of the gravitational collinear singularity, while the absence of the Yang-Mills dimension shift by $-1$ reflects the different energy scaling of the gravitational coupling.

Eqs.(\ref{eq:gluon-celestial-ope}) and (\ref{eq:graviton-celestial-ope}) should be read as leading OPE statements.  Descendants and subleading terms can be organized systematically, for example by BMS symmetry or by worldsheet and ambitwistor methods \cite{Banerjee:2020kaa,Adamo:2021zpw}.  In particular, the first subleading term in the positive-helicity graviton OPE can be expressed in terms of BMS descendants whose coefficients are fixed by the leading OPE coefficient and the BMS algebra \cite{Banerjee:2020kaa}.  The leading terms already contain the key physical message: celestial OPE coefficients are Mellin transforms of four-dimensional splitting functions.

Soft currents and collinear OPEs are not independent structures.  Taking conformally soft limits of celestial OPEs produces current algebras, and taking further mode expansions produces infinite-dimensional symmetry algebras.  In gauge theory this gives Kac-Moody-type structures associated with large gauge transformations.  In gravity it gives the BMS algebra and, in special soft sectors, the $w_{1+\infty}$ algebra of gravitons discussed in Section~\ref{sec:w-infinity}.  Phase-space and gauge-fixing analyses give a complementary derivation of these algebras directly from null-boundary data and boundary conditions \cite{Ruzziconi:2019pzd,Freidel:2021fxf,Donnay:2021wrk}.  A related bulk-reduction approach constructs scalar and gravitational theories directly at future null infinity, identifies Poincare fluxes and their supertranslation and superrotation extensions, and in general dimensions realizes Carrollian diffeomorphisms through quantum flux operators \cite{Liu:2022mne,Liu:2023gwa,Li:2023xrr}.  Double-soft analyses and nested soft limits further clarify which extensions survive as genuine charge algebras \cite{Distler:2018rwu,Freidel:2021dfs}.  For a systematic celestial-CFT treatment of these asymptotic symmetries, see \cite{Donnay:2020guq}.

The balanced interpretation is the following.  Asymptotic symmetries provide universal constraints on the flat-space $S$-matrix, and the Mellin basis turns these constraints into local-looking operator statements on the celestial sphere.  But the resulting celestial theory is not yet a conventional two-dimensional CFT with a completely known spectrum and state space.  The operator algebra is fixed in universal soft and collinear regimes, while its global completion is constrained by analyticity, unitarity, infrared physics and the distributional support of amplitudes.  These issues motivate the block and analytic-correlator discussion of the next section.

Beyond the single-particle primaries produced directly by the external legs of an \(S\)-matrix element, the flat hologram should also contain multiparticle primaries, multiparticle OPE channels and OPE blocks.  This broader viewpoint is supported from several complementary directions: multicollinear limits reveal multiparticle terms and possible obstructions to naive double-residue associativity, explicit two-particle celestial operators and their OPE coefficients can be extracted from multiparticle factorization channels, and boundary constructions of composite celestial operators clarify how these structures fit into a CFT-like operator algebra \cite{Ball:2023sdz,Guevara:2024ixn,Calkins:2026hpg}.  Related work constructs such composite primaries from both Carrollian and celestial viewpoints, emphasizing the compatibility of principal-series data, highest-weight conditions and unitarity in the celestial OPE \cite{Kulp:2024scx}.

\section{Blocks and analytic celestial amplitudes}
\label{sec:blocks}

The preceding section reviewed the universal soft and collinear limits of celestial amplitudes.  The present section asks a different question: once these local limits are known, can one organize the full dependence on celestial cross ratios in a way analogous to conformal field theory?  We first recall the low-point examples where the central difficulty is already visible, and then turn to block decompositions, differential equations and non-distributional completions.

\subsection{Three- and four-point celestial amplitudes}
\label{subsec:low-point-celestial-amplitudes}

The first explicit low-point gluon amplitudes in the celestial basis make the central analytic tension very clear.  They transform with the covariance of two-dimensional spin-one correlators, but four-dimensional momentum conservation forces them to have distributional support \cite{Pasterski:2017ylz}.  Further low-point examples, including amplitudes in gauge theory, gravity and string theory, were worked out in Ref.~\cite{Stieberger:2018edy}.
The momentum-space input for the gluon examples below is the color-ordered MHV tree amplitude.  For negative-helicity gluons \(1,2\), the Parke-Taylor formula reads, up to the overall coupling and phase convention \cite{Parke:1986gb},
\be
  A_n^{\rm MHV}(1^-,2^-,3^+,\ldots,n^+)
  =
  \frac{\langle12\rangle^4}
  {\langle12\rangle\langle23\rangle\cdots\langle n1\rangle},
  \qquad
  \langle ij\rangle=\lambda_i^\alpha\lambda_{j\alpha}.
\label{eq:parke-taylor-mhv}
\ee
The celestial amplitudes are obtained by Mellin transforming this rational spinor-helicity amplitude together with the momentum-conserving delta function.  The latter is responsible for the special support exhibited by the three- and four-point formulas.

We now recall the three- and four-point examples of Ref.~\cite{Pasterski:2017ylz}.  At tree level, the gluon amplitudes considered there are conformally invariant.  Writing \(\omega_i=s\sigma_i\), with \(\sigma_i\in[0,1]\) and \(\sum_i\sigma_i=1\), the overall scale integral gives
\be
  \int_0^\infty ds\,s^{-1+\ii\sum_i\lambda_i}
  =2\pi\,\delta\!\left(\sum_i\lambda_i\right),
\label{eq:pss-scale-delta}
\ee
while the remaining simplex integrals contain the momentum-conservation constraint and the simplex constraint.  For \(n\leq5\) these constraints localize the simplex variables,
\be
  \delta^{(4)}\!\left(\sum_i\epsilon_i\sigma_i q_i\right)
  \delta\!\left(\sum_i\sigma_i-1\right)
  =
  C(z_i,\bar z_i)\prod_i\delta(\sigma_i-\sigma_{*i}),
\label{eq:pss-simplex-localization}
\ee
where \(C(z_i,\bar z_i)\) contains the residual angular support \cite{Pasterski:2017ylz}.  The channel-dependent support of these constraints, and its organization into celestial kinematic patches, was studied geometrically in Ref.~\cite{Mizera:2022sln}.  

Consider first the split-signature three-point MHV branch.  On the support \(z_{ij}\neq0\), the constraints become
\be
\begin{split}
  \delta^{(4)}\!\left(\sum_{i=1}^3\epsilon_i\sigma_i q_i\right)
  \delta\!\left(\sum_{i=1}^3\sigma_i-1\right)
  =&\,
  \frac{\delta(\bar z_{12})\delta(\bar z_{13})}
       {4\sigma_1\sigma_2\sigma_3D_3^2}
  \delta\!\left(\sigma_1-\frac{z_{23}}{D_3}\right)
  \\
  &\times
  \delta\!\left(\sigma_2+\epsilon_1\epsilon_2\frac{z_{13}}{D_3}\right)
  \delta\!\left(\sigma_3-\epsilon_1\epsilon_3\frac{z_{12}}{D_3}\right),
\end{split}
\label{eq:pss-three-point-delta-localization}
\ee
with
\be
  D_3=(1-\epsilon_1\epsilon_2)z_{13}
  +(\epsilon_1\epsilon_3-1)z_{12}.
\label{eq:pss-d3-definition}
\ee
Thus the MHV three-point correlator is supported on \(\bar z_{12}=\bar z_{13}=0\); the anti-MHV branch is obtained by exchanging \(z\leftrightarrow\bar z\).  Inserting the color-ordered MHV numerator with helicities \((1^-,2^-,3^+)\) then gives
\be
  \widetilde{\calA}_{--+}(\lambda_i;z_i,\bar z_i)
  =
  -\pi\,\delta\!\left(\sum_i\lambda_i\right)\,
  \frac{
  \mathrm{sgn}(z_{12}z_{23}z_{31})\,
  \delta(\bar z_{13})\delta(\bar z_{12})
  }{
  |z_{12}|^{-1-\ii\lambda_3}
  |z_{23}|^{1-\ii\lambda_1}
  |z_{13}|^{1-\ii\lambda_2}
  }
  \prod_{i=1}^{3}\mathbf{1}_{[0,1]}(\sigma_{*i}) .
\label{eq:pss-three-gluon-review}
\ee
Here \(\Delta_i=1+\ii\lambda_i\), \(z_i,\bar z_i\in\mathbb R\) are independent in \((-+-+)\) signature, and the \(\sigma_{*i}\) are the solutions fixed by Eq.(\ref{eq:pss-three-point-delta-localization}).  The factor \(\prod_i\mathbf 1_{[0,1]}(\sigma_{*i})\) is the remnant of the simplex domain and selects the allowed crossing channel.  The anti-MHV formula is obtained by exchanging \(z\leftrightarrow\bar z\) and the helicities.

At four points the same logic leads to a different kind of support.  Momentum conservation no longer forces collinearity, but it localizes the celestial correlator on the real cross-ratio locus.  The simplex constraints give
\be
\begin{split}
  &\delta^{(4)}\!\left(\sum_{i=1}^4\epsilon_i\sigma_i q_i\right)
  \delta\!\left(\sum_{i=1}^4\sigma_i-1\right)
  =
  \frac14
  \delta\!\left(
  \left|
  z_{12}z_{34}\bar z_{13}\bar z_{24}
  -z_{13}z_{24}\bar z_{12}\bar z_{34}
  \right|\right)
  \prod_{i=1}^4\delta(\sigma_i-\sigma_{*i}),
  \\
  &\sigma_{*1}=-\frac{\epsilon_1\epsilon_4}{D_4}
  \frac{z_{24}\bar z_{34}}{z_{12}\bar z_{13}},
  \qquad
  \sigma_{*2}=\frac{\epsilon_2\epsilon_4}{D_4}
  \frac{z_{34}\bar z_{14}}{z_{23}\bar z_{12}},
  \\
  &\sigma_{*3}=-\frac{\epsilon_3\epsilon_4}{D_4}
  \frac{z_{24}\bar z_{14}}{z_{23}\bar z_{13}},
  \qquad
  \sigma_{*4}=\frac1{D_4},
\end{split}
\label{eq:pss-four-point-delta-localization}
\ee
where
\be
  D_4=(1-\epsilon_1\epsilon_4)
  \frac{z_{24}\bar z_{34}}{z_{12}\bar z_{13}}
  +(\epsilon_2\epsilon_4-1)
  \frac{z_{34}\bar z_{14}}{z_{23}\bar z_{12}}
  +(1-\epsilon_3\epsilon_4)
  \frac{z_{24}\bar z_{14}}{z_{23}\bar z_{13}} .
\label{eq:pss-d4-definition}
\ee
The first delta function is equivalently the condition \(z=\bar z\), up to the standard nonzero Jacobian factor, where
\[
  z=\frac{z_{12}z_{34}}{z_{13}z_{24}},
  \qquad
  \bar z=\frac{\bar z_{12}\bar z_{34}}{\bar z_{13}\bar z_{24}} .
\]
For the Lorentzian-signature MHV amplitude with helicities \((1^-,2^-,3^+,4^+)\), this gives
\be
\begin{split}
  \widetilde{\calA}_{--++}(\lambda_i;z_i,\bar z_i)
  =&
  -\frac{\pi}{4}\,
  \delta\!\left(\sum_{k=1}^{4}\lambda_k\right)
  \delta\!\left(\frac{|z-\bar z|}{2}\right)
  \left[
  \prod_{i<j}^{4}
  z_{ij}^{h/3-h_i-h_j}
  \bar z_{ij}^{\bar h/3-\bar h_i-\bar h_j}
  \right]
  \\
  &\times
  z^{5/3}(1-z)^{-1/3}
  \prod_{i=1}^{4}\mathbf{1}_{[0,1]}(\sigma_{*i}) .
\end{split}
\label{eq:pss-four-gluon-review}
\ee
Here \(h=\sum_i h_i\), \(\bar h=\sum_i\bar h_i\), and the weights are \(h_{1,2}=\ii\lambda_{1,2}/2\), \(h_{3,4}=1+\ii\lambda_{3,4}/2\), \(\bar h_{1,2}=1+\ii\lambda_{1,2}/2\), \(\bar h_{3,4}=\ii\lambda_{3,4}/2\).  Eq.(\ref{eq:pss-four-point-delta-localization}) explains both the factor \(\delta(|z-\bar z|/2)\) and the channel-dependent indicator functions in Eq.(\ref{eq:pss-four-gluon-review}).

The lesson of Eqs.(\ref{eq:pss-three-gluon-review}) and (\ref{eq:pss-four-gluon-review}) is that global conformal covariance is not the same as being an ordinary smooth Euclidean CFT correlator.  The three-point answer has collinear support, while the four-point answer is localized on the real cross-ratio locus \(z=\bar z\).  The rest of this section therefore has two intertwined analytic tasks: to identify the conformal representations that appear in block or partial-wave decompositions, and to find observables for which these decompositions act on sufficiently regular functions of celestial coordinates.  We review conformal-block constructions \cite{Fan:2021isc,Nandan:2019jas,Atanasov:2021cje,Fan:2021pbp,Guevara:2021tvr,Surubaru:2025qhs}, non-distributional constructions \cite{Fan:2022vbz,Stieberger:2022zyk}, and later developments involving split representations, scalar blocks, leaf amplitudes, soft-algebra actions and positive geometries \cite{Chang:2023ttm,Fan:2023lky,Pacifico:2025emk,Melton:2023bjw,Melton:2024jyq,Duary:2024cqb,Mandal:2024zao,Ball:2023ukj,Dong:2025qiv}.

\subsection{Conformal blocks}

In a CFT, a conformal block packages one exchanged primary together with all of its descendants.  The block expansion therefore separates kinematics, fixed by conformal symmetry, from dynamical data, namely the spectrum and OPE coefficients.  In celestial holography this gives a precise question: can the factorization channels of the four-dimensional S-matrix be reorganized as exchanges of celestial conformal families?  Related work on Poincare constraints, ambidextrous light transforms, celestial diamonds and state-operator maps clarifies what kind of representation space such decompositions act on \cite{Law:2019glh,Sharma:2021gcz,Pasterski:2021fjn,Crawley:2021ivb,Cardona:2017keg}.

For four massless insertions, global conformal covariance fixes the coordinate dependence up to a reduced correlator of the cross ratios \(z,\bar z\).  A celestial amplitude therefore takes the general form
\be
  \widetilde{\calA}_4(\Delta_i,J_i;z_i,\bar z_i)
  = \mathcal{K}_{h_i,\bar h_i}(z_i,\bar z_i)\,
    \mathcal{G}_{\Delta_i,J_i}(z,\bar z) ,
\label{eq:four-point-covariance-review-expanded}
\ee
where $h_i=(\Delta_i+J_i)/2$, $\bar h_i=(\Delta_i-J_i)/2$, and $\mathcal{K}_{h_i,\bar h_i}$ is the standard kinematic prefactor fixed by global conformal symmetry. The dynamical information is in $\mathcal{G}$. In an ordinary two-dimensional CFT one would expand this function in conformal blocks,
\be
  \mathcal{G}(z,\bar z)
  = \sum_{\mathcal{O}} C_{12\mathcal{O}}C_{34\mathcal{O}}\,
    G_{h,\bar h}^{h_i,\bar h_i}(z,\bar z) .
\label{eq:ordinary-block-expansion-review}
\ee
The coefficients are products of OPE coefficients, while the block $G_{h,\bar h}$ sums the descendants of a primary operator of weights $(h,\bar h)$. The physical meaning is clear: the four-point function is decomposed into all possible intermediate conformal families exchanged in a chosen OPE channel. Equivalently, conformal symmetry implies that each block is an eigenfunction of the quadratic conformal Casimir. In the celestial conformal-block analysis of ref.~\cite{Atanasov:2021cje}, the relevant $13$--$24$ channel equation is written as
\be
  \left(\mathcal D_z+\mathcal D_{\bar z}\right) k_{h,\bar h}(z,\bar z)
  =
  \bigl[h(h-1)+\bar h(\bar h-1)\bigr] k_{h,\bar h}(z,\bar z),
\label{eq:celestial-casimir-equation}
\ee
where $k_{h,\bar h}$ denotes the global block and
\be
  \mathcal D_z
  =
  z^2(1-z)\partial_z^2
  -\left(1-h_{13}+h_{24}\right)z^2\partial_z
  +h_{13}h_{24}z,\qquad h_{ij}=h_i-h_j,
\ee
with an analogous expression for $\mathcal D_{\bar z}$. This equation is the representation-theoretic reason why conformal blocks are the natural functions in which to expand celestial four-point amplitudes. The nontrivial physics lies in the spectrum of exchanged representations and in the spectral density multiplying these eigenfunctions \cite{Fan:2021isc,Atanasov:2021cje}.  In this sense, conformal blocks give the global completion of local collinear OPE data and descendant towers \cite{Ebert:2020nqf}.

Celestial correlators require one further step.  Since the external states usually lie on the principal series and the exchanged data need not form a discrete Euclidean spectrum, the natural starting point is a conformal partial-wave expansion.  For massive spinning bosons, massive celestial fermions and Dirac conformal primaries, this first requires choosing the appropriate conformal-primary basis before extracting OPE data \cite{Law:2020tsg,Narayanan:2020amh,Iacobacci:2020por}.  One then asks whether the spectral contour can be deformed into a sum or integral over conformal blocks.  Schematically \cite{Nandan:2019jas},
\be
  \widetilde{\calA}_4
  =
  \sum_{J\in\mathbb Z}\int_{-\infty}^{+\infty}\frac{\dd\nu}{2\pi}\,
  \rho(\nu,J)\,\Psi_{\nu,J}(z_i,\bar z_i),
\label{eq:partial-wave-review-expanded}
\ee
Here $\Psi_{\nu,J}$ is a conformal partial wave, built from a block and its shadow, and the exchanged weights may be parameterized as
\be
  h=\frac{1+J}{2}+\frac{\ii\nu}{2},
  \qquad
  \bar h=\frac{1-J}{2}+\frac{\ii\nu}{2} .
\label{eq:principal-series-exchange-review-expanded}
\ee
The density $\rho(\nu,J)$ is the celestial analogue of a partial-wave amplitude. Its poles and residues should encode the exchanged four-dimensional physics.

A concrete example is provided by ref.~\cite{Atanasov:2021cje}.  Starting from a four-point celestial amplitude of massless scalars, one decomposes the correlator in the $13$--$24$ channel and then takes the infinite-mass limit of a massive exchange to obtain the contact amplitude.  In the notation of that analysis,
\be
\begin{split}
  {\cal A}_4^c
  = & I_{13-24}(z_i,\bar z_i)\,
  \frac{2}{\sin(\pi\beta/2)}
  \lim_{m\to\infty}
  \Bigg[
  \frac12\int_{-\ii\infty}^{\ii\infty}\dd\alpha\,
  c_{13\alpha}^{L}c_{24\alpha}^{L}D^{L,\alpha\alpha}\,
  k_{\frac12+\alpha}(z)k_{\frac12-\alpha}(\bar z)
  \\
  &+\sum_{n=0}^{\infty}
  c_{13n}c_{24n}D^{nn}
  \cos\pi\!\left(h_{13}+\frac n2\right)
  \cos\pi\!\left(h_{24}+\frac n2\right)
  k_{\frac{1+n}{2}}(z)k_{\frac{1+n}{2}}(\bar z)
  \Bigg],
\end{split}
\label{eq:contact-block-review}
\ee
with $\beta=\sum_i(\Delta_i-1)$ and $I_{13-24}$ the conformally covariant prefactor appropriate to the channel. The coefficients are fixed by celestial three- and two-point data,
\be
\begin{split}
  c_{ijn}
  &=
  \sqrt{\lambda}\,
  \frac{m^{\Delta_i+\Delta_j-3}}{2^{\Delta_i+\Delta_j}}\,
  B\!\left(\frac{1+n}{2}+h_{ij},\frac{1+n}{2}-h_{ij}\right),
  \\
  c_{ij\alpha}^{L}
  &=
  -\pi\ii\,\frac{\sqrt{\lambda}}{m^3}
  \left(\frac m2\right)^{2h_i+2h_j}
  \frac1\alpha,
  \qquad
  D^{nn}=m^2\frac{n}{2\pi},
  \qquad
  D^{L,\alpha\alpha}=-\frac{m^2\alpha^2}{\pi^2\ii}.
\end{split}
\label{eq:contact-block-coefficients}
\ee
This formula shows that even the celestial transform of a contact interaction can have an exchange expansion: the contact amplitude is reconstructed from a discrete tower of integer-dimension massive scalar blocks together with a continuum of light-transform blocks.  The coefficients are built from the same beta-function OPE data that appear in celestial three-point amplitudes, and in the contact limit the massive cubic coupling and quartic coupling obey $g=m\sqrt{\lambda/3}$ \cite{Atanasov:2021cje}.  Thus celestial blocks can reveal exchanged conformal families even when the underlying momentum-space process looks point-like.

A complementary solvable example is provided by the large-\(N\) \(O(N)\) sigma model, where the all-loop celestial four-point function admits conformal partial-wave and conformal-block decompositions whose analytic structure encodes metastable resonances, factorization and a celestial version of unitarity \cite{Garcia-Sepulveda:2022lga}.  This example is useful because it shows how block decompositions can organize not only tree-level factorization, but also the resonance structure and imaginary parts required by unitarity.

\subsection{Distributional support and the shadow transform}

As we have seen in Eqs.(\ref{eq:pss-three-gluon-review}) and (\ref{eq:pss-four-gluon-review}), the amplitudes are globally conformally covariant, but the original plane-wave \(S\)-matrix still contains the four-dimensional momentum-conserving delta function.  After the Mellin transform this constraint becomes support on special loci of celestial kinematics: for three massless particles it enforces collinearity in split signature, while for four massless particles in Lorentzian signature it localizes the reduced correlator on the real cross-ratio locus \(z=\bar z\), with additional channel-dependent support from the allowed energy fractions. 

This creates the main challenge of using standard CFT bootstrap methods in celestial amplitudes: the object of interest is a distribution rather than a smooth Euclidean correlator.  The conformal-block analysis of celestial gluon amplitudes in ref.~\cite{Fan:2021isc} gives a useful way to see both the problem and a partial cure.  Its single-valued continuation and later shadow-amplitude/OPE analyses clarify that the shadow transform changes the analytic class in which the OPE and block expansion are formulated \cite{Fan:2021pbp,Chang:2022jut,Himwich:2025bza}.  More recently, the shadow-completion analysis of celestial OPEs sharpened this point by arguing that OPE closure requires both the Mellin-basis exchange and the shadow representative of the same exchanged bulk particle, with the shadow OPE coefficient fixed by a universal shadow factor \cite{Liu:2026yoh}.  The basic operation is to shadow-transform one external operator.  If the fourth operator has weights $(h_4,\bar h_4)$, then schematically
\be
  \widetilde{\calO}_{4}(z_4,\bar z_4)
  = \int \dd^2 w\,
  (z_4-w)^{-2(1-h_4)}(\bar z_4-\bar w)^{-2(1-\bar h_4)}
  \calO_4(w,\bar w) .
\label{eq:shadow-transform-review}
\ee
The shadow transform trades a primary of weights $(h,\bar h)$ for one of weights $(1-h,1-\bar h)$ while preserving covariance.  In the celestial application it also smears the kinematic delta function, making the block expansion less singular.

In channels compatible with both the two-dimensional OPE channel and the four-dimensional scattering channel, the exchanged conformal families have dimensions of the form
\be
  \Delta = 2+M+\ii\lambda,
  \qquad M=0,1,2,\ldots,
\label{eq:compatible-spectrum-review}
\ee
with integer spins
\be
  J=-M,-M+2,\ldots,M-2,M .
\label{eq:compatible-spin-review}
\ee
These exchanges are the group-theoretic representations allowed by the tensor product of the two external color-adjoint gluon operators.  In incompatible channels the decomposition is more exotic: one encounters continuous complex spin and positive integer dimensions.  The shadow transform therefore plays two roles.  Mathematically, it converts the singular four-point object into a form amenable to block technology.  Physically, it exposes which two-dimensional OPE channels are aligned with four-dimensional factorization.

It is useful to make one qualification explicit.  Eq.(\ref{eq:ordinary-block-expansion-review}) should not be read as saying that celestial amplitudes are ordinary Euclidean CFT correlators.  Celestial correlators have two-dimensional conformal covariance, but their external states are boost eigenstates, their dimensions typically lie on the principal series, and their correlators inherit momentum-conservation constraints from the four-dimensional S-matrix.  These differences show up sharply in the distinction between compatible and incompatible channels in the gluon analysis of refs.~\cite{Fan:2021isc,Fan:2021pbp}; they also motivate shadow, ambidextrous and integer bases, in which locality, crossing or discreteness is emphasized differently \cite{Chang:2022jut,Jorge-Diaz:2022dmy,Cotler:2023qwh}.  A pole in $\rho(\nu,J)$ should therefore not automatically be interpreted as an independent local operator of a microscopic Euclidean CFT.  It may represent an exchanged bulk particle, a light-ray transform of four-dimensional data, a shadow contribution or a residue produced by analytic continuation.  With this qualification, conformal blocks remain powerful: they identify exchanged representations, organize collinear limits and make precise which part of four-dimensional factorization appears as celestial OPE data.

\subsection{Differential equations}

A complementary way to probe celestial analyticity is to ask whether amplitudes obey useful differential equations in celestial variables.  The Mellin transform converts powers of energies into shifts of conformal dimensions, while momentum-space homogeneity and symmetry constraints become differential or difference operators.  Soft and collinear limits therefore become Ward identities, current-algebra constraints, null-state relations and recursion relations, a viewpoint sharpened by celestial recursion and all-order OPE constraints in the MHV sector \cite{Hu:2022bpa,Adamo:2022wjo,Ren:2023trv,Banerjee:2023bni}.  These equations provide a bridge between block decompositions and the original scattering problem: they remember translation invariance, little-group scaling, soft-current Ward identities and collinear factorization at the same time.

For tree-level MHV gluon amplitudes in pure Yang--Mills theory, one obtains a system of first-order equations which closely resemble Knizhnik--Zamolodchikov equations, but with an additional term tied to the subleading soft gluon current algebra \cite{Banerjee:2020vnt}.  In the notation used throughout this review, the result may be written as
\begin{align}
0=&\bigg[
\frac{C_A}{2}\,\partial_{z_i}
-h_i\sum_{j\ne i}\frac{T_i^aT_j^a}{z_i-z_j}
\nonumber\\
&\quad
+\frac{1}{2}\sum_{j\ne i}
\frac{\epsilon_j\left(2\bar h_j-1-(\bar z_i-\bar z_j)\partial_{\bar z_j}\right)}{z_i-z_j}
\,T_j^a P_j^{-1}T_i^aP_{-1,-1}(i)
\bigg]
\left\langle\prod_{k=1}^n
\mathcal O^{a_k}_{h_k,\bar h_k}(z_k,\bar z_k)
\right\rangle_{\rm MHV}.
\label{eq:gluon-mhv-differential-equation}
\end{align}
Here the equation is imposed for each positive-helicity gluon labelled by \(i\).  The constant \(C_A\) is the adjoint Casimir of the gauge group, \(T_i^a\) acts on the color index of the \(i\)-th gluon operator, and \(\epsilon_j=\pm1\) distinguishes outgoing and incoming particles.  The operators \(P_j^{-1}\) and \(P_{-1,-1}(i)\) are Mellin-space shift operators: multiplication by powers of the energy in momentum space changes the conformal dimensions of the corresponding celestial operators.  The first two terms in Eq.(\ref{eq:gluon-mhv-differential-equation}) have the form expected from a current algebra on the celestial sphere, while the final term is the correction induced by the subleading conformally soft gluon.  The equation is therefore not the KZ equation of an ordinary WZW model.  It is a scattering-theoretic deformation of such an equation, in which the current algebra remembers both color and four-dimensional soft kinematics.

The same lesson appears from a complementary direction in the analysis of celestial dual superconformal symmetry and MHV amplitudes \cite{Hu:2021lrx}.  A useful general rule is that, if \(\mathcal O\) is an operator acting on a momentum-space amplitude, its celestial representative is obtained by moving \(\mathcal O\) through the Mellin transform,
\begin{equation}
\widetilde{\mathcal O}\,\widetilde{\mathcal A}_n
=\int\left(\prod_{i=1}^n\frac{d\omega_i}{\omega_i}\,\omega_i^{\Delta_i}\right)
\mathcal O\,\mathcal A_n .
\label{eq:celestial-operator-mellin-transform}
\end{equation}
This formula, used in \cite{Hu:2021lrx} to translate dual superconformal generators to the celestial basis, explains why celestial symmetry constraints generally mix derivatives in \(z_i,\bar z_i\) with shifts in \(\Delta_i\).  Derivatives with respect to spinors or momenta act partly on the directions of the null momenta and partly on their energies; the latter action becomes a dimension shift after Mellin transformation.  Thus a familiar momentum-space differential equation can become a difference-differential equation on the celestial sphere.  The same mechanism also appears in the Carrollian basis: differential equations for Carrollian amplitudes provide a parallel organization of the same flat-space data and clarify how celestial dimension shifts are traded for retarded-time dependence at null infinity \cite{Ruzziconi:2024zkr}.

The tree-level MHV gluon amplitudes provide a particularly transparent example.  After using momentum conservation to solve for four energies, the celestial MHV amplitude can be written as an Aomoto--Gelfand hypergeometric integral \cite{Hu:2021lrx},
\begin{align}
\widetilde{\mathcal M}_n
=&\;2\pi\,\frac{z_{st}^4}{z_{12}z_{23}\cdots z_{n1}}\,
\frac{\delta\!\left(\sum_i(\Delta_i-1)\right)}{2^{n-4}}\,
\frac{1}{U}\prod_{a,b}{\bf 1}(x_{a,b}>0)
\nonumber\\
&\times\int\left(\prod_{c=1}^{n-4}du_c\right)
\delta\!\left(1-\sum_{d=1}^{n-4}u_d\right)
\left(\prod_{a=1}^{n-4}u_a^{\Delta_a-J_a-1}\right)
\prod_{b=n-3}^{n}\left(\sum_{a=1}^{n-4}x_{a,b}u_a\right)^{\Delta_b-J_b-1} \nonumber\\
=&\;\mathcal N\,\delta\!\left(\sum_i(\Delta_i-1)\right)
F(x_{a,b},\Delta_i).
\label{eq:celestial-mhv-aomoto-gelfand}
\end{align}
Here \(s,t\) label the two negative-helicity gluons, \(J_i\) is the helicity, \(U\) is the Jacobian minor obtained when the momentum-conserving delta function is used to eliminate four energies, and the variables \(x_{a,b}\) are conformally invariant combinations of the celestial coordinates and in/out signs determined by this choice of elimination.  The step functions record the physical domain in which the solved energies are positive.  The reduced function \(F\) is an Aomoto--Gelfand hypergeometric function; consequently it obeys homogeneity and Grassmannian covariance equations.  In the notation of \cite{Hu:2021lrx}, if
\begin{equation}
F(Z)=\int\prod_{b=1}^{m}\ell_b(u)^{\alpha_b}
\frac{du_1\cdots du_k}{\mathrm{Vol}\,GL(1)},
\qquad
\ell_b(u)=\sum_{a=1}^k u_a z_{a,b},
\label{eq:aomoto-gelfand-integral}
\end{equation}
then the transformation properties of \(F\) imply
\begin{align}
\sum_{b=1}^{m}z_{a,b}\frac{\partial F}{\partial z_{c,b}}&=-\delta_{a,c}F,
\qquad 1\leq a,c\leq k,
\label{eq:aomoto-glk-equation}\\
\sum_{a=1}^{k}z_{a,b}\frac{\partial F}{\partial z_{a,b}}&=\alpha_bF,
\qquad 1\leq b\leq m .
\label{eq:aomoto-homogeneity-equation}
\end{align}
Eqs.(\ref{eq:aomoto-glk-equation}) and (\ref{eq:aomoto-homogeneity-equation}) are the hypergeometric equations behind the celestial MHV gluon amplitude.  They are not imposed as abstract Euclidean CFT constraints; rather, they arise because the Mellin transform of a four-dimensional amplitude with momentum conservation naturally produces a projective integral over energy fractions.  This hypergeometric viewpoint has become a useful bridge to conformal-block analyses of celestial Yang--Mills, scalar and graviton amplitudes \cite{Fan:2022kpp,Fan:2023lky,Surubaru:2025qhs}.  In four-point kinematics these equations reduce to ordinary differential equations in the conformal cross ratio, while for higher multiplicity they become a system of partial differential equations in the variables \(x_{a,b}\).  Related PDE structures have also reappeared in the Liouville-theory organization of MHV celestial amplitudes, where they help connect the Aomoto--Gelfand representation to a two-dimensional correlator problem \cite{Mol:2024qct}.  This is closely related to the conformal Casimir equations discussed in the block decomposition: both identify preferred functions of cross ratios, but the hypergeometric equations remember the scattering origin of the correlator and the distributional restrictions coming from real kinematics.

There is an analogous, and in some respects stronger, structure in gravity.  In the MHV graviton sector, the leading positive-helicity soft graviton gives supertranslation generators, while the subleading soft graviton theorem gives an \(\overline{SL(2,\mathbb C)}\) current algebra on the celestial sphere.  These generators close on the MHV sector and imply null-state decoupling relations for graviton celestial correlators \cite{Banerjee:2020zlg}.  Subsequent work extended this logic to subsubleading soft graviton symmetry, reinforcing the idea that graviton MHV amplitudes are controlled by a hierarchy of soft-current constraints rather than by global conformal covariance alone \cite{Banerjee:2021cly}.  A representative relation is
\begin{align}
\left[
\mathcal J^1_{-1}(n)\mathcal P_{-1,-1}(n)
-(\Delta_n-1)\mathcal P_{-2,0}(n)
\right]
\left\langle
G^-_{\Delta_1}(1)G^-_{\Delta_2}(2)
\prod_{i=3}^{n-2}G^+_{\Delta_i}(i)G^+_{\Delta_n}(n)
\right\rangle=0 .
\label{eq:graviton-null-state-differential-equation}
\end{align}
Here \(G^\pm_{\Delta_i}(i)\) denotes a graviton primary of helicity \(\pm2\) and dimension \(\Delta_i\) inserted at \((z_i,\bar z_i)\), \(\mathcal J^1_{-1}\) is a mode of the subleading-soft current, and \(\mathcal P_{m,n}\) denotes a supertranslation or dimension-shift operator.  The equation says that a particular descendant built from the soft-graviton symmetry is null inside the MHV sector.  In this sense the gravitational equations are closer to null-vector decoupling equations than to ordinary hypergeometric equations, although both play the same organizing role: they restrict the admissible celestial correlators and their OPE data.

The comparison between the gluon and graviton cases is instructive.  The gluon equations are KZ-like and color-current driven, with corrections that remember subleading soft gluons.  The hypergeometric equations identify the reduced MHV amplitudes as special functions on a Grassmannian-like parameter space.  The graviton equations are controlled by supertranslations and subleading soft graviton currents, and their shift operators encode the fact that soft factors change Mellin weights.  In all cases the equations are most transparent in sectors, such as MHV tree amplitudes, where soft and collinear limits close on a restricted set of celestial operators.  Outside such sectors, loop effects, non-MHV amplitudes and the singular support of momentum conservation make the analytic problem substantially more delicate.

These equations explain why non-distributional celestial amplitudes are more than a technical redefinition.  If all correlators remained distributions supported on the momentum-conserving locus, the role of differential equations, conformal blocks and crossing would remain partly formal.  The singularity structure of leaf amplitudes and proposals for regular celestial amplitudes aim to make these analytic constraints act on better-behaved objects \cite{Mandal:2024zao,Liu:2025voe}.  The constructions reviewed next should be read in this light: they seek observables on which the block, OPE and differential-equation structures become visible as statements about functions of celestial coordinates.

\subsection{Non-distributional celestial amplitudes from background fields}

The distributional nature of conventional celestial amplitudes motivates a second line of development: construct observables that remain conformally covariant but are not supported only on the special kinematic locus of plane-wave scattering.  Refs.~\cite{Fan:2022vbz,Stieberger:2022zyk} give a concrete realization in Yang-Mills theory.  Related developments include light-ray and light-transformed correlators with ordinary CFT power-law behavior \cite{Hu:2022syq,Banerjee:2022hgc}, the ambidextrous basis in split signature \cite{Jorge-Diaz:2022dmy}, massive-background OPE tests \cite{Banerjee:2023rni}, scalar-graviton observables \cite{Ball:2023ukj}, and leaf amplitudes that resolve MHV amplitudes into smooth correlators on hyperbolic leaves before the radial scale integral restores the usual boost-weight delta function \cite{Melton:2023bjw,Melton:2024jyq,Duary:2024cqb,Mandal:2024zao}.  We return in Section~\ref{subsec:leaf-liouville-dictionary} to the role of leaf objects in the Liouville description.  They also provide a useful arena for half-collinear and Klein-space analytic continuations, where single-minus gluon and graviton tree amplitudes can be constrained by soft limits or $\mathcal Lw_{1+\infty}$ Ward identities \cite{Guevara:2026qzd,Guevara:2026qwa}.  The basic move is to evaluate scattering in a nontrivial background rather than around the empty translationally invariant vacuum.

The physical motivation is straightforward.  A background field can supply or absorb momentum, so the celestial correlator is not forced to live on the planarity locus $z=\bar z$.  If the background is chosen carefully, the resulting amplitude can retain enough conformal structure to be interpreted as a celestial correlator.

In the constructions of refs.~\cite{Fan:2022vbz,Stieberger:2022zyk}, self-dual or related gauge-theory backgrounds lead to celestial Yang-Mills amplitudes with a Liouville-type structure. A useful schematic form is
\be
  \widetilde{\calA}^{\rm bg}_n
  = \prod_{i=1}^{n}\left(\int_0^{+\infty}\dd\omega_i\,\omega_i^{\Delta_i-1}\right)
  \calA^{\rm bg}_n(\omega_i,z_i,\bar z_i) .
\label{eq:bg-celestial-transform-review}
\ee
The explicit low-point formulas make the improvement over the vacuum Mellin transform very concrete.  For the MHV helicity assignment with two negative-helicity and one positive-helicity gluon, the single-valued three-point correlator obtained from the collinear limit of the single-valued four-point solution is 
\begin{align}
&\left\langle
\phi^{a_1,-\epsilon}_{\Delta_1,-}(z_1,\bar z_1)
\phi^{a_2,-\epsilon}_{\Delta_2,-}(z_2,\bar z_2)
\phi^{a_3,+\epsilon}_{\Delta_3,+}(z_3,\bar z_3)
\right\rangle_{\rm SV}
\nonumber\\
&\quad = i f^{a_1a_2a_3}\,
\delta(\lambda_1+\lambda_2+\lambda_3)
(1+i\lambda_1+i\lambda_2)B(-i\lambda_1,-i\lambda_2)
\nonumber\\
&\qquad\times
z_{12}^{1-i\lambda_1-i\lambda_2}\,
\bar z_{12}^{-i\lambda_1-i\lambda_2-2}
 z_{23}^{i\lambda_1-1}\,\bar z_{23}^{i\lambda_1}
 z_{13}^{i\lambda_2-1}\,\bar z_{13}^{i\lambda_2} .
\label{eq:sv-three-gluon-correlator}
\end{align}
Here $\Delta_i=1+i\lambda_i$, $B(x,y)$ is the Euler beta function, and $f^{abc}$ are the Yang--Mills structure constants.  This formula, derived in ref.~\cite{Fan:2022vbz}, has the coordinate dependence required by global conformal covariance, but unlike the ordinary plane-wave three-gluon celestial amplitude it is not killed by four-dimensional momentum conservation.  The delta function now enforces the residual scaling condition rather than a codimension constraint on insertion points.

For four gluons, the same construction gives a single-valued correlator of the form \cite{Fan:2022vbz}
\begin{align}
&\left\langle
\phi^{a_1,-\epsilon}_{\Delta_1,-}(z_1,\bar z_1)
\phi^{a_2,-\epsilon}_{\Delta_2,-}(z_2,\bar z_2)
\phi^{a_3,+\epsilon}_{\Delta_3,+}(z_3,\bar z_3)
\phi^{a_4,+\epsilon}_{\Delta_4,+}(z_4,\bar z_4)
\right\rangle_{\rm SV}
\nonumber\\
&\quad =
 z_{12}^{-i\lambda_1-i\lambda_2+2}
 z_{24}^{i\lambda_1-2}
 z_{14}^{-i\lambda_1-i\lambda_4}
 z_{13}^{-2-i\lambda_3}
 \bar z_{12}^{-i\lambda_1-i\lambda_2-2}
 \bar z_{24}^{i\lambda_1}
 \bar z_{14}^{-i\lambda_1-i\lambda_4}
 \bar z_{13}^{-i\lambda_3}
\nonumber\\
&\qquad\times
\left[
 f^{a_1a_2b}f^{a_3a_4b}G_{\rm SV}(x,\bar x)
 + f^{a_1a_3b}f^{a_2a_4b}\widetilde G_{\rm SV}(x,\bar x)
\right],
\qquad
\widetilde G_{\rm SV}(x,\bar x)=-xG_{\rm SV}(x,\bar x) .
\label{eq:sv-four-gluon-correlator}
\end{align}
The nontrivial function $G_{\rm SV}$ is the single-valued solution of the Banerjee--Ghosh differential equations in the different OPE regions.  Eq.(\ref{eq:sv-four-gluon-correlator}) is useful because it separates the conformal prefactor and color structures from the genuinely dynamical cross-ratio dependence.  The latter is a smooth single-valued function rather than a distribution supported on $z=\bar z$.

A later analysis showed that the same four-gluon object admits an especially transparent factorized form \cite{Fan:2022kpp},
\begin{align}
&\left\langle
\phi^{a_1,-\epsilon}_{\Delta_1,-}
\phi^{a_2,-\epsilon}_{\Delta_2,-}
\phi^{a_3,+\epsilon}_{\Delta_3,+}
\phi^{a_4,+\epsilon}_{\Delta_4,+}
\right\rangle
\nonumber\\
&\quad =
\left(
\frac{f^{a_1a_2b}f^{a_3a_4b}}{z_{12}z_{23}z_{34}z_{41}}
+
\frac{f^{a_1a_3b}f^{a_2a_4b}}{z_{13}z_{32}z_{24}z_{41}}
\right)
\frac{z_{12}^{2}}{\bar z_{12}^{2}}
S(z_i,\bar z_i) .
\label{eq:sv-four-gluon-factorization}
\end{align}
The first factor is the group-dependent current correlator of four holomorphic WZW currents, while $S(z_i,\bar z_i)$ is a scalar celestial correlator.  In the same conventions,
\begin{align}
S(z_i,\bar z_i)
=&\; |z_{12}|^{-i\lambda_1-i\lambda_2}|z_{34}|^{-i\lambda_3-i\lambda_4}
\left|\frac{z_{14}}{z_{13}}\right|^{i\lambda_3-i\lambda_4}
\left|\frac{z_{24}}{z_{14}}\right|^{i\lambda_1-i\lambda_2}
\nonumber\\
&\times
\sum_{\Delta=i\lambda_3+i\lambda_4,\,4+i\lambda_1+i\lambda_2}^{\prime}
(1-\Delta)
B\!\left(\frac{\Delta-d_{12}}{2},\frac{\Delta+d_{12}}{2}\right)
B\!\left(\frac{\Delta-d_{34}}{2},\frac{\Delta+d_{34}}{2}\right)
 g_{\Delta}(x,\bar x) .
\label{eq:sv-scalar-block-factor}
\end{align}
Here $d_{ij}=d_i-d_j$ are four-dimensional scalar-dimension differences in the notation of ref.~\cite{Fan:2022kpp}, and $g_\Delta(x,\bar x)$ is the dimensionally reduced four-dimensional scalar conformal block.  The primed sum contains the block with $\Delta=i\lambda_3+i\lambda_4$ and its four-dimensional shadow with $\Delta=4+i\lambda_1+i\lambda_2$.  The prime over the sum accounts for a relative minus sign between the two terms.  The non-distributional four-gluon correlator does not merely resemble a CFT correlator kinematically: it factorizes into a holomorphic current sector and a scalar block sector.  This is the precise sense in which the background-field construction supplies a cleaner arena for OPE and conformal-block data.

The key difference from the vacuum amplitude is that $\calA^{\rm bg}_n$ no longer contains only the standard overall momentum-conservation delta function.  The background carries spacetime dependence, and the transformed correlator becomes a genuine function, or at least a less singular distribution, of the insertion points.  This is precisely the improvement needed for CFT reasoning: OPE limits, conformal blocks and Ward identities can now be discussed before the full distributional machinery of four-point plane-wave scattering is imposed.  The Liouville language in ref.~\cite{Stieberger:2022zyk} should be read in this spirit.  It is not merely an analogy between formulas; the relevant Yang-Mills amplitudes produce structures naturally described by a two-dimensional conformal field theory with Liouville-type dependence, while the related current-algebra and scalar-block factorization was analyzed in ref.~\cite{Fan:2022kpp}.  Some of the dynamics hidden by the distributional vacuum S-matrix therefore becomes visible after the correct background or dressing is chosen.

This viewpoint explains why conformal blocks and non-distributional amplitudes are naturally paired.  Blocks ask which celestial representations are exchanged; non-distributional amplitudes provide cleaner correlators from which such data can be extracted.  Recent proposals to reconstruct amplitudes from crossing-symmetric celestial OPE data and to define regular celestial amplitudes sharpen this point by treating the OPE as a constructive principle rather than only a limit of known momentum-space amplitudes \cite{Liu:2025dhh,Liu:2025voe}.  Shadow-OPE approaches make the same point from a complementary direction: sufficiently constrained celestial OPE data may reconstruct amplitudes, provided one uses a class of correlators regular enough for crossing and block methods to apply \cite{Liu:2025dhh,Himwich:2025bza,Liu:2025voe}.

\section{Celestial Liouville theory}
\label{sec:celestial-liouville}

The preceding section emphasized a central tension in celestial holography. Mellin transforms make Lorentz symmetry manifest as two-dimensional conformal symmetry, but ordinary plane-wave amplitudes still carry the rigid imprint of four-dimensional translation invariance through momentum-conserving delta functions. This is why many basic celestial amplitudes are distributions rather than ordinary functions of points on the celestial sphere. Liouville theory enters this story as a surprisingly concrete way of separating these two features. It supplies an intrinsic two-dimensional conformal dynamics whose light vertex operators reproduce the non-holomorphic part of certain celestial Yang-Mills amplitudes, while the holomorphic Parke-Taylor or current-algebra factor keeps track of spin and color. In this sense, celestial Liouville theory should not be viewed as a replacement for the Mellin-transform definition of celestial amplitudes, but as an organizing framework for those deformations and representations in which celestial correlators become closer to ordinary CFT correlators.

The construction was first made explicit for Yang-Mills theory coupled to a dynamical complex dilaton, evaluated in a special dilaton shockwave background \cite{Stieberger:2022zyk}, and was subsequently extended to supersymmetric multiplets and to a more explicit Yang-Mills/Liouville operator dictionary \cite{Taylor:2023bzj,Stieberger:2023fju}. The background absorbs momentum, so that the resulting celestial correlators are not supported only on the codimension-four locus imposed by the flat-space momentum delta function. At the same time, the source is chosen so that the celestial sphere retains global conformal covariance. This is conceptually close to the non-distributional amplitudes discussed in Section~\ref{sec:blocks}: one modifies the scattering problem just enough to obtain genuine two-dimensional correlators, but in a way that preserves the celestial interpretation of the external states.

\subsection{Liouville data and the semiclassical regime}

The Liouville sector used in celestial applications is the standard two-dimensional Liouville theory on the sphere. In the conventions of Ref.~\cite{Stieberger:2022zyk}, its local Lagrangian density is
\be
  \mathcal L_L=\frac{1}{\pi}\partial\phi\,\bar\partial\phi+\mu e^{2b\phi},
  \qquad
  Q=b+\frac{1}{b},
  \qquad
  c_L=1+6Q^2 .
\label{eq:liouville-action-central-charge}
\ee
Here $b$ is the Liouville coupling, $\mu$ is the cosmological constant, $Q$ is the background charge and $c_L$ is the central charge. The background-charge term is implicit in this sphere notation; it is what makes the stress tensor have central charge $1+6Q^2$. The celestial limit emphasized in Refs.~\cite{Stieberger:2022zyk,Stieberger:2023fju} is the large central charge or semiclassical regime $b\to 0$, for which $c_L\sim 6/b^2$. This is the regime in which Liouville correlators are dominated by a classical saddle, and it is also the regime that reproduces tree-level celestial Yang-Mills amplitudes in the dilaton background.

Liouville primary fields are exponential vertex operators,
\be
  V_{\alpha}(z,\bar z)=e^{2\alpha \phi(z,\bar z)},
  \qquad
  h_{\alpha}=\bar h_{\alpha}=\alpha(Q-\alpha) .
\label{eq:liouville-vertex-dimension}
\ee
This formula is the basic bridge to celestial kinematics. In celestial amplitudes, the conformal weights of external operators are Mellin dual to energies. In the Liouville description, the corresponding weights are carried by light operators with $\alpha_i=\sigma_i b$. Their total scaling dimension is
\be
  d_i=2h_i=2\sigma_i+2b^2\sigma_i(1-\sigma_i) .
\label{eq:liouville-light-dimension}
\ee
Eq.(\ref{eq:liouville-light-dimension}), taken from Ref.~\cite{Stieberger:2022zyk}, shows why the parameter $\sigma_i$ naturally plays the role of a Mellin variable: at $b=0$ the dimension is linear in $\sigma_i$, while finite $b$ introduces a controlled nonlinear correction. This nonlinear correction is the source of the finite-central-charge effects that later reappear as logarithmic deformations of celestial amplitudes and OPEs.

For the particular spherical shockwave background relevant to the original construction, the Liouville path integral is dominated by the classical solution \cite{Stieberger:2022zyk}
\be
  2b\phi_0(z,\bar z)=
  \log \left[\frac{A}{\pi(1+z\bar z)^2}\right] .
\label{eq:liouville-classical-saddle}
\ee
The constant $A$ fixes the total area of the corresponding round metric on the celestial sphere. The physical point is that the saddle is not an arbitrary auxiliary field: it is precisely the two-dimensional geometry that packages the angular dependence produced by the dilaton shockwave. The large-$c_L$ Liouville saddle therefore gives a geometric representation of the non-holomorphic celestial data.

\subsection{Mellin transforms, dilaton backgrounds and factorization}

The starting point remains the celestial Mellin transform of a momentum-space amplitude. For massless particles with momenta $p_i^\mu=\omega_i q^\mu(z_i,\bar z_i)$, the celestial amplitude is
\be
  \widetilde{\calA}_n(\Delta_i,z_i,\bar z_i)
  = \prod_{i=1}^n \int_0^\infty \dd \omega_i\,\omega_i^{\Delta_i-1}
  \calA_n(\omega_i,z_i,\bar z_i) .
\label{eq:liouville-mellin-transform}
\ee
This is the same transform reviewed earlier in Eq.(\ref{eq:celestial-amplitude-definition}); it is repeated here because the Liouville construction changes the object being Mellin transformed. Instead of the translation-invariant plane-wave amplitude alone, one evaluates Yang-Mills theory in a prescribed dilaton background. In Ref.~\cite{Stieberger:2022zyk} the corresponding MHV celestial amplitude factorizes as
\be
  \calM_N(z_i,\bar z_i|\Delta_i)
  =\mathcal{J}_N(z_i)\,\mathcal{S}_N(z_i,\bar z_i|\Delta_i) .
\label{eq:liouville-factorization}
\ee
The holomorphic factor $\mathcal{J}_N$ is the current-algebra or Parke-Taylor part, including color and helicity information, while $\mathcal{S}_N$ is a scalar Mellin integral controlled by the dilaton source. In a fixed color ordering one may think of $\mathcal{J}_N$ as the familiar holomorphic MHV factor $z_{12}^4/(z_{12}z_{23}\cdots z_{N1})$, with the full amplitude obtained by summing over color orderings. The scalar part contains
\be
  \mathcal{S}_N \sim \prod_{i=1}^N \int_0^\infty \dd \omega_i\,\omega_i^{\Delta_i-1}
  \int \dd^4X\,J_{\Phi}(X)\,\frac{e^{iX\cdot Q}}{Q^2},
  \qquad
  Q^\mu=\sum_{i=1}^N \omega_i q_i^\mu .
\label{eq:liouville-scalar-source-integral}
\ee
This equation is a compact version of the scalar factor derived in Ref.~\cite{Stieberger:2022zyk}. The pole $1/Q^2$ is the massless dilaton propagator connecting the gauge-theory system to the external source. The source is
\be
  J_{\Phi}(X)=-\Box\left[\delta(X^2)\delta(X^+-1)\theta(X^++X^-)\right] .
\label{eq:liouville-dilaton-source}
\ee
This source explicitly breaks ordinary four-dimensional translation invariance, but it does so in a controlled conformal way: its support is tied to the null cone and to a fixed projective slice. The breaking is what removes the momentum-conserving delta function from the celestial transform; the conformal choice of support is what preserves the interpretation as a correlator on the celestial sphere. Thus the Liouville construction should be read as a mechanism for trading distributional momentum conservation for a background-induced two-dimensional conformal correlator. The price is equally important: the resulting object is no longer the vacuum plane-wave S-matrix, but a closely related observable designed to expose the two-dimensional conformal data more cleanly.

For the mostly-plus MHV sector, the Liouville momenta are chosen as \cite{Stieberger:2022zyk}
\be
  \sigma_1=\frac{1+i\lambda_1}{2},
  \qquad
  \sigma_2=\frac{1+i\lambda_2}{2},
  \qquad
  \sigma_k=\frac{i\lambda_k}{2}\quad (k\geq 3),
  \qquad
  \sum_{i=1}^N \lambda_i=0 .
\label{eq:liouville-mhv-sigma-map}
\ee
The two negative-helicity gluons are distinguished by the shift by one unit in $\sigma$, matching the helicity dependence of the MHV amplitude. This is a useful place to see how celestial and Carrollian viewpoints complement each other. In the Carrollian basis the retarded time coordinate keeps energy dependence local; in the Mellin-Liouville basis, energy dependence is converted into Liouville momenta, and the loss of manifest translation invariance becomes the price paid for two-dimensional conformal covariance.

\subsection{Dotsenko-Fateev integrals and celestial correlators}

The technical reason Liouville theory is useful here is that its light-operator correlators reduce to integrals of Dotsenko-Fateev or Selberg type. In the Coulomb-gas representation, nonzero correlators with screening insertions obey the charge-balance condition \cite{Dotsenko:1984nm,Dotsenko:1984ad,Stieberger:2022zyk}
\be
  \sum_i \alpha_i = Q-\frac{m}{b}-nb,
\label{eq:liouville-charge-balance}
\ee
where $m$ and $n$ count the two types of Liouville screening insertions. In the light limit $\alpha_i=\sigma_i b$, the one-screening sector with $\sum_i\sigma_i=1$ gives the characteristic integral
\be
  \left\langle \prod_i V_{\sigma_i b}(z_i,\bar z_i)\right\rangle_L
  \sim \prod_{i<j}|z_{ij}|^{-4b^2\sigma_i\sigma_j}
  \int \dd^2u\,\prod_i |z_i-u|^{-4\sigma_i} .
\label{eq:liouville-df-light-integral}
\ee
The prefactor is the free-field contraction of the light operators; the integral over $u$ is the screening integral. For celestial amplitudes this expression is important because the same type of integral appears after the Mellin transform of the dilaton-source scalar factor. More concretely, with the MHV identification in Eq.(\ref{eq:liouville-mhv-sigma-map}) one obtains integrals of the form \cite{Stieberger:2022zyk}
\be
  I_N=
  \int \dd^2u\,
  |z_1-u|^{-2(1+i\lambda_1)}
  |z_2-u|^{-2(1+i\lambda_2)}
  \prod_{k=3}^N |z_k-u|^{-2i\lambda_k} .
\label{eq:liouville-celestial-selberg-integral}
\ee
This is the concrete point at which the Liouville and celestial descriptions meet: the celestial integral over energies and the Liouville screening integral are not merely analogous, but can be identified in the shockwave construction. The Selberg/Dotsenko-Fateev structure also suggests why conformal blocks and OPEs should be natural languages for these amplitudes; related Selberg dualities imply nontrivial relations among celestial amplitudes with shifted conformal dimensions \cite{Giribet:2024vnk}. The singularities of the integrals as insertion points collide encode collinear limits, while their monodromy and analytic continuation properties organize the same data that appear in celestial conformal block decompositions.

The operator form of the identification is especially transparent. For the MHV gluon sector, Ref.~\cite{Stieberger:2022zyk} introduced celestial operators
\be
  \calO_{\lambda}^{-a}(z,\bar z)
  =\Gamma(1+i\lambda)\,\widehat J^a(z)\,e^{(1+i\lambda)b\phi(z,\bar z)},
  \qquad
  \calO_{\lambda}^{+a}(z,\bar z)
  =\Gamma(i\lambda)\,J^a(z)\,e^{i\lambda b\phi(z,\bar z)} .
\label{eq:liouville-gluon-operators}
\ee
Here $J^a$ and $\widehat J^a$ are holomorphic current-sector operators carrying color and helicity information, while the exponential factors are light Liouville operators carrying Mellin energy. Eq.(\ref{eq:liouville-gluon-operators}) explains why the celestial OPEs reviewed in Section~\ref{sec:symmetry-ope} contain beta functions: the product of the Mellin gamma factors and the exponential Liouville operators turns collinear energy sharing into Euler beta functions. In the supersymmetric extension, the same factorization becomes \cite{Taylor:2023bzj}
\be
  O^a_{\Delta,J}(z,\bar z)
  =O^a_J(z)\,\Gamma(\Delta-J)\,
  e^{(\Delta-J)b\phi(z,\bar z)} ,
\label{eq:liouville-supersym-operator-map}
\ee
where $J$ is the four-dimensional helicity and $O_J^a(z)$ belongs to the holomorphic current or supercurrent sector. This formula makes clear that the Liouville factor is universal across the gauge multiplet; supersymmetry acts in the chiral current sector, while Mellin-energy dependence is carried by the same light Liouville exponential.

\subsection{Finite central charge and PDEs}

The original Liouville construction is semiclassical. A more ambitious proposal is that finite-$b$ Liouville theory gives a systematic completion of the celestial representation of Yang-Mills amplitudes, a direction sharpened by recent perturbative studies of the light-operator DOZZ expansion and the three-gluon amplitude \cite{Ferrari:2025vrx,Ferrari:2025mmm,Biskowski:2026qsu}. Ref.~\cite{Stieberger:2023fju} formulates this idea by writing gluon operators as WZW-like current operators tensored with light Liouville exponentials,
\be
  O^{+a}_{\Delta}=F_+(\Delta,\mu,b)\,J^a(z)\,
  e^{2\sigma(\Delta-1)b\phi},
  \qquad
  O^{-a}_{\Delta}=F_-(\Delta,\mu,b)\,J^a(z)\,
  e^{2\sigma(\Delta+1)b\phi} .
\label{eq:liouville-finite-b-operators}
\ee
The normalizations $F_{\pm}$ contain gamma functions and powers of $\pi\mu\gamma(b^2)b^{-2b^2}$, as in the DOZZ normalization of Liouville theory. Their role is not cosmetic: they are needed for the inverse Mellin transform of the WZW-Liouville correlator to reproduce the expected three-gluon amplitude in the small-$Q^2$ limit. In this formulation, tree-level Yang-Mills corresponds to $b\to 0$, while the first finite-$b$ correction is related to the one-loop running of the gauge coupling through \cite{Stieberger:2023fju}
\be
  b^2=\frac{\beta_0 g^2(M)}{8\pi^2} .
\label{eq:liouville-b-beta-function}
\ee
Here $g(M)$ is the Yang-Mills coupling at renormalization scale $M$ and $\beta_0$ is the one-loop beta-function coefficient. This relation is suggestive rather than a full derivation of perturbative Yang-Mills from Liouville theory, but it gives a quantitative meaning to the finite-central-charge expansion: $b^2$ counts a class of logarithmic corrections that resemble loop-level celestial data.

A complementary development derives partial differential equations for celestial Liouville amplitudes from Liouville Ward identities \cite{Mol:2024qct}. For MHV gluons, one defines a finite-$b$ Liouville-dressed amplitude
\be
  \widehat\calA_n=
  \frac{1}{(2\pi)^4\mathcal N_{\mu,b}}
  \frac{z_{12}^4}{z_{12}z_{23}\cdots z_{n1}}
  \int_0^\infty \dd \tau\,\tau^3
  \left\langle
  \prod_{i=1}^n e^{-2\rho_i\log \tau}
  \Gamma(2\rho_i)V_{b\rho_i}(z_i,\bar z_i)
  \right\rangle_L ,
\label{eq:liouville-pde-amplitude}
\ee
with
\be
  \rho_i^{(0)}=\frac{1}{2}(\Delta_i+\alpha'_i),
  \qquad
  \rho_i(b)=\rho_i^{(0)}+b^2\big((\rho_i^{(0)})^2-\rho_i^{(0)}\big)+\calO(b^4) .
\label{eq:liouville-rho-expansion}
\ee
The constants $\alpha'_i$ encode the helicity weights in the momentum-space MHV amplitude. After inverse Mellin transformation to energies, the Ward identity for the Liouville correlator gives the first-order PDE \cite{Mol:2024qct}
\be
  \sum_{i=1}^n
  \left(z_i\partial_{z_i}-\frac{1}{2}\omega_i\partial_{\omega_i}\right)\calA_n
  =\frac{4-n}{2}\calA_n+\calO(b^4) .
\label{eq:liouville-gluon-pde}
\ee
This equation is conceptually useful because it turns finite-$b$ Liouville covariance into a differential constraint on the energy-space amplitude. It also makes contact with the Carrollian viewpoint: $\omega_i\partial_{\omega_i}$ is conjugate to scaling in retarded time, while $z_i\partial_{z_i}$ acts on the celestial sphere. The PDE therefore relates angular conformal covariance to energy scaling, precisely the relation that is split differently in Mellin and Carrollian representations.

The same analysis proposes a renormalized celestial Liouville amplitude whose first correction is governed by \cite{Mol:2024qct}
\be
  \mathcal{R}_n=1-\frac{b^2}{4}\sum_{1\leq i<j\leq n}
  \log \frac{\omega_i\omega_j |z_{ij}|^2}{m^2},
  \qquad
  \calA_n=\mathcal{R}_n A_n^{(0)}+\calO(b^4) .
\label{eq:liouville-renormalization-factor}
\ee
Here $m$ is an infrared scale and $A_n^{(0)}$ is the leading MHV amplitude. The logarithm is exactly the kind of structure expected from loop-level celestial OPEs: it becomes derivatives with respect to Mellin dimensions after transforming back to the celestial basis. This is why finite-$b$ Liouville theory is potentially more than a convenient integral representation. It may provide a controlled two-dimensional language for loop-induced deformations of celestial OPE coefficients and conformal blocks.

Three caveats should be kept in view.  First, the best-developed Liouville correspondence is for Yang-Mills MHV sectors and special dilaton backgrounds.  Closely related Liouville-based constructions of MHV leaf amplitudes and graviton amplitudes suggest possible extensions beyond gauge-theory shockwaves, but the gravitational interpretation is still less complete than the Yang-Mills one \cite{Melton:2024gyu,Melton:2024jyq,Mol:2024etg,Mol:2024qct}.  Second, finite-\(b\) Liouville theory has the strong-weak duality \(b\leftrightarrow 1/b\), whose celestial meaning is not yet understood; the appearance of \(H_3^+\) WZW models in proposed celestial CFT constructions suggests that known Liouville/WZW dualities may eventually be relevant \cite{Ogawa:2024nhx}.  Third, the relation to asymptotic symmetries remains indirect: soft currents and current algebras live naturally in the holomorphic sector, whereas Liouville controls the non-holomorphic Mellin-energy dependence.  A complete celestial theory should explain whether these two pieces arise from a single boundary dynamics or from a useful factorized description.

These questions connect the Liouville discussion to the later analysis of loop effects, OPE data and nonperturbative structures. The value of the Liouville perspective is that it gives these questions a common language: Mellin energies become Liouville momenta, collinear limits become OPE limits of light operators, and the passage from tree level to loops becomes a finite-central-charge deformation of a two-dimensional conformal theory.

\subsection{Leaf amplitudes and the Liouville dictionary}
\label{subsec:leaf-liouville-dictionary}

The Liouville construction above is closely related to another way of isolating the non-distributional part of a celestial amplitude.  Instead of introducing a background source, one can decompose the ordinary MHV amplitude before performing the last scale integral responsible for the usual boost-weight delta function.  This is the idea behind celestial leaf amplitudes, introduced in Ref.~\cite{Melton:2023bjw}.  A closely related ``outside-in'' construction shows how two-dimensional gluon operators and a suitable scalar source can reproduce celestial gluon amplitudes, with Liouville vertex operators providing one concrete realization of the boundary three-point data \cite{Melton:2023lnz}.  The leaf construction is most transparent in split signature, or Klein space, where the relevant region of spacetime can be foliated by hyperbolic leaves.  The conformal boundary of the basic leaf geometry is a Lorentzian torus, whose two- and three-point conformal correlators already exhibit features that are invisible in ordinary Euclidean celestial kinematics \cite{Melton:2023hiq}.  The leaf amplitude is the correlator associated with one such leaf.  It is not the full plane-wave celestial amplitude, but it contains the angular and conformal data that are usually obscured by the distribution enforcing translation invariance.

Starting from the Parke-Taylor amplitude in Eq.(\ref{eq:parke-taylor-mhv}) and the conformal primary wavefunctions introduced in Eq.(\ref{eq:scalar-conformal-primary-wavefunction}), the leaf construction of Ref.~\cite{Melton:2023bjw} restricts products of wavefunctions to hyperbolic leaves.  The Mellin transform then isolates angular and conformal data before the final scale integral produces the usual boost-weight delta function.

For the color-ordered MHV configuration \(1^-2^-3^+\cdots n^+\), Ref.~\cite{Melton:2023bjw} writes the celestial amplitude in the form
\begin{align}
  \calA_n^{\rm MHV}
  &=
  \frac{\delta(\beta)}{8\pi^3}
  \left[
  \mathcal L_n(\sigma_i,\bar\sigma_i)
  +\mathcal L_n(\sigma_i,-\bar\sigma_i)
  \right],
  \nonumber\\
  \mathcal L_n
  &=
  \frac{s_{12}^{3}}{s_{23}s_{34}\cdots s_{n1}}\,
  \mathcal C_n ,
  \qquad
  \mathcal C_n
  =
  \int_{{\rm AdS}_3/\mathbb Z}\dd^3\hat x\,
  \prod_{i=1}^{n}\Phi_{2\bar h_i}(\hat x,\hat p_i) .
\label{eq:leaf-amplitude-factorization}
\end{align}
Here \(s_{ij}=\sin\sigma_{ij}\) in the angular variables used in the split-signature parametrization, \(\bar h_i=(\Delta_i-J_i)/2\), and
\(\beta=\sum_i(\Delta_i-1)\) is the total boost weight.  The two terms in the first line correspond to the two analytic continuations of the leaf data.  The first factor in \(\mathcal L_n\) is the familiar holomorphic MHV structure, while \(\mathcal C_n\) is a scalar contact integral on \( {\rm AdS}_3/\mathbb Z\).  Thus the same amplitude separates into a chiral current-like factor and a non-holomorphic scalar factor, exactly the kind of separation that appeared in the Liouville-background construction.  The distributional part of the ordinary celestial amplitude is isolated in the radial integral
\be
  \int_0^\infty \dd\tau\,\tau^{-\beta-1}
  =2\pi\,\delta(\beta) ,
\label{eq:leaf-radial-delta}
\ee
again in the conventions of Ref.~\cite{Melton:2023bjw}.  Eq.(\ref{eq:leaf-radial-delta}) explains why the leaf object is smoother than the final celestial amplitude: the hyperbolic integral \(\mathcal C_n\) exists before the scale integral imposes the boost-weight constraint.
Subsequent work has made this improvement more concrete by developing spectral representations for leaf amplitudes and by analyzing the singularity structure of the four-point leaf correlator \cite{Duary:2024cqb,Mandal:2024zao}.  These analyses support the view that leaf amplitudes are not merely a change of variables, but a setting in which ordinary analytic questions about conformal correlators can be asked before the final distributional projection.

The connection to Liouville theory was sharpened in Ref.~\cite{Melton:2024gyu}, where the scalar contact integral in Eq.(\ref{eq:leaf-amplitude-factorization}) was identified with the semiclassical limit of a Liouville correlator.  In the small-\(b\) limit, with light Liouville insertions \(V_{b\sigma_i}\), one finds a relation of the schematic form
\begin{align}
  \left\langle\prod_{i=1}^{n}
  V_{b\sigma_i}(z_i,\bar z_i)
  \right\rangle_L
  &=
  \frac{e^{-2\gamma_E+2/b^2}\lambda^{1/b^2}}
  {\pi b^3}\,
  \csc\!\left[\pi\left(b^{-2}-\frac{\beta}{2}\right)\right]
  \lambda^{-1-\beta/2}
  \mathcal C_{2\sigma_1,\ldots,2\sigma_n}
  +\cdots ,
  \nonumber\\
  \mathcal C_{2\sigma_1,\ldots,2\sigma_n}
  &=
  \int_{H^3}D^3x\,
  \prod_{i=1}^{n}G_{2\sigma_i}(z_i,\bar z_i;x) .
\label{eq:liouville-leaf-contact-integral}
\end{align}
Here \(G_{2\sigma_i}\) is the bulk-to-boundary propagator on Euclidean \(H^3\), \(\lambda\) is the Liouville cosmological constant in the normalization of Ref.~\cite{Melton:2024gyu}, and the ellipsis denotes terms subleading in the semiclassical limit or dependent on normalization conventions.  The precise prefactor is less important than the structural statement: light Liouville correlators compute \(H^3\) contact diagrams, and these contact diagrams are the Euclidean counterparts of the hyperbolic leaf integrals appearing in MHV celestial amplitudes.

This observation leads to an operator dictionary for MHV celestial gluons.  Ref.~\cite{Melton:2024gyu} proposes that the positive- and negative-helicity celestial gluon operators are represented by a chiral current sector dressed by light Liouville vertex operators,
\begin{align}
  \calO_{\Delta}^{+a,\varepsilon}(z,\bar z)
  &=
  e^{-i\varepsilon\pi(\Delta-1)/2}
  \lim_{b\to0}N_{\Delta}^{+}\,
  J^a(z)\,
  V_{b(\Delta-1)/2}(z,\bar z),
  \nonumber\\
  \calO_{\Delta}^{-a,\varepsilon}(z,\bar z)
  &=
  e^{-i\varepsilon\pi(\Delta+1)/2}
  \lim_{b\to0}N_{\Delta}^{-}\,
  \bar J^a(z)\,
  V_{b(\Delta+1)/2}(z,\bar z) .
\label{eq:dressed-liouville-gluon-operators}
\end{align}
The sign \(\varepsilon=\pm1\) records outgoing or incoming kinematics, \(J^a\) and \(\bar J^a\) are chiral current-sector operators carrying the color and helicity dependence, and \(N_\Delta^\pm\) are normalization factors fixed so that the correlator reproduces the celestial amplitude.  With two negative-helicity and \(n-2\) positive-helicity insertions, this gives
\begin{align}
  &\left\langle
  \calO_{\Delta_1}^{-a_1,\varepsilon_1}
  \calO_{\Delta_2}^{-a_2,\varepsilon_2}
  \prod_{j=3}^{n}
  \calO_{\Delta_j}^{+a_j,\varepsilon_j}
  \right\rangle
  \nonumber\\
  &\qquad =
  \prod_{j=1}^{n}
  e^{-i\pi\varepsilon_j\bar h_j}
  \Gamma(2\bar h_j)\,
  \frac{z_{12}^{4}}{z_{12}z_{23}\cdots z_{n1}}\,
  \mathcal C_{2\bar h_1,\ldots,2\bar h_n}(z_i,\bar z_i) ,
\label{eq:dressed-liouville-mhv-correlator}
\end{align}
where \(\bar h_i=(\Delta_i-J_i)/2\).  Eq.(\ref{eq:dressed-liouville-mhv-correlator}) makes the logic of the construction especially clear.  The Parke-Taylor factor is supplied by the chiral current sector, while the Liouville correlator supplies the non-holomorphic \(H^3\) contact integral.  Passing from the Euclidean contact integral to the Lorentzian leaf and finally to the usual celestial amplitude then amounts to analytic continuation followed by the radial Mellin integral that produces \(\delta(\beta)\).

The leaf and dressed-Liouville descriptions therefore extend the shockwave construction without changing the scattering problem itself: they reorganize the ordinary MHV amplitude into smoother hyperbolic pieces before the radial scale integral imposes the boost-weight delta function.  The dressed-Liouville proposal supplies a two-dimensional CFT representation of those pieces, so leaf amplitudes sit naturally between the block and differential-equation material of Section~\ref{sec:blocks} and the soft-algebra structures of Section~\ref{sec:w-infinity}.  Subsequent work has developed this bridge through soft algebras acting on leaf amplitudes \cite{Melton:2024jyq}, parafermionic and supersymmetric realizations \cite{Donnay:2025yoy,Mol:2024vok}, extensions to \(N^k\)MHV sectors in minitwistor superspace \cite{Mol:2026ylx}, and gravitational MHV leaf amplitudes generated by an \(Lw_{1+\infty}\) Ward identity \cite{Guevara:2025tsm}.  These results indicate that the leaf/Liouville picture is a controlled sector in which regular conformal structure, soft symmetry and MHV scattering can be studied before the final distributional projection.

\section{UV/IR structure}
\label{sec:uv-ir}

In a momentum-space effective field theory, one normally separates soft physics from short-distance physics by introducing scales and matching Wilson coefficients.  The celestial transform does something quite different: it integrates each external energy over the entire positive real axis.  As a result, the same conformal correlator probes the soft endpoint, the hard endpoint, and any non-perturbative completion that controls the interpolation between them.  This is not a technical nuisance but one of the main conceptual lessons of the celestial basis.  The Mellin variables are natural for Lorentz covariance, but they make the celestial correlator an intrinsically scale-integrated observable.

This section reviews this UV/IR structure of celestial amplitudes.  We first isolate the endpoint sensitivity of the Mellin transform, then explain how string theory and eikonal exponentiation provide two complementary mechanisms for improving the high-energy behavior.  We finally discuss how infrared exponentiation appears in celestial correlators as dimension renormalization, dimension-shift operators and logarithmic structures in the celestial OPE.  

\subsection{Mellin endpoint sensitivity}

For massless four-point scattering the kinematics can be reduced to one energy scale and one real cross ratio.  After extracting the universal conformal prefactor, the dynamical part of the celestial amplitude takes the form \cite{Arkani-Hamed:2020gyp,Adamo:2024mqn}
\be
\widetilde{\mathcal M}_4(\Delta_i,z_i,\bar z_i)=X(\Delta_i,z_i,\bar z_i)\,\mathcal A(\beta,z),\qquad
\beta=\sum_{i=1}^4(\Delta_i-1),
\label{eq:uvir-four-point-reduction}
\ee
with
\be
\mathcal A(\beta,z)=\int_0^\infty \frac{d\omega}{ \omega}\,\omega^\beta\,
\mathcal M\bigl(s=\omega^2,t=-z\omega^2\bigr).
\label{eq:uvir-beta-transform}
\ee
Here $z$ is the conformal cross ratio, $X$ contains the standard $SL(2,\mathbb C)$ weights together with the kinematic support enforcing real scattering, and $\mathcal M(s,t)$ is the stripped momentum-space amplitude.  Eq.(\ref{eq:uvir-beta-transform}) is the basic reason why celestial amplitudes are sensitive to both endpoints of energy space: the same Mellin integral samples $\omega\to0$ and $\omega\to\infty$.

This feature is especially sharp for amplitudes with power-law high-energy growth.  If the fixed-angle stripped amplitude behaves as $\mathcal M(\omega)\sim \omega^p$, the celestial transform contains \cite{Arkani-Hamed:2020gyp}
\be
\mathcal A(\beta)\sim \int_0^\infty \frac{d\omega}{ \omega}\,\omega^{\beta+p},
\label{eq:uvir-power-growth}
\ee
which is not an ordinary function of $\beta$.  On the principal continuous series, such integrals are at best distributions; away from it they require analytic continuation or a physical completion of the amplitude.  This is the sense in which celestial amplitudes are anti-Wilsonian: a local low-energy expansion does not decouple from the ultraviolet endpoint of the Mellin integral.

A useful diagnostic is provided by a single massive exchange.  For
\be
\mathcal M(s)=\lambda \frac{M^2}{ s-M^2},
\label{eq:uvir-massive-exchange}
\ee
one obtains, with the usual $i\epsilon$ prescription, \cite{Arkani-Hamed:2020gyp}
\be
\mathcal A(\beta)=\lambda M^\beta \frac{i\pi}{e^{-i\pi\beta}-1}.
\label{eq:uvir-massive-mellin}
\ee
The result is meromorphic in $\beta$.  Its poles encode the low-energy expansion of the exchange amplitude, while the exponential scale $M^\beta$ remembers the mass threshold.  Thus celestial analyticity in the $\beta$-plane organizes EFT data and threshold data in a single object.  Conversely, the absence of a sensible analytic structure in $\beta$ is a sign that the momentum-space amplitude has not been completed in a way compatible with locality, causality and unitarity.

The same lesson applies to perturbative gravity.  Early loop-level celestial computations made this endpoint problem concrete: finite rational one-loop amplitudes already give nontrivial celestial correlators in gauge theory and gravity, while infrared-divergent planar amplitudes can be reorganized as operators acting on tree-level celestial correlators \cite{Albayrak:2020saa,Gonzalez:2020tpi}.  A finite number of loop corrections only produces a finite sum of powers of $\omega$ and logarithms of $\omega$.  Terms of the schematic form \cite{Adamo:2024mqn}
\be
\int_0^\infty d\omega\,\omega^{\beta+2J-3+m}(\log\omega)^n
\label{eq:uvir-loop-distribution}
\ee
remain ill-defined as ordinary functions and become derivatives of delta functions only on special distributional loci.  What is needed is not another finite order in perturbation theory, but an infinite resummation or a UV completion that changes the endpoint behavior of the integrand.

This endpoint perspective also connects to analytic questions usually phrased in terms of Regge behavior and effective-field-theory corrections. Celestial Regge theory reformulates the Mellin transform so that dispersion relations, poles and discontinuities of bulk amplitudes become celestial CFT data in a Regge-like limit \cite{Casali:2026epz}. In a complementary effective-field-theory setting, quadratic curvature corrections to gravity modify the phase-dressed celestial amplitude, its shadow transform and the associated OPE coefficients, providing a useful test of how UV-sensitive bulk corrections appear in celestial block data \cite{Bhattacharyya:2025nfp}. These developments feed back into the conformal-block perspective of Section~\ref{sec:blocks}: celestial blocks are not merely kinematic decorations, but probes of the analytic structure and high-energy completion of the underlying S-matrix.

\subsection{String softness and UV completion}

String amplitudes give the cleanest perturbative example of such a completion.  Subsequent work has developed this observation beyond the original tree-level four-point setting, including one-loop open strings, field-theory expansions of string amplitudes, higher-point celestial strings, celestial string integrands and Carrollian string amplitudes \cite{Donnay:2023kvm,Saha:2024qpt,Castiblanco:2024hnq,Bockisch:2024bia,Stieberger:2024shv}.  In the open superstring four-gluon amplitude, the Yang--Mills partial amplitude is multiplied by the Veneziano form factor \cite{Stieberger:2018edy}
\be
F_I(s,u)=-s\,B(-s,1-u)=
\frac{\Gamma(1-s)\Gamma(1-u)}{ \Gamma(1-s-u)},
\qquad s=\alpha' s_{12},\quad u=-\alpha' s_{23}.
\label{eq:uvir-open-string-formfactor}
\ee
At small energy $F_I\to1$, so the infrared behavior agrees with field theory.  At fixed angle and large energy, however, the ratio of Gamma functions is exponentially soft away from the string poles.  The corresponding celestial integral is \cite{Stieberger:2018edy}
\be
J_I=\int_0^\infty d\omega_4\,\omega_4^{i\sum_i\lambda_i-1}F_I(s,u),
\qquad
s=\alpha'(r-1)\frac{|z_{14}|^2|z_{34}|^2}{|z_{13}|^2}\omega_4^2,
\qquad u=-\frac{s}{ r}.
\label{eq:uvir-string-celestial-integral}
\ee
Here $r$ is the real cross ratio related to the scattering angle by $r^{-1}=\sin^2(\theta/2)$.  Eq.(\ref{eq:uvir-string-celestial-integral}) makes the role of the string scale transparent: the Mellin transform runs over all massive string thresholds, and the soft high-energy behavior of $F_I$ renders the integral much better behaved than its field-theory counterpart.

The same analysis also shows why the celestial field-theory limit is subtle.  In momentum space, the limit $\alpha'\to0$ decouples massive string modes at fixed energy.  In the celestial transform the energy has already been integrated out, and the dependence on $\alpha'$ is largely reorganized into an overall Mellin scale and a series over string thresholds.  The open-string energy integral can be written as \cite{Stieberger:2018edy}
\be
I(r,\beta)=-\Gamma(1-\beta)\frac{r}{2}
\int_0^1 \frac{dx}{ x}\,\left[r\log x-\log(1-x)\right]^{\beta-1},
\label{eq:uvir-string-integral-representation}
\ee
after analytic continuation from the beta-function representation.  This formula exhibits two important structures at once.  First, the first term in the small-angle expansion reproduces the field-theory distribution in $\sum_i\lambda_i$.  Second, the remaining terms are controlled by polylogarithmic periods and by residues at the massive string poles.  Thus a UV-complete celestial amplitude is not obtained by erasing heavy states; rather, the full tower is reorganized into the analytic structure of the celestial correlator.

For closed-string or heterotic amplitudes the same logic persists, with single-valued structures replacing the open-string periods.  The heterotic form factor \cite{Stieberger:2018edy}
\be
F_H(s,u)=-\frac{\Gamma(-s)\Gamma(-t)\Gamma(-u)}{ \Gamma(s)\Gamma(t)\Gamma(u)}
\label{eq:uvir-heterotic-formfactor}
\ee
again suppresses the UV endpoint and converts otherwise divergent gravitational celestial transforms into well-defined stringy correlators.  This provides a useful benchmark: any proposed non-perturbative celestial theory should explain how high-energy softness, massive thresholds and analyticity in Mellin space are encoded intrinsically on the celestial sphere.  Recent work on high-energy string theory and the celestial sphere, as well as S-matrix bootstrap constraints on superpolynomial softness and Regge trajectories, sharpens this point by treating UV softness as a structural constraint rather than a technical regulator \cite{Kervyn:2025wsb,Haring:2023zwu}.

\subsection{Infrared exponentiation and celestial factorization}

The opposite endpoint of the Mellin transform is governed by soft physics.  In momentum space, amplitudes in QED, non-abelian gauge theory and gravity are infrared divergent because infinitely many soft quanta can be emitted at negligible energy.  After introducing an IR cutoff $\Lambda_{\rm IR}$, these divergences exponentiate; in celestial variables, the non-abelian generalization is naturally described in terms of color-dipole soft operators, Wilson-line correlators and dimension shifts \cite{Magnea:2021fvy,Gonzalez:2021dxw,Nastase:2021izh}.  The celestial transform converts this familiar statement into a conformal hard/soft factorization \cite{Arkani-Hamed:2020gyp}
\be
\mathcal A=\mathcal A_{\rm soft}\,\mathcal A_{\rm hard}.
\label{eq:uvir-soft-factorization}
\ee
The content of this formula is basis-dependent.  In the conformal basis, the soft factor is not merely an overall number; it changes how the external states transform under $SL(2,\mathbb C)$.

For massless scalar QED, the Weinberg soft exponent contains the cusp anomalous dimension times $\log\Lambda_{\rm IR}$.  After Mellin transformation, the energy-dependent part of the soft factor can be absorbed into a shift of conformal dimensions \cite{Arkani-Hamed:2020gyp}
\be
\Delta_k\longrightarrow \Delta_k+\alpha Q_k^2,
\qquad
\alpha=\frac{e^2}{4\pi^2}\log\Lambda_{\rm IR},
\label{eq:uvir-qed-dimension-shift}
\ee
while the angular dependence remains as a two-dimensional correlator of the Goldstone mode $\Phi$ for large gauge transformations,
\be
\mathcal A_{\rm soft}=\Big\langle e^{iQ_1\Phi(z_1,\bar z_1)}\cdots e^{iQ_n\Phi(z_n,\bar z_n)}\Big\rangle.
\label{eq:uvir-qed-goldstone}
\ee
The shift in Eq.(\ref{eq:uvir-qed-dimension-shift}) is the celestial avatar of the soft anomalous dimension.  It tells us that IR logarithms in momentum space become anomalous conformal dimensions and logarithmic dependence on the celestial coordinates.  Effective-action and Wilson-line derivations of soft dressings make this statement more structural: the soft modes can be understood as boundary degrees of freedom or generalized Wilson-line dressings rather than as a regulator-dependent artifact \cite{Nguyen:2023ibj,He:2024ddb}.

Gravity is similar but more operator-valued.  If $G_{\Delta_k}(z_k,\bar z_k)$ denotes a graviton conformal primary, the energy multiplication operator becomes a translation operator on the conformal dimension \cite{Arkani-Hamed:2020gyp}
\be
P_kG_{\Delta_k}(z_k,\bar z_k)=\eta_k G_{\Delta_k+1}(z_k,\bar z_k),
\qquad
P_k\mathcal A(\ldots,\Delta_k,\ldots)=\eta_k\mathcal A(\ldots,\Delta_k+1,\ldots),
\label{eq:uvir-gravity-shift-operator}
\ee
where $\eta_k=+1$ for outgoing and $-1$ for incoming particles.  The gravitational soft factor can then be represented as a correlator of the supertranslation Goldstone mode $C(z,\bar z)$, a viewpoint that has been sharpened by deriving the corresponding supertranslation effective action directly from the infrared structure of the gravitational phase space \cite{Nguyen:2021ydb},
\be
\mathcal A_{\rm soft}=\Big\langle e^{iP_1C(z_1,\bar z_1)}\cdots e^{iP_nC(z_n,\bar z_n)}\Big\rangle.
\label{eq:uvir-gravity-goldstone}
\ee
The important difference from QED is that the leading gravitational soft factor shifts the Mellin dimensions rather than simply multiplying by powers of $|z_{ij}|$.  This is why loop-level gravitational celestial amplitudes naturally contain dimension-shift operators and derivatives with respect to conformal weights.  These structures are expected to feed directly into celestial OPE data: logarithms of energies become derivatives in $\Delta$, and soft factors become operator insertions associated with asymptotic symmetries.  The same loop-sensitive soft sector also underlies the loop-corrected celestial stress tensor and logarithmic soft graviton theorems derived from superrotation Ward identities \cite{Donnay:2020lur,Agrawal:2023zea,Choi:2024ygx,Choi:2026cyh}.  Early loop-level analyses already showed how celestial amplitudes can acquire nontrivial conformal structure beyond tree level \cite{Banerjee:2017jeg}, while the infrared dressing problem is closely related to the revised understanding of QED infrared divergences \cite{Kapec:2017tkm}.

Infrared-safe celestial amplitudes can be obtained by dressing the external states.  In the conformal-primary version of Faddeev--Kulish dressing, the soft factor is removed and one is left with \cite{Arkani-Hamed:2020gyp}
\be
\mathcal A_{\rm dressed}=\mathcal A_{\rm hard}.
\label{eq:uvir-dressed-hard}
\ee
This formula is conceptually useful for holography: the celestial theory should distinguish universal soft data, controlled by asymptotic symmetries, from hard scattering data, which contains the genuinely dynamical information.

\subsection{Eikonal exponentiation}

A second mechanism for improving the high-energy endpoint is eikonal exponentiation.  The celestial eikonal regime was first formulated as a weight-shifting resummation on the celestial sphere, and it is the celestial counterpart of the standard gravitational eikonal program reviewed in detail in \cite{deGioia:2022fcn,DiVecchia:2023frv}.  In trans-Planckian small-angle gravitational scattering, ladder and crossed-ladder exchanges resum into a phase.  The parametric regime is \cite{Adamo:2024mqn}
\be
\frac{\hbar}{ E}\ll G E\ll \frac{\hbar}{ q},
\label{eq:uvir-eikonal-hierarchy}
\ee
where $E$ is the center-of-mass energy and $q$ is the transverse momentum transfer.  The left inequality says that the Schwarzschild radius is large compared with the de Broglie wavelength, while the right inequality says that the impact parameter remains large compared with the Schwarzschild radius.

For massless scalar scattering through gravity, the eikonal amplitude may be written as the Born amplitude times a phase.  In the conventions of \cite{Adamo:2024mqn},
\be
\mathcal M_{\rm eik}=\frac{8\pi Gs^2}{ t}\,
\frac{\Gamma(-iGs)}{\Gamma(iGs)}
\left(\frac{4\mu^2}{ -t}\right)^{-iGs},
\label{eq:uvir-eikonal-momentum}
\ee
where $\mu$ is an infrared scale.  Substituting $s=\omega^2$ and $t=-z\omega^2$ gives the celestial eikonal transform \cite{Adamo:2024mqn}
\be
\mathcal A_{\rm eik}(\beta,z)=-\frac{G}{ z}\int_0^\infty d\omega\,\omega^{\beta+1}
\frac{\Gamma(-iG\omega^2)}{\Gamma(iG\omega^2)}
\left(\frac{4\mu^2}{ z\omega^2}\right)^{-iG\omega^2}.
\label{eq:uvir-eikonal-celestial}
\ee
The crucial point is that the Gamma-function ratio and the phase generate oscillations of the schematic form $\omega^{-iG\omega^2}$ at large energy.  These oscillations are absent at any finite order in perturbation theory, but they are precisely what makes the Mellin transform analytically controllable.

The analytic mechanism is captured by the eikonal Gamma function introduced in \cite{Adamo:2024mqn},
\be
\Gamma_E(\beta,\lambda)=\int_0^\infty dx\,x^{\beta-1}e^{-x}x^{-\lambda x}.
\label{eq:uvir-eikonal-gamma}
\ee
For $\lambda>0$ this integral can be analytically continued in $\beta$ by the recursion relation
\be
\Gamma_E(\beta,\lambda)=-\frac{1}{\beta}
\left[(1+\lambda)\Gamma_E(\beta+1,\lambda)
-\lambda\partial_\beta\Gamma_E(\beta+1,\lambda)\right].
\label{eq:uvir-eikonal-gamma-continuation}
\ee
The continuation has poles at negative integers, with increasing pole order as one moves left in the $\beta$-plane.  For the four-point gravitational eikonal amplitude this becomes the statement that \cite{Adamo:2024mqn}
\be
\mathcal A_{\rm eik}(\beta,z)\underset{\beta\to-2(n+1)}{\sim}
-\frac{G^{n+1}}{2z}
\left(\frac{(-i)^n}{(\frac{\beta}{2}+n+1)^{n+1}}+\text{lower-order poles}\right)+\text{regular}.
\label{eq:uvir-eikonal-poles}
\ee
These poles arise from the small-energy expansion of the eikonal integrand, while regularity on the positive real $\beta$ axis reflects the improved high-energy behavior.  The same Mellin amplitude therefore separates two pieces of physics in its analytic structure: infrared and low-energy data appear as poles at negative dimensions, whereas UV softness controls the large positive-$\beta$ behavior.

The eikonal result is also naturally connected with Carrollian holography.  In a Carrollian description, retarded time is conjugate to energy; in the celestial description, Mellin dimension is conjugate to scale.  Eikonal phases shift retarded-time dependence on null infinity and, after Mellin transformation, become non-perturbative functions of $\Delta$.  This is one reason the eikonal regime provides a useful bridge between celestial correlators and Carrollian amplitudes: it is simple enough to compute, but non-perturbative enough to avoid the endpoint pathologies of fixed-order gravity.  The same bridge is visible in the recent development of Carrollian amplitudes, their Feynman rules and loop structure, and their differential equations as non-distributional correlators at null infinity \cite{Mason:2023mti,Liu:2024nfc,Ruzziconi:2024zkr}.

\subsection{Loop corrections}

The UV and IR lessons above should be read together.  Infrared loops exponentiate into soft anomalous dimensions, Goldstone correlators and dimension-shift operators.  Ultraviolet completion or eikonal resummation changes the analytic behavior of the Mellin transform itself.  At loop level, these two effects are entangled: logarithms of momentum invariants become logarithms of celestial cross ratios together with derivatives in conformal dimensions.  Consequently, celestial loop amplitudes need not look like ordinary Euclidean CFT correlators with fixed external dimensions.  They naturally involve mixing in $\Delta$, distributional support from momentum conservation, and logarithmic structures reminiscent of logarithmic CFT.  This interpretation has been sharpened by viewing universal IR factors as marginal deformations of the celestial theory and by identifying logarithmic doublets in the supertranslation sector \cite{He:2023lvk,Bissi:2024brf}.  A related direction deforms the soft theorem itself by a two-dimensional $T\bar T$ deformation, illustrating how irrelevant deformations of a putative celestial theory can be formulated directly at the level of four-dimensional soft-graviton constraints \cite{He:2022zcf}.

This perspective clarifies the role of the celestial OPE.  Collinear limits are local limits on the celestial sphere, but loop corrections dress the corresponding OPE coefficients with soft factors and anomalous dimensions.  In gauge theory this dressing can often be represented as multiplicative angular factors and dimension shifts.  Explicit computations of loop-level gluon and graviton OPEs show that logarithms, double poles in the conformal dimension and mixing among conformally soft operators are not optional decorations but genuine loop effects \cite{Bhardwaj:2022anh,Krishna:2023ukw}.  In gravity the leading soft sector acts through the shift operators in Eq.(\ref{eq:uvir-gravity-shift-operator}), so the OPE algebra is expected to receive corrections that move operators along the conformal-dimension direction.  This is closely related to the appearance of towers of conformally soft gravitons and to the deformation of symmetry algebras discussed later in the review.

The UV and IR discussion therefore supplies the quantum corrections against which any proposed celestial operator algebra or flat-space holographic dictionary must be tested.  The next section turns to one of the sharpest proposed algebraic structures in this direction: the tower of conformally soft gravitons and the associated celestial $w_{1+\infty}$ algebra.

\section{\texorpdfstring{Celestial $w_{1+\infty}$ algebras}{Celestial w1+infinity algebras}}
\label{sec:w-infinity}
Infinite-dimensional symmetry algebras arise in celestial holography because the soft expansion of four-dimensional gauge and gravitational scattering is naturally organized as a tower of two-dimensional currents on the celestial sphere.  The ordinary BMS group already indicates that the infrared sector of gravity is not controlled by a finite-dimensional spacetime group.  The stronger statement, developed in the conformally soft basis, is that the leading holomorphic collinear behavior of celestial graviton operators closes into a chiral algebra \cite{Guevara:2021abz,Strominger:2021mtt}.  In the self-dual sector this algebra is the wedge algebra of $w_{1+\infty}$, the algebra of polynomial area-preserving diffeomorphisms of a two-plane.  This is important for flat-space holography because it gives a concrete algebraic structure which is simultaneously visible in soft theorems, celestial OPEs, twistor geometry and canonical phase-space charges.  Later work has sharpened this picture by deriving the same structure from twistor space, from all-spin OPE constraints, and from asymptotic phase-space charges \cite{Adamo:2021lrv,Himwich:2021dau,Freidel:2021ytz}.

The conceptual role of this section is therefore narrower than a general review of infinite-dimensional algebras, but broader than a derivation of one commutator.  The celestial $w$-algebra should be viewed as a consistency test for any proposed flat-space hologram.  It asks whether soft graviton theorems, local OPE limits, radiative phase-space charges and twistor symmetries can be interpreted as different presentations of the same boundary symmetry.  This is why the section first reviews the flat-space tree-level wedge algebra, then the phase-space derivation, and only then the deformations: the deformed algebras are meaningful only after the undeformed structure and its physical origin are clear.

The point of the construction is not that $w_{1+\infty}$ replaces the BMS algebra.  Rather, the BMS generators sit inside a larger tower of conformally soft graviton modes.  The tower is obtained by taking residues of celestial graviton primaries at special integer conformal dimensions.  Following the original derivation of the holographic symmetry algebras in gauge theory and gravity \cite{Guevara:2021abz} and Strominger's identification of the resulting wedge algebra \cite{Strominger:2021mtt}, define positive-helicity conformally soft gravitons by
\begin{equation}
  H^k(z,\bar z)=\lim_{\varepsilon\to0}\varepsilon\,
  G^+_{k+\varepsilon}(z,\bar z),
  \qquad k=2,1,0,-1,\ldots .
  \label{eq:winf-conformally-soft}
\end{equation}
Here $G^+_\Delta$ is the outgoing positive-helicity graviton conformal primary of dimension $\Delta$.  The residue in Eq.(\ref{eq:winf-conformally-soft}) is nonzero because Mellin-transformed soft theorems produce simple poles at integer values of $\Delta$.  The corresponding two-dimensional weights are \cite{Guevara:2021abz,Strominger:2021mtt}
\begin{equation}
  (h,\bar h)=\left(\frac{k+2}{2},\frac{k-2}{2}\right) .
  \label{eq:winf-soft-weights}
\end{equation}
For $k\leq 2$, $\bar h$ is non-positive.  This is why the conformally soft modes are not ordinary unitary CFT primaries; they are symmetry currents extracted from the infrared singularities of the four-dimensional $S$-matrix.  The finite antiholomorphic mode expansion
\begin{equation}
  H^k(z,\bar z)=\sum_{n=(k-2)/2}^{(2-k)/2}
  \frac{H^k_n(z)}{\bar z^{n+(k-2)/2}}
  \label{eq:winf-soft-mode-expansion}
\end{equation}
then follows from the fact that the operator has antiholomorphic conformal weight $(k-2)/2$ \cite{Guevara:2021abz,Strominger:2021mtt}.  The range of $n$ is finite, and this finiteness is what ultimately produces the wedge, rather than the full centrally extended quantum $W_{1+\infty}$ algebra.

\subsection{From the graviton OPE to the wedge algebra}

The most economical derivation of the celestial $w_{1+\infty}$ algebra begins with the universal collinear OPE of two positive-helicity graviton primaries.  At tree level in Einstein gravity, the leading holomorphic singularity is \cite{Pate:2019lpp,Guevara:2021abz,Strominger:2021mtt,Himwich:2021dau,Ren:2023trv}
\begin{equation}
  G^+_{\Delta_1}(z_1,\bar z_1)G^+_{\Delta_2}(z_2,\bar z_2)
  \sim -\frac{\kappa}{2}\frac{1}{z_{12}}
  \sum_{n=0}^{\infty}B(\Delta_1-1+n,\Delta_2-1)
  \frac{\bar z_{12}^{n+1}}{n!}\,
  \bar\partial^n G^+_{\Delta_1+\Delta_2}(z_2,\bar z_2) .
  \label{eq:winf-graviton-ope}
\end{equation}
Here $z_{12}=z_1-z_2$, $\bar z_{12}=\bar z_1-\bar z_2$, $B(x,y)=\Gamma(x)\Gamma(y)/\Gamma(x+y)$, and $\kappa=\sqrt{32\pi G_N}$.  Eq.(\ref{eq:winf-graviton-ope}) is the celestial transform of the universal holomorphic collinear splitting of graviton amplitudes.  Its worldsheet origin has also been derived in string and ambitwistor formulations, where the celestial OPE is inherited from a short-distance OPE of vertex operators \cite{Jiang:2021csc,Adamo:2021zpw,Bu:2021avc}.  Its importance is that the coefficient is a beta function in the Mellin dimensions.  Taking residues at the conformally soft poles of the two external gravitons therefore turns a kinematic collinear singularity into a current algebra on the celestial sphere.

Using Eq.(\ref{eq:winf-conformally-soft}) in Eq.(\ref{eq:winf-graviton-ope}), one obtains the OPE of the conformally soft modes, and equivalently the commutator of their holomorphic modes.  In the normalization of \cite{Guevara:2021abz},
\begin{equation}
\begin{split}
  [H^k_m,H^l_n]
  ={}& -\frac{\kappa}{2}\,[n(2-k)-m(2-l)]
  \frac{\left(\frac{2-k}{2}-m+\frac{2-l}{2}-n-1\right)!}
       {\left(\frac{2-k}{2}-m\right)!\left(\frac{2-l}{2}-n\right)!} \\
  &\times
  \frac{\left(\frac{2-k}{2}+m+\frac{2-l}{2}+n-1\right)!}
       {\left(\frac{2-k}{2}+m\right)!\left(\frac{2-l}{2}+n\right)!}
  H^{k+l}_{m+n} .
\end{split}
  \label{eq:winf-H-commutator}
\end{equation}
The factorial factors encode the finite mode ranges in Eq.(\ref{eq:winf-soft-mode-expansion}).  They are somewhat unnatural from the viewpoint of the final symmetry algebra, but they are natural from the celestial OPE because they remember the normalization of the conformally soft operators.

The simplicity of the answer appears after the rescaling introduced in \cite{Strominger:2021mtt},
\begin{equation}
  w^p_m(z)=\frac{1}{\kappa}(p-m-1)!(p+m-1)!
  H^{-2p+4}_m(z),
  \qquad p=1,\frac32,2,\frac52,\ldots,
  \label{eq:winf-rescaling}
\end{equation}
with wedge range
\begin{equation}
  1-p\leq m\leq p-1 .
  \label{eq:winf-wedge-range}
\end{equation}
The commutator becomes
\begin{equation}
  [w^p_m,w^q_n]=\bigl[m(q-1)-n(p-1)\bigr]w^{p+q-2}_{m+n} .
  \label{eq:winf-wedge-algebra}
\end{equation}
This is the wedge algebra of $w_{1+\infty}$ in the conventions of \cite{Strominger:2021mtt}.  It is the algebra of polynomial Hamiltonian vector fields on a two-dimensional phase space.  Indeed, if
\begin{equation}
  w^p_m=u^{p+m-1}v^{p-m-1},\qquad
  \{f,g\}=\partial_u f\,\partial_v g-\partial_v f\,\partial_u g,
  \label{eq:winf-poisson-realization}
\end{equation}
then the Poisson bracket reproduces Eq.(\ref{eq:winf-wedge-algebra}), up to a convention-dependent overall factor \cite{Strominger:2021mtt,Bu:2022iak}.  The overall factor is not physical; it can be absorbed into a universal rescaling of the generators, but it should be kept in mind when comparing the amplitude, twistor and deformation literature.  Twistor constructions make this phase-space interpretation geometrical: the loop algebra $Lw_{1+\infty}$ acts by Poisson diffeomorphisms on twistor space, and its charges can be obtained from twistor actions or represented on Carrollian data at null infinity \cite{Adamo:2021lrv,Mason:2022hly,Kmec:2024nmu,Donnay:2024qwq}.  This realization explains the frequent description of the celestial algebra as an algebra of area-preserving diffeomorphisms.  The adjective ``wedge'' is essential: the finite range Eq.(\ref{eq:winf-wedge-range}) keeps only the globally well-defined polynomial modes.  Possible central extensions and the full quantum $W$-algebra require additional input, and are not fixed by the tree-level celestial OPE alone.  A complementary covariant approach defines wedge algebras directly on cuts of null infinity, clarifying how the familiar loop algebra depends on the topology and geometry of the cut \cite{Geiller:2024bgf,Cresto:2024fhd}.

The same construction also organizes gauge-theory currents.  It is useful to spell out the gluon case because it gives the simplest example of an infinite-dimensional soft algebra, often called the $S$-algebra, that is parallel to but distinct from the gravitational $w_{1+\infty}$ wedge algebra.  Let $O^{a,+}_\Delta(z,\bar z)$ denote an outgoing positive-helicity gluon primary with adjoint color index $a$.  The holomorphic collinear singularity of the Yang--Mills amplitude implies the celestial OPE \cite{Guevara:2021abz}
\begin{equation}
  O^{a,+}_{\Delta_1}(z_1,\bar z_1)O^{b,+}_{\Delta_2}(z_2,\bar z_2)
  \sim
  \frac{-i f^{ab}{}_{c}}{z_{12}}\,
  B(\Delta_1-1,\Delta_2-1)\,
  O^{c,+}_{\Delta_1+\Delta_2-1}(z_2,\bar z_2),
  \label{eq:soft-gluon-primary-ope}
\end{equation}
where $z_{12}=z_1-z_2$, $B(x,y)=\Gamma(x)\Gamma(y)/\Gamma(x+y)$, and $f^{ab}{}_{c}$ are the Lie-algebra structure constants.  The poles of the beta function at $\Delta=k=1,0,-1,\ldots$ define the conformally soft gluon currents
\begin{equation}
  R^{k,a}_{n}(z):=\lim_{\varepsilon\to0}\varepsilon\,
  O^{a,+}_{k+\varepsilon,n}(z),
  \qquad
  \frac{k-1}{2}\le n\le \frac{1-k}{2},
  \qquad
  k=1,0,-1,\ldots ,
  \label{eq:soft-gluon-R-definition}
\end{equation}
after expanding the right-moving dependence as
\begin{equation}
  R^{k,a}(z,\bar z)
  =\sum_{n=(k-1)/2}^{(1-k)/2}
  \frac{R^{k,a}_{n}(z)}{\bar z^{\,n+(k-1)/2}},
  \qquad
  (h,\bar h)=\left(\frac{k+1}{2},\frac{k-1}{2}\right).
  \label{eq:soft-gluon-R-modes}
\end{equation}
The finite range of $n$ is the finite-dimensional $SL(2)_R$ multiplet carried by the conformally soft pole.  Including the antiholomorphic descendants in Eq.(\ref{eq:soft-gluon-primary-ope}) gives a closed OPE for these soft currents \cite{Guevara:2021abz},
\begin{equation}
  R^{k,a}(z_1,\bar z_1)R^{\ell,b}(z_2,\bar z_2)
  \sim
  \frac{-i f^{ab}{}_{c}}{z_{12}}
  \sum_{r=0}^{1-k}
  \binom{2-k-\ell-r}{1-\ell}
  \frac{(\bar z_{12})^r}{r!}\,
  \bar\partial^{\,r}R^{k+\ell-1,c}(z_2,\bar z_2).
  \label{eq:soft-gluon-R-ope}
\end{equation}
Taking the holomorphic contour commutator and extracting the finite $SL(2)_R$ modes then gives
\begin{equation}
  \begin{aligned}
  [R^{k,a}_{n},R^{\ell,b}_{n'}]
  ={}&-i f^{ab}{}_{c}\,
  \frac{
    \left(\frac{1-k}{2}-n+\frac{1-\ell}{2}-n'\right)!
  }{
    \left(\frac{1-k}{2}-n\right)!
    \left(\frac{1-\ell}{2}-n'\right)!
  }
  \\[2pt]
  &\times
  \frac{
    \left(\frac{1-k}{2}+n+\frac{1-\ell}{2}+n'\right)!
  }{
    \left(\frac{1-k}{2}+n\right)!
    \left(\frac{1-\ell}{2}+n'\right)!
  }\,
  R^{k+\ell-1,c}_{n+n'} .
  \end{aligned}
  \label{eq:soft-gluon-R-algebra}
\end{equation}
Eq.(\ref{eq:soft-gluon-R-algebra}) is the explicit soft-current algebra obtained from the collinear OPE.  Its factorial coefficients are the Clebsch--Gordan data for the finite $SL(2)_R$ multiplets.  In the basis introduced in \cite{Strominger:2021mtt},
\begin{equation}
  S^{q,a}_{m}
  =(q-m-1)!(q+m-1)!\,R^{3-2q,a}_{m},
  \qquad
  1-q\le m\le q-1,
  \qquad
  q=1,\frac{3}{2},2,\ldots ,
  \label{eq:soft-gluon-S-basis}
\end{equation}
these coefficients are absorbed into the normalization of the modes and the algebra collapses to the simple form
\begin{equation}
  [S^{p,a}_{m},S^{q,b}_{n}]
  =-i f^{ab}{}_{c}\,S^{p+q-1,c}_{m+n}.
  \label{eq:soft-gluon-S-algebra}
\end{equation}
The leading current $S^{1,a}_0$ is the global color generator, while the tower with $q>1$ contains the conformally soft descendants of the leading and subleading gluon theorems.  Thus the gauge-theory soft algebra is a color current algebra graded by the same $SL(2)_R$ quantum number that organizes the gravitational wedge modes.  The gravitational algebra acts on this tower as
\begin{equation}
  [w^p_m,S^{q,a}_n]
  =\bigl[m(q-1)-n(p-1)\bigr]S^{p+q-2,a}_{m+n},
  \label{eq:winf-gauge-module}
\end{equation}
so the gluon soft currents form a module for the celestial $w_{1+\infty}$ wedge algebra \cite{Guevara:2021abz,Strominger:2021mtt,Alday:2026rso}.  This illustrates a general lesson: the $w_{1+\infty}$ structure is not a separate symmetry added to BMS or large gauge symmetry, but a higher tower into which the familiar soft symmetries embed.  The same logic extends to arbitrary helicity and to non-minimal or supersymmetric deformations, where the Jacobi identity imposes nontrivial constraints on the spectrum and couplings \cite{Himwich:2021dau,Mago:2021wje,Bu:2021avc,Crawley:2024cak}.  In supersymmetric theories this viewpoint can be made geometrically explicit by replacing the celestial sphere with a celestial supersphere, which organizes conformally soft theorems, OPEs and supersymmetric \(Lw_{1+\infty}^{\wedge}\)-type algebras in a unified language \cite{Tropper:2024css}.  Further supersymmetric and deformed realizations of celestial \(w_{1+\infty}\)-type algebras have been studied in celestial CFT, celestial holography and extended supergravity, including recent \({\cal N}=8\) soft-algebra and multiparticle contributions \cite{Ahn:2021erj,Ahn:2022oor,Ahn:2024kpv,Ahn:2025sid,Ahn:2025hoy,Ahn:2026err}.

\subsection{Phase-space charges}
\label{subsec:w-phase-space-einstein}

The derivation of the celestial $w_{1+\infty}$ algebra from OPEs is powerful because it ties the algebra directly to factorization and conformally soft graviton insertions.  It is not, however, the only way in which the algebra appears.  A conceptually different route was developed in \cite{Freidel:2021ytz}, where the same structure is extracted from the large-radius expansion of Einstein's equations and then represented by charges on the gravitational phase space.  This is important because it shows that the celestial algebra is not merely an artifact of the Mellin basis or of a particular OPE limit.  It is also encoded in the dynamics of asymptotically flat gravity itself.

The starting point is the Bondi description of the asymptotic gravitational field.  In the notation of \cite{Freidel:2021ytz}, the metric is written as
\be
 ds^2=-2e^{2\beta}du\left(dr+\Phi du\right)+r^2\gamma_{AB}
  \left(d\sigma^A-\frac{\Upsilon^A}{r^2}du\right)
   \left(d\sigma^B-\frac{\Upsilon^B}{r^2}du\right),
   \label{eq:fpr-bondi-metric}
   \ee
   where $u$ is retarded time, $r$ is an areal radial coordinate, $\sigma^A$ are coordinates on a cut of null infinity, and $\gamma_{AB}$ is the metric on that cut.  The radiative data are encoded in the shear $C$ and the news $\hat N=\partial_u\bar C$ (up to the conventions used in \cite{Freidel:2021ytz}).  Expanding the Newman--Penrose Weyl scalars at large $r$ identifies an infinite sequence of quantities $\mathcal Q_s$ whose evolution is governed by the vacuum Einstein equations.  At the linearized level this tower obeys
   \be
   \dot{\mathcal Q}_{s+1}=D\mathcal Q_s,
   \qquad s\geq 2,
   \label{eq:fpr-linear-higher-spin-evolution}
   \ee
   where the dot denotes $\partial_u$ and $D$ is the spin-raising covariant derivative on the celestial sphere \cite{Freidel:2021ytz}.  Eq.(\ref{eq:fpr-linear-higher-spin-evolution}) is the first sign of the higher-spin structure: successive charges are not independent but are linked by the angular derivative and time evolution dictated by Einstein's equations.
   
The nonlinear construction refines this into a recursive hierarchy.  Writing $\mathcal Q_s^k$ for the component of order $k$ in the radiative data, one obtains \cite{Freidel:2021dfs}
   \be
   \mathcal Q_s^k=D\partial_u^{-1}\mathcal Q_{s-1}^k+\frac{s+1}{2}\partial_u^{-1}\left(C\mathcal Q_{s-2}^{k-1}\right),
   \label{eq:fpr-charge-recursion}
   \ee
   with $\partial_u^{-1}$ defined by integrating from future timelike infinity back to the point $u$ \cite{Freidel:2021ytz}.  The first term is the linear propagation of the tower, while the second term is the nonlinear gravitational correction.  This recursion is the phase-space analogue of the celestial OPE recursion: in the OPE derivation the algebra follows from collinear factorization; here it follows from repeatedly solving the Einstein equations near null infinity.
   
   The associated soft charge density is obtained by taking an early-time limit of a specific linear combination of the $\mathcal Q_s$.  Its linear piece is
   \be
   q_s^1(z)=D^{s+2}N_s(z),
   \qquad
   N_s(z)=\frac{1}{2}\frac{(-1)^{s+1}}{s!}
   \int_{-\infty}^{+\infty}du\,u^s\hat N(u,z).
   \label{eq:fpr-soft-charge-density}
   \ee
   The moments $N_s$ are precisely the conformally soft graviton modes appearing in the celestial construction \cite{Freidel:2021ytz}.  Thus the charge density built from the radiative phase space knows about the same residues in conformal dimension that appear in the OPE analysis.  In celestial variables this relation is made explicit by
   \be
   G^-_\Delta(z)=-\frac{\Gamma(\Delta-1)}{2}\int_{-\infty}^{+\infty}du\,u^{-\Delta+1}\hat N(u,z),
   \qquad
   {\rm Res}_{\Delta=1-s}G^-_\Delta(z)=N_s(z),
   \label{eq:fpr-conformally-soft-residue}
   \ee
   which identifies the soft moments with poles of the conformal primary graviton \cite{Freidel:2021ytz}.  This equation is the bridge between the two languages: the phase-space charge is a moment of the Bondi news, while the celestial current is the same moment viewed as a residue of a Mellin-transformed operator.
   
   To act on the gravitational phase space, the local charge density is smeared with a function $\tau_s$ on the celestial sphere,
   \be
   Q_s(\tau)=\frac{8}{\kappa^2}\int_S d^2z\sqrt q\,\tau_s(z)q_s(z),
   \label{eq:fpr-smeared-charge}
   \ee
   where $q$ is the determinant of the metric on the cut and $\kappa^2=32\pi G$ \cite{Freidel:2021ytz}.  The quadratic part of the charge generates a canonical transformation of the shear.  A compact form of the action is
   \be
   \{Q_s^2(\tau),C(u,z)\}=\sum_{p=0}^s\frac{u^{s-p}}{(s-p)!}\,\delta^p_{D^{s-p}\tau_s}C(u,z),
   \label{eq:fpr-charge-action-on-shear}
   \ee
   with
   \be
   \delta^p_\tau C=
   \sum_{k=0}^{\min(3,p)}\binom{3}{k}(p+1-k)(D^k\tau)
   D^{p-k}\partial_u^{1-p}C.
   \label{eq:fpr-pseudodifferential-action}
   \ee
   These formulae show why the symmetry is higher-spin-like.  For $p=0$ the transformation is a time translation of the shear; for $p=1$ it resembles a chiral diffeomorphism; for larger $p$ it becomes pseudo-differential in retarded time and angular derivatives.  The celestial $w_{1+\infty}$ algebra is therefore not just a current algebra on the sphere, but a canonical action on radiative gravitational data.
   
   The Poisson brackets of the charges then reproduce the loop algebra.  To linear order in the radiative fields one finds
   \be
   \{Q_s(\tau),Q_{s'}(\tau')\}^1
   =(s'+1)Q^1_{s+s'-1}(\tau'D\tau)
   -(s+1)Q^1_{s+s'-1}(\tau D\tau'),
   \label{eq:fpr-smeared-charge-bracket}
   \ee
   and, after expanding the smearing functions in modes,
   \be
   [Q^s_{m,n},Q^{s'}_{m',n'}]
   =i\left[m(1+s')-m'(1+s)\right]
   Q^{s+s'-1}_{m+m'-1,n+n'}.
   \label{eq:fpr-loop-w-algebra}
   \ee
   This is the $Lw_{1+\infty}$ structure on the gravitational phase space \cite{Freidel:2021ytz}.  The result matches the wedge algebra obtained from celestial OPEs, but its meaning is different: the bracket is a canonical bracket of gravitational charges, not an operator product assumed in a putative celestial CFT.
   
   This phase-space derivation also clarifies the relation between several approaches to celestial symmetry.  The OPE derivation identifies the algebra by taking conformally soft limits of graviton operators.  The soft-theorem derivation identifies the same modes through Ward identities of the S-matrix.  The phase-space derivation explains why these Ward identities should exist in gravity: the charges are constructed from the radiative data and their conservation follows from the large-$r$ Einstein equations.  Later work has extended this viewpoint by analyzing how the celestial charges probe the subleading structure of asymptotically flat spacetimes \cite{Geiller:2024bgf}, by giving covariant formulations of the gravitational wedge algebra on cuts of null infinity \cite{Cresto:2024fhd}, and by connecting the same $Lw_{1+\infty}$ structure to twistor-action and Carrollian descriptions \cite{Kmec:2024nmu,Donnay:2024qwq}.  These developments reinforce the idea that the celestial algebra is simultaneously an OPE algebra, a soft algebra, and a gravitational phase-space algebra.
   
   The construction above is classical and, in its cleanest form, tied to the radiative phase space and the wedge sector.  Its purpose in the present section is therefore not to settle the quantum completion of the algebra, but to establish a more invariant origin for the same current algebra seen in celestial OPEs.  The lesson is decisive: the celestial $w_{1+\infty}$ algebra is not only a property of collinear OPEs.  It can be read directly from Einstein's equations and represented by canonical charges acting on gravitational radiation.
\subsection{Deformations and backgrounds}

The original algebra Eq.(\ref{eq:winf-wedge-algebra}) is a tree-level, flat-space, self-dual structure.  Several recent developments ask which parts of it survive once one changes the background, includes a cosmological constant, or quantizes the self-dual sector.  This question has also been studied through one-loop associativity tests of the celestial OPE and through explicit self-dual gravity computations \cite{Bittleston:2022jeq,Ball:2021tmb,Banerjee:2023jne,Krishna:2023ukw}.  These works are best understood as deformations of the phase-space bracket Eq.(\ref{eq:winf-poisson-realization}) and of the celestial OPE Eq.(\ref{eq:winf-graviton-ope}), rather than as unrelated algebraic proposals.

A first controlled deformation is the Moyal deformation of self-dual gravity.  The Poisson bracket realization of the wedge algebra can be promoted to a star-commutator \cite{Bu:2022iak}; this deformation is closely related to chiral higher-spin theory and its deformed celestial operator algebras \cite{Monteiro:2022xwq},
\begin{equation}
  \{f,g\}_q=q^{-1}(f\star g-g\star f),\qquad
  f\star g=f\exp\!\left[q\left(\overleftarrow\partial_u\overrightarrow\partial_v
  -\overleftarrow\partial_v\overrightarrow\partial_u\right)\right]g .
  \label{eq:winf-moyal-bracket}
\end{equation}
The parameter $q$ controls the noncommutativity of the two-dimensional phase space.  Eq.(\ref{eq:winf-moyal-bracket}) is written in the convention of \cite{Bu:2022iak}.  With the monomial basis Eq.(\ref{eq:winf-poisson-realization}), this convention differs from Eq.(\ref{eq:winf-wedge-algebra}) by the same harmless overall factor already noted above; equivalently, one may rescale all generators or replace $q^{-1}$ by $(2q)^{-1}$ if one wants the $q\to0$ limit to reproduce exactly the normalization of Eq.(\ref{eq:winf-wedge-algebra}).  The Jacobi identity is nevertheless automatic, because the bracket is the commutator of an associative star product.  At finite $q$ the algebra is a Moyal deformation of the wedge algebra, closely related to a wedge subalgebra of a $W$-algebra.  This provides a precise sense in which noncommutative or higher-spin-like corrections can deform celestial $w_{1+\infty}$ while preserving associativity.

Another important test is scattering on nontrivial self-dual radiative backgrounds, defects and gravitational instanton backgrounds \cite{Bittleston:2023bzp,Bogna:2024gnt}.  The leading holomorphic collinear splitting functions, and hence the associated celestial chiral algebra, are remarkably robust on such backgrounds \cite{Adamo:2023zeh}.  In the corresponding twistor-space description the soft graviton currents obey
\begin{equation}
  w^p_m(z)w^q_n(z')\sim
  \frac{2\,[m(q-1)-n(p-1)]}{z-z'}
  w^{p+q-2}_{m+n}(z') .
  \label{eq:winf-radiative-ope}
\end{equation}
Expanding the holomorphic currents as
\begin{equation}
  w^p_m(z)=\sum_r z^{r-1}g^p_{m,r},
  \qquad
  \{g^p_{m,r},g^q_{n,s}\}
  =2[m(q-1)-n(p-1)]g^{p+q-2}_{m+n,r+s},
  \label{eq:winf-radiative-modes}
\end{equation}
one obtains a loop-algebra version of the same celestial $w_{1+\infty}$ structure \cite{Adamo:2023zeh}.  The factor of two is a convention.  Conceptually, the result suggests that the $w_{1+\infty}$ algebra is tied to the integrable self-dual sector and to holomorphic collinear factorization, rather than to the strict vacuum background alone.

A qualitatively different deformation arises for nonzero cosmological constant.  From the viewpoint of celestial amplitudes, a cosmological constant modifies the relation between translation generators and the celestial sphere.  A curvature-corrected celestial graviton OPE is consistent with a deformation of the wedge algebra \cite{Taylor:2023ajd}.  Related AdS$_4$ analyses of chiral boundary conditions, light-ray operators in CFT$_3$ and self-dual dynamics show that curvature naturally deforms the Poisson bracket and the associated chiral algebra \cite{Gupta:2022mdt,Lipstein:2023pih,Sheta:2025oep,Bittleston:2024rqe,Chowdhury:2024dcy}.  The leading terms in the positive-helicity graviton OPE take the form
\begin{equation}
\begin{split}
  G^+_{\Delta_3}(z_3,\bar z_3)G^+_{\Delta_4}(z_4,\bar z_4)
  ={}&-\frac{\kappa}{2}\frac{\bar z_{34}}{z_{34}}
  B(\Delta_3-1,\Delta_4-1)G^+_{\Delta_3+\Delta_4}(z_4,\bar z_4)\\
  &+\frac{\kappa\Lambda}{2}\frac{\Delta_3+\Delta_4}{z_{34}^{2}}
  B(\Delta_3-2,\Delta_4-2)G^+_{\Delta_3+\Delta_4-2}(z_4,\bar z_4)\\
  &+\frac{\kappa\Lambda}{2}\frac{\Delta_3}{z_{34}}
  B(\Delta_3-2,\Delta_4-2)\partial G^+_{\Delta_3+\Delta_4-2}(z_4,\bar z_4) .
\end{split}
  \label{eq:winf-lambda-ope}
\end{equation}
The first line is the flat-space OPE Eq.(\ref{eq:winf-graviton-ope}) at leading order in the antiholomorphic expansion.  The remaining terms are proportional to the cosmological constant $\Lambda$ and shift the conformal dimension of the exchanged graviton.  Their presence is forced by the deformed action of translations and by associativity constraints on the OPE.

The resulting algebra can be written as \cite{Taylor:2023ajd}
\begin{equation}
\begin{split}
  [w^p_{a,m},w^q_{b,n}]
  ={}&[m(q-1)-n(p-1)]w^{p+q-2}_{a+b,m+n}\\
  &-\Lambda[a(q-2)-b(p-2)]w^{p+q-1}_{a+b,m+n} .
\end{split}
  \label{eq:winf-lambda-algebra}
\end{equation}
The additional label $a$ is conjugate to the extra mode structure required at nonzero curvature.  The relative sign and the dependence $a(q-2)-b(p-2)$ in the second line are not arbitrary: in the OPE derivation they are the minimal modification needed for associativity, while in the twistor derivation they follow from the $\Lambda$-deformed Poisson bracket and hence satisfy the Jacobi identity automatically \cite{Taylor:2023ajd,Bittleston:2024rqe}.  When $\Lambda\to0$, the second line disappears and Eq.(\ref{eq:winf-lambda-algebra}) reduces to the flat wedge algebra Eq.(\ref{eq:winf-wedge-algebra}).  A useful check is that the finite-dimensional subalgebra generated by
\begin{equation}
  w^1_{a,0}=\Lambda L_a,
  \qquad
  w^2_{0,m}=\bar L_m,
  \qquad
  w^{3/2}_{k,l}=P_{k,l}
  \label{eq:winf-lambda-identification}
\end{equation}
contains the curvature-deformed translation commutator
\begin{equation}
  [P_{i,j},P_{k,l}]
  =\Lambda j\,\delta_{j,-l}L_{i+k}
  +\Lambda i\,\delta_{i,-k}\bar L_{j+l} .
  \label{eq:winf-curved-translations}
\end{equation}
This is precisely what one expects from the isometry algebra of de Sitter or anti-de Sitter space: translations commute in the Poincare limit, but close onto Lorentz generators at nonzero curvature.

The same deformation has a particularly transparent twistor interpretation.  For self-dual gravity with cosmological constant the relevant twistor bracket is \cite{Bittleston:2024rqe}
\begin{equation}
  \{f,g\}_{\Lambda}
  =\epsilon^{\dot\alpha\dot\beta}
  \frac{\partial f}{\partial\mu^{\dot\alpha}}
  \frac{\partial g}{\partial\mu^{\dot\beta}}
  +\Lambda\epsilon_{\alpha\beta}
  \frac{\partial f}{\partial\lambda_{\alpha}}
  \frac{\partial g}{\partial\lambda_{\beta}} .
  \label{eq:winf-twistor-bracket}
\end{equation}
Here $(\lambda_\alpha,\mu^{\dot\alpha})$ are homogeneous twistor coordinates.  The first term is the familiar flat twistor Poisson bracket, while the second term is the curvature correction.  Evaluated on the appropriate monomial basis, Eq.(\ref{eq:winf-twistor-bracket}) reproduces Eq.(\ref{eq:winf-lambda-algebra}).  This derivation makes the Jacobi identity automatic, since it follows from the Poisson structure, and clarifies why the cosmological deformation is geometric rather than merely a formal modification of structure constants.

The deformed algebra also acts on matter or gauge-theory modules.  For a field of helicity label $s$, the twistor analysis gives \cite{Bittleston:2024rqe}
\begin{equation}
\begin{split}
  \{w^p_{m,a},x^q_{n,b}\}_{\Lambda}
  ={}&[m(q-1)-n(p-1)]x^{p+q-2}_{m+n,a+b}\\
  &-\Lambda[a(q-s)-b(p-2)]x^{p+q-1}_{m+n,a+b} .
\end{split}
  \label{eq:winf-lambda-module}
\end{equation}
This formula shows that curvature deforms not only the gravitational algebra, but also its representation on celestial matter operators.  It is therefore directly relevant to the question of how celestial symmetries should be formulated in AdS$_4$ and dS$_4$, and how a flat-space celestial algebra may emerge as a limit of more conventional holographic settings.  Recent CFT$_3$ analyses go further by realizing the $\mathcal L_\Lambda w_{1+\infty}$ action through light-ray or ANEC-type operators, suggesting that the deformed soft algebra may have a direct boundary-CFT avatar rather than only a bulk twistor or amplitude realization \cite{Sheta:2025oep,Strominger:2026yrh}.

The results reviewed above point to a coherent picture: the undeformed celestial $w_{1+\infty}$ algebra is the chiral algebra of conformally soft gravitons in flat-space tree-level scattering, while Moyal, background and cosmological-constant deformations show how this algebra can be modified in controlled settings.  The remaining conceptual issues are collected in Section~
\ref{sec:outlook}.

\section{Twistor methods}
\label{sec:twistor-ambitwistor}

Twistor methods enter celestial holography because they are adapted to the same null geometry that underlies the celestial basis.  A point on the celestial sphere labels a null direction; a Mellin weight labels the scaling of the corresponding null momentum; helicity is most naturally expressed in spinor variables.  Twistor and ambitwistor constructions package these ingredients in a way that makes conformal covariance, soft limits, chiral current algebras and scattering-equation formulae unusually transparent.  The aim of this section is therefore not to review twistor theory as a subject in its own right (for a useful review of twistor theory, see \cite{Adamo:2017qyl}), but to explain how twistorial variables formulate celestial amplitudes, derive celestial OPEs and expose the symmetry structures discussed in the previous sections.  Direct twistor-space formulations, supersymmetric and stringy worldsheet extensions, all-order OPE constructions in special helicity sectors, and twistor-action derivations of celestial chiral algebras develop this perspective in complementary directions.  The curved-background side has an important precursor in the twistor construction of gravity with a cosmological constant from rational curves \cite{Adamo:2015ina}, which provides useful context for the AdS/dS and cosmological-constant deformations discussed below.

The reader should also keep in mind a limitation.  Twistor methods are most powerful in sectors where holomorphy, self-duality, MHV structure or worldsheet localization controls the dynamics.  This is why they are so effective for celestial OPEs and chiral symmetry algebras, but it is not yet known whether the same language gives a complete description of generic non-self-dual, loop-level gravitational scattering.  In the present review twistor methods are therefore treated as an efficient window into protected structures, and as a guide to possible holographic organization, rather than as a finished nonperturbative definition of celestial holography.

\subsection{Null directions, spinors and twistor data}
\label{subsec:twistor-null-directions}

The starting point is the spinor-helicity form of a null momentum.  In four dimensions one writes
\be
  p_{\alpha\dot\alpha}=\lambda_\alpha\tilde\lambda_{\dot\alpha},
  \qquad p^2=0 .
\label{eq:twistor-spinor-momentum}
\ee
In the celestial parametrization used throughout this review, the scale of the spinors is separated from their projective direction.  For an outgoing or incoming massless particle one may take
\be
  \lambda_\alpha=\sqrt{\omega}\,z_\alpha,
  \qquad
  \tilde\lambda_{\dot\alpha}=\varepsilon\sqrt{\omega}\,\bar z_{\dot\alpha},
  \qquad
  z_\alpha=(1,z),
  \quad \bar z_{\dot\alpha}=(1,\bar z),
\label{eq:twistor-celestial-spinors}
\ee
with $\varepsilon=+1$ for outgoing and $\varepsilon=-1$ for incoming states.  This is the spinorial form of the momentum dictionary in Section~\ref{sec:celestial-amplitudes}; it is also the parametrization used in the worldsheet derivation of celestial OPEs in ref.~\cite{Adamo:2021zpw}.  The Mellin transform is therefore a transform along the common scaling of $\lambda$ and $\tilde\lambda$, while $(z,\bar z)$ remain projective spinor coordinates on the celestial sphere.  This simple observation is the technical reason twistors are so well matched to celestial amplitudes: the transformation from plane waves to conformal primaries acts on the radial scale of spinor-helicity variables, not on the projective twistor data.

A twistor is written as $Z^A=(\lambda_\alpha,\mu^{\dot\alpha})$, with the incidence relation
\be
  \mu^{\dot\alpha}=x^{\alpha\dot\alpha}\lambda_\alpha .
\label{eq:twistor-incidence-relation}
\ee
In the celestial setting this relation should be viewed less as new kinematics than as a geometric reorganization of familiar scattering data.  The projective spinor $\lambda_\alpha$ carries the holomorphic celestial coordinate, while $\mu^{\dot\alpha}$ is the variable conjugate to spacetime position along the corresponding null direction.  This geometric reorganization has been developed both in direct celestial twistor-amplitude constructions \cite{Brown:2022miw} and in scaling reductions of twistor space that relate celestial holography to lower-dimensional chiral structures \cite{Bu:2023cef}.  In the twistor-space account of celestial $w_{1+\infty}$ symmetries, the radiative data at null infinity define an asymptotic twistor space, and the self-dual sector of gravity is represented by deformations of its complex structure \cite{Adamo:2021lrv}.  This is why twistors are useful for celestial holography: they put the null, chiral and conformal aspects of the problem into the same geometric language.

There is also a useful contrast with the Carrollian picture.  Carrollian variables keep the retarded time $u$ at null infinity explicit, while twistor variables emphasize the complex geometry of null rays and their helicity data.  They are therefore complementary rather than competing descriptions.  The twistor approach is particularly efficient when the physics is controlled by holomorphicity, self-duality, soft residues or collinear limits; the Carrollian approach is better suited for questions involving flux, memory and time evolution at $\mathscr I$.

\subsection{Ambitwistor strings and conformally soft vertex operators}
\label{subsec:ambitwistor-soft-vertices}

Ambitwistor strings provide a worldsheet representation of field-theory scattering amplitudes in which the scattering equations arise dynamically.  The relevance to celestial holography was made explicit in ref.~\cite{Adamo:2019ipt}, where the usual momentum-eigenstate vertex operators were replaced by vertex operators adapted to the conformal primary basis.  This worldsheet viewpoint was subsequently extended to celestial OPEs in string theory, supersymmetric soft algebras and celestial string integrands, making it one of the main systematic ways to derive operator-algebraic data rather than postulate it \cite{Jiang:2021csc,Jiang:2021xzy,Bu:2021avc,Castiblanco:2024hnq,Bockisch:2024bia}.  The key technical point is that the Mellin integral can be included inside the vertex operator, so the worldsheet correlator computes a celestial object directly.  For Yang-Mills theory, the integrated conformal-basis vertex operator takes the schematic but precise form
\be
  U=\int_0^\infty \frac{\dd t}{t}\,t^\Delta
  \int_\Sigma j\cdot\mathsf T\,
  \left(\epsilon\cdot P\pm \ii t\,\epsilon\cdot\Psi\,k\cdot\Psi\right)
  e^{\pm \ii t k\cdot X-\varepsilon_{\rm reg} t}\,
  \bar\delta(t k\cdot P) .
\label{eq:ambitwistor-celestial-ym-vertex}
\ee
Here $\Sigma$ is the worldsheet, $X^\mu$ and $P_\mu$ are the ambitwistor string fields, $j$ is the worldsheet current algebra, $\mathsf T$ is a color generator, $k^\mu$ is the null direction associated with $(z,\bar z)$, and $t$ is the Mellin variable.  Eq.(\ref{eq:ambitwistor-celestial-ym-vertex}) is the conformal-basis version of the standard ambitwistor vertex operator in ref.~\cite{Adamo:2019ipt}.  The factor $\bar\delta(t k\cdot P)$ imposes the scattering equation associated with this external state, while $t^\Delta$ implements the Mellin transform.  The regulator $e^{-\varepsilon_{\rm reg} t}$ specifies the convergence prescription and is the worldsheet counterpart of the usual celestial Mellin regulator.

The same construction gives gravity vertex operators by replacing the single current-algebra factor with the appropriate product of left- and right-moving polarization structures.  In particular, the integrated graviton vertex contains
\be
  U_{\rm grav}=\int_0^\infty \frac{\dd t}{t}\,t^\Delta\int_\Sigma
  \left(\epsilon\cdot P\pm \ii t\,\epsilon\cdot\Psi\,k\cdot\Psi\right)
  \left(\tilde\epsilon\cdot P\pm \ii t\,\tilde\epsilon\cdot\tilde\Psi\,k\cdot\tilde\Psi\right)
  e^{\pm \ii t k\cdot X-\varepsilon_{\rm reg} t}\,
  \bar\delta(t k\cdot P) .
\label{eq:ambitwistor-celestial-gravity-vertex}
\ee
This expression, again from ref.~\cite{Adamo:2019ipt}, is important because it shows that the Mellin transform can be built into the worldsheet theory before doing the amplitude calculation.  The resulting celestial amplitude is not obtained by first computing a momentum-space expression and only then transforming it; it is computed directly as a correlator of conformal-basis vertex operators.

The worldsheet formulation also gives a clean interpretation of conformally soft limits.  In the Yang-Mills case, the leading conformally soft gluon at $\Delta=1$ is represented by the contour operator
\be
  U^{\rm soft}_{\rm YM}= \frac{1}{2\pi\ii}\oint \frac{j\cdot\mathsf T\,\epsilon\cdot P}{k\cdot P} ,
\label{eq:ambitwistor-soft-gluon-current}
\ee
while in gravity the leading soft graviton gives
\be
  U^{\rm soft}_{\rm grav}= \frac{1}{2\pi\ii}\oint \frac{\epsilon\cdot P\,\epsilon\cdot P}{k\cdot P} .
\label{eq:ambitwistor-soft-graviton-current}
\ee
These are Eqs.(3.2) and (3.10) of ref.~\cite{Adamo:2019ipt}.  Their meaning is direct: a pole in celestial conformal dimension becomes a contour insertion on the worldsheet.  Acting on the remaining vertex operators, this contour reproduces the large-gauge or gravitational soft charge.  This is the worldsheet origin of the current-algebra statements reviewed in Section~\ref{sec:symmetry-ope}.  Related spacetime and holographic formulations of the same soft-current algebra, including the gauge-theory S-algebra, have clarified which parts of the construction are genuinely twistorial and which are universal consequences of asymptotic soft charges \cite{Kmec:2025ftx}.

At tree level this setup leads to compact CHY-like formulae in the conformal basis.  For gravity, ref.~\cite{Adamo:2019ipt} obtains a celestial version of the scattering-equation formula of the form
\be
\begin{split}
A_n=&\prod_{i=1}^n\int_0^\infty \frac{\dd t_i}{t_i}\,t_i^{\Delta_i}e^{-\varepsilon_{\rm reg} t_i}
\,\delta^{d+2}\!\left(\sum_{j=1}^n \alpha_j t_j k_j\right) \\
&\times\int\dd\mu_n\,\mathrm{Pf}'M(\sigma,\alpha t k,\epsilon)\,
\mathrm{Pf}'M(\sigma,\alpha t k,\tilde\epsilon)
\prod_j{}'\bar\delta\!\left(\sum_{l\ne j}\frac{\alpha_j\alpha_l t_jt_l\,k_j\cdot k_l}{\sigma_{jl}}\right) .
\end{split}
\label{eq:ambitwistor-celestial-scattering-equations}
\ee
Here $\sigma_i$ are punctures on the Riemann sphere, $\dd\mu_n$ is the quotient measure on their moduli space, $\alpha_i=\pm1$ distinguishes incoming from outgoing states, and $\mathrm{Pf}'M$ is the reduced Pfaffian containing the polarization data.  The primed product removes the redundant scattering equations associated with worldsheet Mobius invariance.  This equation makes explicit why ambitwistor strings are useful for celestial amplitudes: the Mellin variables $t_i$ and the scattering equations coexist in a single worldsheet integral.

\subsection{Worldsheet OPEs and celestial OPEs}
\label{subsec:worldsheet-celestial-opes}

The relation between worldsheet OPEs and celestial OPEs is one of the most concrete outputs of the ambitwistor approach.  The main observation of ref.~\cite{Adamo:2021zpw} is that the short-distance singularity of two worldsheet vertex operators localizes precisely on the short-distance limit $z_{ij},\bar z_{ij}\to0$ of two celestial insertions.  This result was complemented by ambitwistor-string derivations of supersymmetric celestial OPEs and soft algebras \cite{Bu:2021avc}, by string-worldsheet analyses of the $w_{1+\infty}$ algebra \cite{Jiang:2021csc}, and by later all-order results in the MHV sector and on-shell recursion approaches \cite{Adamo:2022wjo,Ren:2023trv}.  Thus the worldsheet OPE dynamically reproduces the celestial OPE coefficients that were originally obtained by Mellin transforming collinear splitting functions.  This is a useful check on the celestial dictionary: the same beta functions and descendant towers arise whether one starts from four-dimensional factorization or from the local worldsheet operator product.

For example, the same-helicity graviton OPE including the antiholomorphic descendant tower takes the form
\be
  \mathcal V^{\varepsilon}_{+,\Delta_i}\,\mathcal V^{\varepsilon}_{+,\Delta_j}
  \sim \frac{\bar z_{ij}}{z_{ij}}
  \sum_{m=0}^{\infty}B(\Delta_i+m-1,\Delta_j-1)
  \frac{\bar z_{ij}^{m}}{m!}\,
  \bar\partial_j^m\mathcal V^{\varepsilon}_{+,\Delta_i+\Delta_j}(z_j,\bar z_j) .
\label{eq:worldsheet-graviton-descendant-ope}
\ee
This is the outgoing same-helicity result of ref.~\cite{Adamo:2021zpw}.  The $m=0$ term reproduces the leading celestial graviton OPE discussed in Section~\ref{sec:symmetry-ope}; the higher terms give the $\overline{SL(2,\mathbb R)}$ descendant contributions that are invisible if one keeps only the leading collinear singularity.  The beta function is the Mellin transform of the energy-fraction dependence of the collinear splitting amplitude, while the derivatives $\bar\partial_j^m$ organize the antiholomorphic Taylor expansion around the second insertion.

This result explains why the OPE data, conformal block decomposition and differential equations of Section~\ref{sec:blocks} are not separate pieces of technology.  They are different ways of organizing the same analytic information.  The OPE coefficient knows about collinear factorization; the descendant tower knows about conformal covariance; and the worldsheet origin explains why these structures appear in a controlled perturbative expansion.  In this sense, the ambitwistor string gives a microscopic derivation of a substantial part of the celestial operator algebra, at least at tree level.

\subsection{\texorpdfstring{Twistor space and the celestial $w_{1+\infty}$ algebra}{Twistor space and the celestial w1+infinity algebra}}
\label{subsec:twistor-w-algebra}

Twistor methods also clarify why the $w_{1+\infty}$ algebra appears in the gravitational celestial OPE.  In the self-dual sector of gravity, Penrose's non-linear graviton construction represents the geometry in terms of deformations of twistor space \cite{Penrose:1976js}.  The celestial application of this idea is that conformally soft positive-helicity gravitons act as Hamiltonians for holomorphic symplectomorphisms of the twistor fibers, so the relevant symmetry algebra is the loop algebra of Hamiltonian vector fields on a complex two-plane \cite{Adamo:2021lrv}.  Compared with the OPE derivation in Section~\ref{sec:w-infinity}, the twistor derivation makes the geometric origin of the algebra explicit.

This also fixes a useful piece of terminology.  In gauge theory, the conformally soft positive-helicity gluons generate an infinite-dimensional chiral algebra that may be identified with the loop algebra of polynomial maps from $\mathbb C^2$ to the gauge Lie algebra $\mathfrak g$,
\(\mathcal L\mathfrak g[\mathbb C^2]\); this is often called the $\mathcal S$-algebra.  In gravity, the analogous conformally soft positive-helicity gravitons generate
\(\mathcal L\mathfrak{ham}(\mathbb C^2)\), the loop algebra of holomorphic Hamiltonian vector fields on $\mathbb C^2$, equivalently the loop algebra of the wedge algebra \(w^\wedge_{1+\infty}\) \cite{Adamo:2021lrv,Bittleston:2022jeq,Adamo:2023zeh}.  The gravitational generators can therefore be represented by polynomial Hamiltonians on the twistor fiber.

A convenient basis for the fiber functions is
\be
  w^p_m=(\mu^{\dot 0})^{p+m-1}(\mu^{\dot 1})^{p-m-1},
  \qquad p=1,\frac{3}{2},2,\frac{5}{2},\ldots .
\label{eq:twistor-w-basis-functions}
\ee
This basis appears in the twistor-space construction of ref.~\cite{Adamo:2021lrv}.  The Poisson bracket on the twistor fiber gives
\be
  \{w^p_m,w^q_n\}
  =2\bigl(m(q-1)-n(p-1)\bigr)w^{p+q-2}_{m+n} .
\label{eq:twistor-poisson-w-algebra}
\ee
The same structure is obtained from the ambitwistor worldsheet currents in ref.~\cite{Adamo:2021zpw}.  Eq.(\ref{eq:twistor-poisson-w-algebra}) is the classical wedge form of the $w_{1+\infty}$ algebra discussed in Section~\ref{sec:w-infinity}: the labels $p$ and $m$ encode spin and mode number, while the coefficient is the structure constant of area-preserving diffeomorphisms of the twistor fiber.

The twistor interpretation is conceptually useful because it explains why the algebra is especially natural in the self-dual and MHV sectors.  These are precisely the sectors in which the holomorphic geometry of twistor space is most powerful.  Soft graviton vertex operators become currents generating fiber symplectomorphisms; their OPEs with hard gravitons reproduce the celestial soft action; and the wedge algebra arises before one has to confront the full non-chiral complexity of quantum gravity.  This does not mean that the algebra is automatically exact in the full theory, but it does explain why it is robust in the tree-level chiral regime.  The same chiral geometry has been sharpened through holomorphic-disc and twistor-action descriptions of celestial $Lw_{1+\infty}$ charges \cite{Mason:2022hly,Kmec:2024nmu}, through Carrollian representations of the twistor-space algebra \cite{Donnay:2024qwq}, and through connections between celestial chiral algebras, color-kinematics duality and integrability \cite{Monteiro:2022lwm}.

\subsection{Twisted holography and celestial chiral correlators}
\label{subsec:twisted-holography-celestial}

The preceding subsections emphasized that twistor space is not merely a convenient parametrization of four-dimensional null kinematics.  It also gives a natural home for the chiral algebras generated by conformally soft modes.  A particularly concrete realization of this idea relates celestial holography to twisted holography: four-dimensional scattering data are encoded in conformal blocks of a two-dimensional chiral algebra associated with a holomorphic theory on twistor space \cite{Costello:2022wso}.\footnote{Here ``conformal blocks'' means conformal blocks of the chiral/vertex algebra in the sense of Ref.~\cite{Costello:2022wso}: prescriptions for chiral correlation functions compatible with the vertex-algebra OPEs, equivalently the chiral correlators associated with local operators of the four-dimensional theory.  They should be distinguished from the celestial conformal blocks of Section~\ref{sec:blocks}, which are conformal partial waves used to decompose Mellin-transformed scattering amplitudes into intermediate celestial primary exchanges.}  This construction complements the ambitwistor-string approach reviewed above.  Instead of starting from a worldsheet formula for the full S-matrix and then Mellin transforming external states, it starts from a twistor-space field theory whose boundary algebra already has the operator-product structure expected of the chiral sector of celestial holography.

The basic twistor-space field theory appearing in the self-dual Yang--Mills sector is holomorphic BF theory on projective twistor space.  In the conventions of \cite{Costello:2022wso}, the fields are a $(0,1)$ connection $\mathcal A$ and a $(3,1)$-form $\mathcal B$ valued in the gauge algebra, with action
\be
S_{\rm hBF}=\int_{\mathbb{PT}} {\rm Tr}\left(\mathcal B\,F^{0,2}(\mathcal A)\right).
\label{eq:twisted-hbf-action}
\ee
Here $\mathbb{PT}$ denotes projective twistor space and $F^{0,2}(\mathcal A)$ is the $(0,2)$ part of the curvature of $\mathcal A$.  The equation $F^{0,2}=0$ says that $\mathcal A$ defines a holomorphic bundle on twistor space, which is the twistor transform of the self-dual Yang--Mills equations.  The gauge transformations
\be
\delta\mathcal A=\bar\partial\chi+[\mathcal A,\chi],\qquad
\delta\mathcal B=\bar\partial\nu+[\mathcal B,\chi]
\label{eq:twisted-hbf-gauge}
\ee
make explicit why celestial symmetry algebras can be interpreted as twistor-space gauge transformations \cite{Costello:2022wso}.  The parameters $\chi$ and $\nu$ are holomorphic gauge data on twistor space.  When expanded near the celestial sphere, their modes become the currents generated by conformally soft gluons and gravitons.  This gives a geometric explanation of a recurring theme in celestial holography: the soft algebra is chiral because it is the boundary shadow of holomorphic gauge symmetry.

The chiral algebra is organized by currents whose modes remember the twistor fiber coordinates.  If $\mu^{\dot\alpha}$ are the dotted spinor coordinates along the twistor fiber, the modes can be packaged into the generating function \cite{Costello:2022wso}
\be
J(\mu,z)=\sum_{r,s\geq 0}\frac{(\mu^{\dot 1})^r(\mu^{\dot 2})^s}{r!s!}\,J[r,s](z),
\label{eq:twisted-current-generating-function}
\ee
with an analogous expansion for the conjugate currents \cite{Costello:2022wso}.  The celestial coordinate $z$ labels the base of the twistor fibration, while the integers $(r,s)$ label polynomial dependence on the null momentum spinor.  This formula is useful because it displays, in one object, the two pieces of information that celestial amplitudes usually separate: the position on the celestial sphere and the energy or spinor-helicity dependence carried by the Mellin transform.  Later work on celestial chiral algebras and color-kinematics duality sharpened this perspective by showing how such current algebras organize kinematic numerators and integrable structures in self-dual sectors \cite{Monteiro:2022lwm}.

The bridge to amplitudes is made through conformal blocks.  Given a local operator $\mathcal O$ in the four-dimensional theory, one defines chiral correlators of vertex-algebra insertions by
\be
\left\langle \mathcal O\vert V_1(z_1)\cdots V_n(z_n)\right\rangle,
\label{eq:twisted-conformal-block}
\ee
where the state $\langle\mathcal O|$ is the conformal block determined by $\mathcal O$ \cite{Costello:2022wso}.  The important point is not only that such correlators can reproduce known amplitudes, but that their associativity is ordinary vertex-algebra associativity.  This turns factorization and OPE consistency into algebraic constraints on the celestial chiral algebra.  The same logic underlies subsequent analyses of loop corrections from Koszul duality, where one asks which quantum deformations of the chiral OPE are compatible with associativity \cite{Fernandes:2023ibv}.

For form factors, the four-dimensional OPE supplies the second ingredient.  The product of local operators is written schematically as \cite{Costello:2022wso}
\be
{\rm tr}(B^2)(0){\rm tr}(B^2)(x_1)\cdots {\rm tr}(B^2)(x_{n-1})
\sim \sum_i F^i(x_1,\ldots,x_{n-1})\,\mathcal O_i(0),
\label{eq:twisted-form-factor-ope}
\ee
where the functions $F^i$ are four-dimensional OPE coefficients \cite{Costello:2022wso}.  Scattering data arise by combining these coefficients with the two-dimensional chiral correlators in Eq.(\ref{eq:twisted-conformal-block}).  This is the conceptual novelty of the construction: amplitudes are controlled by a product of two CFT-like structures, four-dimensional OPE coefficients and two-dimensional chiral conformal blocks.  In celestial language, this is close in spirit to conformal block decompositions of amplitudes, but it is more microscopic because the chiral algebra and its conformal blocks are derived from twistor-space dynamics rather than postulated as an abstract celestial CFT.

The simplest check is the Parke--Taylor structure of MHV Yang--Mills amplitudes.  In the color-ordered sector the chiral correlator of the lowest modes of the currents has the form
\be
\left\langle {\rm tr}(B^2)\vert\widetilde J_{a_1}[0,0](z_1)\cdots
\widetilde J_{a_n}[0,0](z_n)\right\rangle_{\rm c.o.}
\propto \frac{ {\rm tr}(t_{a_1}\cdots t_{a_n})}{z_{12}z_{23}\cdots z_{n1}},
\label{eq:twisted-parke-taylor-correlator}
\ee
where $z_{ij}=z_i-z_j$ and $t_a$ are color generators \cite{Costello:2022wso}.  This is the holomorphic Parke--Taylor denominator written as a chiral-algebra correlator.  The formula should be read as more than a reproduction of a familiar tree amplitude: it identifies the rational worldsheet factor that appears in twistor-string and CHY representations with an OPE-controlled object in a celestial chiral algebra.  Direct celestial-twistor constructions and scaling reductions of twistor space develop this point from the amplitude side, clarifying how twistor variables can be matched to celestial bases and lower-dimensional holographic descriptions \cite{Brown:2022miw,Bu:2023cef}.

This viewpoint also helps explain why the twistor method is naturally tied to the $w_{1+\infty}$ story discussed in Section~\ref{sec:w-infinity}.  The soft algebra arises from holomorphic gauge transformations on twistor space, while conformal blocks compute correlators of the corresponding currents.  Holomorphic-disc and twistor-action approaches to celestial $Lw_{1+\infty}$ charges, together with Carrollian representations of the same twistor-space algebra, can be viewed as complementary ways of making this geometric statement more explicit \cite{Mason:2022hly,Kmec:2024nmu,Donnay:2024qwq}.  The same twisted-holography framework also points toward quantum corrections through anomaly cancellation, axion exchange and one-loop all-plus amplitudes \cite{Costello:2022wso}; in the present review these controlled chiral corrections provide a useful comparison point for the loop and infrared effects discussed in Section~\ref{sec:uv-ir}.

\subsection{CHY formulae, double copy and twisted cohomology}
\label{subsec:twistor-chy-double-copy}

A second ambitwistor development concerns the double-copy structure of celestial amplitudes.  Tree-level celestial amplitudes for biadjoint scalars, Yang-Mills theory and gravity can be written in CHY-like form with operator-valued scattering equations \cite{Casali:2020uvr}.  The key replacement is that multiplication by energy in momentum space becomes a shift operator in conformal dimension:
\be
  \mathcal K_{i\mu}=s_i q_{i\mu}e^{\partial_{\Delta_i}},
  \qquad
  \mathcal E_i=\sum_{j\ne i}\frac{\mathcal K_i\cdot\mathcal K_j}{z_i-z_j} .
\label{eq:operator-valued-scattering-equations}
\ee
Here $q_i^\mu$ is the null direction of the $i$th particle and $s_i=\pm1$ denotes whether the particle is outgoing or incoming.  The exponential $e^{\partial_{\Delta_i}}$ shifts the Mellin weight $\Delta_i$, implementing the fact that energy factors in momentum-space amplitudes become dimension-shift operators in the celestial basis.  The equations $\mathcal E_i=0$ are therefore the celestial analogue of the usual scattering equations.

With these definitions, the gravity formula takes the compact form
\be
  \mathcal A_n^{\rm grav}
  =\int_{\mathcal M_{0,n}}\frac{\dd^n z}{{\rm vol}\,SL(2,\mathbb C)}
  \prod_i{}'\bar\delta(\mathcal E_i)\,
  \mathrm{Pf}'\Psi_n(\{\epsilon,\mathcal K\})\,
  \mathrm{Pf}'\Psi_n(\{\tilde\epsilon,\mathcal K\})\,
  \mathcal S_n(\{\Delta,q,s\}) .
\label{eq:celestial-chy-gravity}
\ee
This is the celestial version of the CHY gravity formula derived in ref.~\cite{Casali:2020uvr}.  The factor $\mathcal S_n$ is the Mellin transform of the scalar contact interaction, while the two reduced Pfaffians are the usual double-copy pair of polarization numerators, now evaluated on the operator-valued momenta $\mathcal K_i$.  Replacing one or both Pfaffians by Parke-Taylor factors gives the corresponding Yang-Mills or biadjoint scalar formulae.

The physical message is that the double copy survives the celestial transform, but in a less naive form than simply Mellin transforming two momentum-space numerators.  The celestial basis turns energies into shift operators, so color-kinematics duality becomes a statement about operator-valued numerators acting on a common Mellin kernel.  This structure admits an interpretation in terms of twisted cohomology on the moduli space of punctured spheres \cite{Casali:2020uvr}.  Such an interpretation is important for holography because it suggests that celestial amplitudes may be governed by geometric structures analogous to those controlling ordinary CHY amplitudes, but with conformal dimensions playing an active spectral role.  Recent celestial-string constructions and integrand-level analyses pursue the same idea from the opposite direction, asking how much of the worldsheet and stringy organization survives directly in the celestial basis \cite{Castiblanco:2024hnq,Bockisch:2024bia}.

A closely related development pushes the twistor and chiral-algebra viewpoint toward a more explicitly top-down form of flat-space holography.  A holographic duality was proposed for a four-dimensional WZW model coupled to scalar-flat K\"ahler gravity on Burns space, an asymptotically flat self-dual background, with a two-dimensional chiral algebra built from gauged beta-gamma systems as the dual description \cite{Costello:2022jpg}.  The construction was further developed from the perspective of the type I topological B-model on twistor space, where Burns space arises from the backreaction of D1 branes and the boundary chiral algebra is obtained from the brane worldvolume theory \cite{Costello:2023hmi}.  This is not the same as the flat limit of ordinary AdS/CFT discussed in the next section, but it is conceptually important: it shows that twistor methods can lead not only to efficient representations of amplitudes and soft algebras, but also to a concrete top-down holographic dual in an asymptotically flat setting.  It therefore provides a useful bridge between the chiral and twistor constructions reviewed in this section and the AdS/CFT-inspired approaches to flat-space holography reviewed below.

The twistor-based approach ties together several themes that otherwise look separate: Mellin transforms become scaling transforms in spinor variables, conformally soft particles arise as residues of worldsheet vertex operators, celestial OPEs descend from worldsheet OPEs, and CHY or double-copy structures survive after energies are replaced by dimension-shift operators.  These methods give one of the sharpest perturbative languages for celestial holography, especially in chiral, self-dual and tree-level regimes where holomorphic structure controls the dynamics.  They also provide a useful bridge to AdS/CFT-inspired constructions, although this bridge is presently best understood in protected sectors rather than as a full nonperturbative boundary theory.

\section{Connection to AdS/CFT}
\label{sec:ads-flat-limits}

AdS/CFT provides the most precise realization of holography, so it is natural to ask whether flat-space holography can be approached by taking the AdS radius to infinity.  The expectation is physically compelling: a sufficiently small region of AdS should look like Minkowski space, and high-energy scattering in the bulk should be encoded in singular Lorentzian regimes of boundary correlators.  Earlier analyses emphasized, however, that the flat-space limit is not a simple pointwise limit of Euclidean CFT data.  Rather, it depends on how the boundary correlator is represented, smeared and scaled, with momentum-space, Mellin-space, coordinate-space and partial-wave descriptions making different aspects of the limit manifest \cite{Li:2021snj}.  The boundary also changes character: the AdS boundary is timelike, whereas the boundary of asymptotically flat spacetime is null infinity, whose intrinsic geometry is Carrollian.  Thus the flat limit must reorganize boundary time, angular directions and energy variables before the resulting observables can resemble scattering data at $\mathscr I$.

This section reviews this connection from a deliberately limited perspective.  We do not attempt a general review of Carrollian holography.  Instead, we explain how Carrollian structures enter the flat limit of AdS/CFT, how bulk-point kinematics and Witten diagrams can lead to celestial amplitudes, and why celestial and Carrollian descriptions should be treated as related but distinct representations of flat-space observables.   Recent work bridging the two descriptions and relating Carrollian amplitudes to celestial symmetry structures makes this complementarity precise \cite{Donnay:2022wvx,Mason:2023mti}.

It is useful to distinguish three operations: taking a large-radius limit of an AdS geometry or Witten diagram, extracting flat-space scattering data from singular Lorentzian regimes of boundary correlators, and choosing a celestial or Carrollian transform of the resulting null-boundary data.  Keeping these operations separate clarifies why AdS/CFT remains an important guide while not yet supplying a complete celestial dual.

\subsection{AdS boundaries, null infinity and the Carrollian limit}
\label{subsec:ads-boundary-null-infinity}

The geometric reason the flat limit is delicate is already visible at the level of the boundary.  In global AdS, the conformal boundary is timelike, and boundary correlators are functions on a Lorentzian cylinder.  In asymptotically flat spacetime, by contrast, radiation reaches future or past null infinity, with coordinates $(u,z,\bar z)$ adapted to retarded or advanced time and angles.  The metric degenerates at $\mathscr I$, leaving a Carrollian structure rather than an ordinary Lorentzian metric.  This is the only Carrollian input needed for the present section: the flat limit changes the causal character of the boundary, and therefore the natural boundary observables.

A useful way to compare the two limits is to write AdS$_4$ in Bondi-like coordinates.  In the conventions of ref.~\cite{Alday:2024yyj}, the AdS metric can be written as
\be
  ds^2_{\rm AdS}
  =-\frac{r^2}{\ell^2}du^2-2\,du\,dr+2r^2 dzd\bar z ,
\label{eq:ads-bondi-metric}
\ee
where $\ell$ is the AdS radius and $(z,\bar z)$ are complex coordinates on the celestial sphere in a local patch.  Sending $\ell\to\infty$ at fixed $(u,r,z,\bar z)$ gives the flat Bondi metric
\be
  ds^2_{\rm flat}=-2\,du\,dr+2r^2 dzd\bar z .
\label{eq:flat-bondi-metric-from-ads}
\ee
Eqs.(\ref{eq:ads-bondi-metric}) and (\ref{eq:flat-bondi-metric-from-ads}) summarize the main geometric point of the flat limit: the term that made the AdS boundary timelike disappears, and the limiting boundary is naturally null.  The same geometric degeneration underlies the derivation of Carrollian correlators from AdS Witten diagrams and related attempts to construct a flat-space Carrollian hologram directly from AdS$_4$/CFT$_3$ \cite{Bagchi:2023fbj,Lipstein:2025jfj}.

The symmetry contraction has the same content.  On null infinity, the global conformal Carrollian vector fields take the form
\be
  \xi=(\mathcal T+u\alpha)\partial_u+\mathcal Y\partial_z+\bar{\mathcal Y}\partial_{\bar z},
  \qquad
  \alpha=\frac12\left(\partial_z\mathcal Y+\partial_{\bar z}\bar{\mathcal Y}\right),
\label{eq:carrollian-vector-field}
\ee
with $\partial_{\bar z}\mathcal Y=\partial_z\bar{\mathcal Y}=0$.  The four ordinary translations correspond to $\mathcal T(z,\bar z)=1,z,\bar z,z\bar z$ in this planar patch \cite{Alday:2024yyj}.  This is the finite-dimensional seed of the BMS enhancement discussed earlier in the review.  For the present section the important point is not the full Carrollian representation theory, but the fact that the flat limit of the AdS boundary naturally produces the same null-boundary kinematics that underlie asymptotic symmetries and celestial scattering.

Recent representation-theoretic work has made this codimension-two viewpoint sharper by constructing highest-weight modules and characters for BMS$_3$ and BMS$_4$ algebras realized on cuts of the asymptotic boundary.  In these modules, supertranslations shift the conformal weight rather than furnishing an independent quantum number \cite{Chen:2025fcc}.  Such results are useful for flat-limit holography because they probe what kind of boundary Hilbert-space data could replace ordinary CFT representations once the boundary becomes null.

A related subtlety appears in string-theoretic realizations of AdS/CFT.  In the best understood example the bulk geometry is not pure AdS, but \(\mathrm{AdS}_5\times S^5\).  If the AdS radius is sent to infinity while the internal radius remains tied to it, the internal space flattens at the same time.  The naive large-radius limit then gives a ten-dimensional flat geometry, whereas the boundary theory is naturally described by a Carrollian limit of four-dimensional \(\mathcal N=4\) super Yang--Mills theory.  This creates a boundary-mismatch problem: the null boundary of \(\mathrm{Mink}_{10}\) is not the same object as the four-dimensional Carrollian spacetime on which the gauge theory lives.  A recent proposal addresses this issue by incorporating the Carroll limit directly into the AdS/CFT construction and suggesting that the flat-space side should instead be viewed after a compactification to \(\mathrm{Mink}_5\times X^5\), with \(X^5\) locally flat, leading to a possible triality between Carroll string theory, relativistic string theory in flat spacetime and Carroll gauge theory \cite{Fontanella:2025carroll}.  This reinforces the main lesson of this subsection: flat-space holography may be connected to AdS/CFT limits, but the relation is not simply a naive large-radius limit of a fixed boundary theory.

\subsection{Extrapolate dictionaries}
\label{subsec:extrapolate-celestial-correlators}

The preceding discussion explains why a flat limit of AdS/CFT naturally leads to null-boundary and Carrollian structures.  There is, however, a complementary way of using the AdS/CFT intuition which does not begin from a large-radius limit of a known AdS dual.  An extrapolate dictionary for celestial holography defines celestial correlators directly from bulk time-ordered correlators in Minkowski space, Mellin transformed with respect to the radial coordinate in a hyperbolic slicing and then extrapolated to the celestial sphere \cite{Sleight:2023ojm}.  This is conceptually different from the original celestial-amplitude prescription reviewed in Section~\ref{sec:celestial-amplitudes}: the latter Mellin transforms on-shell S-matrix elements with respect to external energies, whereas the extrapolate prescription starts from off-shell bulk correlators and is closer in spirit to the AdS/CFT extrapolate dictionary.  It should therefore be viewed as an AdS/CFT-inspired prescription for flat-space observables, not as a derivation of the standard celestial S-matrix basis.  A related bulk-reduction dictionary instead starts from asymptotically flat bulk fields and reduces them to null infinity, producing boundary fields, descendants and Carrollian correlators without first passing through the celestial energy-Mellin basis \cite{Liu:2024nkc}.

The basic geometric input is the decomposition of Minkowski space into regions adapted to hyperbolic slices.  In the conventions of \cite{Sleight:2023ojm,Iacobacci:2024nhw}, timelike regions are described by $X^2=-t^2$ while spacelike regions are described by $X^2=R^2$; the corresponding unit vectors lie on Euclidean AdS or de Sitter slices, depending on the region.  The celestial sphere is the projective light cone
\be
  Q^2=0,\qquad Q\sim \lambda Q,\qquad \lambda\in \mathbb R_+,
  \label{eq:extrapolate-projective-lightcone}
\ee
which is the same Lorentz-covariant space of null directions used in the standard celestial transform.  The novelty is not the boundary on which the operators live, but the bulk operation used to define them.  Ref.~\cite{Iacobacci:2024nhw} makes the radial Mellin transform explicit by introducing
\be
  \phi_\Delta(\hat X)=\int_0^\infty \frac{dt}{t}\,t^\Delta\phi(t\hat X),
  \qquad
  \phi(X)=\int_{c-i\infty}^{c+i\infty}\frac{d\Delta}{2\pi i}\,t^{-\Delta}\phi_\Delta(\hat X),
  \label{eq:radial-mellin-extrapolate}
\ee
where $X=t\hat X$ and the contour parameter $c$ is chosen in the fundamental strip of the Mellin transform \cite{Iacobacci:2024nhw}.  Eq.(\ref{eq:radial-mellin-extrapolate}) should be compared with the energy Mellin transform of scattering states.  In both cases dilations are diagonalized, but here the dilation acts on the bulk radial coordinate rather than on the light-cone energy of an asymptotic particle.  This is why the construction is better thought of as a prescription for celestial correlators of bulk fields, not merely as a rewriting of the S-matrix.

The extrapolate dictionary is then the direct flat-space analogue of the AdS/CFT extrapolate map.  For a time-ordered Minkowski correlator of bulk fields, the proposed celestial correlator is \cite{Sleight:2023ojm,Iacobacci:2024nhw}
\be
\left\langle {\cal O}_{\Delta_1}(Q_1)\cdots {\cal O}_{\Delta_n}(Q_n)\right\rangle_{\rm ext}
=\prod_{i=1}^n \lim_{\hat Y_i\to Q_i}
\int_0^\infty \frac{dt_i}{t_i}\,t_i^{\Delta_i}
\left\langle \phi_1(t_1\hat Y_1)\cdots \phi_n(t_n\hat Y_n)\right\rangle .
\label{eq:extrapolate-celestial-correlator}
\ee
Here $\hat Y_i\to Q_i$ denotes the approach to the conformal boundary of the hyperbolic slice along the null direction $Q_i$.  The subscript ``ext'' is used here only to distinguish this prescription from the S-matrix celestial amplitude.  The definition has two important consequences.  First, the correlator is conformally covariant because $Q_i$ transforms linearly under the Lorentz group and the Mellin weights become celestial conformal dimensions.  Second, translation invariance is not imposed through an overall momentum-conserving delta function, as in ordinary celestial amplitudes.  It is encoded instead in the position-space bulk correlator before the radial Mellin transform.  This shift of viewpoint is one reason the prescription can produce smoother, AdS-like objects even when ordinary celestial amplitudes have distributional support.

A useful way to see the AdS-like structure is to Mellin transform a bulk Feynman propagator with one point taken to the celestial boundary.  The resulting flat bulk-to-boundary object is \cite{Sleight:2023ojm}
\be
G^{\rm flat}_\Delta(X,Q)
=\lim_{\hat Y\to Q}\int_0^\infty \frac{dt}{t}\,t^\Delta\,G_F(X,t\hat Y),
\label{eq:flat-bulk-boundary-extrapolate}
\ee
where $G_F$ is the Minkowski Feynman propagator.  For a massive scalar this object factorizes into a radial Bessel kernel times an Euclidean-AdS bulk-to-boundary propagator,
\begin{align}
G^{(m)}_\Delta(X,Q)
&=c^{\rm flat-AdS}_\Delta\,
{\cal K}^{(m)}_{i(d/2-\Delta)}\!\left(\sqrt{X^2+i\epsilon}\right)
G^{\rm AdS}_\Delta(\hat X_\epsilon,Q),
\qquad \nonumber\\
G^{\rm AdS}_\Delta(\hat X_\epsilon,Q)
&=C^{\rm AdS}_\Delta\frac{(\sqrt{X^2+i\epsilon})^\Delta}{(-2X\cdot Q+i\epsilon)^\Delta} .
\label{eq:flat-to-eads-propagator}
\end{align}

This factorization, derived in \cite{Sleight:2023ojm} and used systematically in \cite{Iacobacci:2024nhw}, is the technical bridge between the flat-space prescription and Witten diagrams on $\mathrm{EAdS}_{d+1}$.  The function ${\cal K}^{(m)}_\nu$ contains the radial dependence and is built from Bessel functions, while $G^{\rm AdS}_\Delta$ carries the angular and conformal dependence on the hyperbolic slice.  Thus the celestial conformal structure is not imposed after the fact; it appears because the radial Mellin transform separates the flat-space propagator into a radial transform and an AdS harmonic problem.

At tree level this separation turns flat-space perturbation theory into a sum of EAdS Witten diagrams multiplied by radial factors.  For example, the contact interaction considered in \cite{Sleight:2023ojm,Iacobacci:2024nhw} gives
\be
\left\langle {\cal O}_1(Q_1)\cdots {\cal O}_n(Q_n)\right\rangle_{\rm contact}
=g\,c^{\rm flat-AdS}_{\Delta_1\ldots\Delta_n}(m_1,\ldots,m_n)
D_{\Delta_1\ldots\Delta_n}(Q_1,\ldots,Q_n),
\label{eq:extrapolate-contact-witten}
\ee
with
\be
D_{\Delta_1\ldots\Delta_n}(Q_1,\ldots,Q_n)
=\int_{{\cal A}_+}d^{d+1}\hat X\,
\prod_{i=1}^n K^{\rm AdS}_{\Delta_i}\!\left(s_{\rm AdS}(\hat X,Q_i)-i\epsilon\right).
\label{eq:eads-contact-d-function}
\ee
The $D$-function is the usual EAdS contact Witten diagram, while $c^{\rm flat-AdS}_{\Delta_1\ldots\Delta_n}$ is a radial coefficient depending on the masses and Mellin weights.  This formula is the cleanest illustration of what the new prescription buys: celestial correlators inherit a familiar conformal integral representation without requiring an a priori two-dimensional CFT Hilbert space or a conventional S-matrix Mellin transform.  Subsequent work has compared this extrapolate dictionary with massless scattering prescriptions \cite{Jorstad:2023ajr}, studied its relation to boundary operators in asymptotically flat spacetime \cite{Banerjee:2024yir}, connected related conformal-correlator limits to bulk-point kinematics \cite{deGioia:2024yne}, and extended the celestial Mellin-amplitude technology from scalar contact diagrams to spin-1 and spin-2 exchange \cite{Malherbe:2025qex}.  The same Witten-diagram technology also clarifies the relation to Carrollian correlators obtained from AdS diagrams \cite{Bagchi:2023fbj}, although the two frameworks keep different boundary data: Carrollian correlators retain retarded-time dependence at null infinity, whereas the celestial extrapolate correlators live on the projective light cone after the radial Mellin transform.

Ref.~\cite{Iacobacci:2024nhw} extends the discussion beyond tree-level contact diagrams by giving Feynman rules for radial reduction and by studying the nonperturbative information contained in two-point functions.  The basic radially reduced two-point object is
\be
G_{\Delta_1,\Delta_2}(\hat X,\hat Y)
=\int_0^\infty \frac{dt_1}{t_1}\frac{dt_2}{t_2}\,
t_1^{\Delta_1}t_2^{\Delta_2}
G_T(t_1\hat X,t_2\hat Y),
\label{eq:radially-reduced-two-point}
\ee
where $G_T$ is the exact time-ordered two-point function.  If the original theory admits a K\"allen--Lehmann spectral representation with spectral density $\rho(\mu^2)$, radial reduction maps this data to a spectral representation on the hyperbolic slices.  In the notation of \cite{Iacobacci:2024nhw}, the reduced spectral density contains the Mellin transform of the ordinary spectral density and a meromorphic product of Gamma functions,
\be
\begin{aligned}
\rho_{\Delta_1+\frac d2,\Delta_2+\frac d2}(\nu)
=&\;{\cal M}[\rho]\!\left(-\frac{\Delta_1+\Delta_2}{2}+1\right)
\frac{2^{\Delta_1+\Delta_2}}{
16\pi(\sqrt{\hat X^2+i\epsilon})^{\Delta_1+d/2}
(\sqrt{\hat Y^2+i\epsilon})^{\Delta_2+d/2}}  \\
&\times
\frac{\Gamma\!\left(\frac{\Delta_1-i\nu}{2}\right)
\Gamma\!\left(\frac{\Delta_1+i\nu}{2}\right)
\Gamma\!\left(\frac{\Delta_2-i\nu}{2}\right)
\Gamma\!\left(\frac{\Delta_2+i\nu}{2}\right)}
{\Gamma(i\nu)\Gamma(-i\nu)} .
\end{aligned}
\label{eq:radial-kl-density}
\ee
This equation is conceptually important because it shows that the analytic structure of the celestial correlator is controlled by two pieces of physical data: the ordinary Minkowski spectral density, through ${\cal M}[\rho]$, and the representation theory of the hyperbolic slices, through the $\nu$-dependent Gamma functions.  The result suggests a possible nonperturbative organization of celestial observables which is not visible in the purely on-shell Mellin transform of fixed scattering amplitudes.  Recent work on celestial Mellin amplitudes develops this viewpoint further by treating the Mellin representation of extrapolate correlators as an analogue of AdS Mellin amplitudes adapted to the celestial sphere \cite{Pacifico:2024dyo}.

The relation between this extrapolate prescription and the original celestial-amplitude program should therefore be stated carefully.  It is not simply a replacement of the S-matrix basis by another notation.  It defines different observables, built from time-ordered bulk correlators and radial Mellin transforms, and it is especially natural when one wants AdS-like analytic tools, Witten-diagram decompositions, or spectral representations.  At the same time, the standard celestial amplitudes remain the most direct transform of scattering experiments at null infinity, and their soft limits, OPEs and asymptotic symmetries are tied closely to the S-matrix.  This is why it is useful to discuss the extrapolate dictionary in the present section: it is one of the clearest examples of how AdS/CFT ideas can be imported into flat-space holography without taking the flat limit of a complete AdS dual.

A related set of constructions uses analytic continuation and quotient geometries to reorganize flat-space data.  In Klein signature, null infinity is replaced by a single null boundary and the scattering matrix can be repackaged as a state on that boundary; this provides another way to understand why hyperbolic leaves and leaf amplitudes can carry smoother celestial data \cite{Melton:2024pre}.  Time-periodic AdS$_3/\mathbb Z$ and equal-radius Lorentzian CFTs supply lower-dimensional laboratories in which principal-series representations, periodic time and celestial leaf geometries can be studied without assuming the standard highest-weight AdS/CFT setup \cite{Melton:2025ecj,Melton:2025jee}.  These examples are not replacements for the four-dimensional celestial S-matrix, but they emphasize that the space of AdS-inspired boundary prescriptions is broader than the strict large-radius limit of an ordinary global AdS boundary theory.

\subsection{Bulk-point kinematics and celestial amplitudes}
\label{subsec:bulk-point-celestial-flat-limit}

The flat limit of AdS/CFT is most directly visible in Lorentzian regimes of boundary correlators.  The classic bulk-point idea is that a boundary correlator develops a singularity when the insertion points can be connected by null geodesics meeting at a bulk point; this singularity encodes local bulk scattering.  This point of view has been refined in the celestial language by studies that move back and forth between AdS correlators, conformal partial waves and flat-space celestial data \cite{Iacobacci:2022yjo,Sleight:2023ojm}.  In the conformal basis, this observation becomes especially close to celestial holography because external energies are Mellin transformed.  The goal is not to identify an ordinary CFT correlator with a celestial amplitude at finite AdS radius, but to isolate the scaling regime in which AdS boundary insertions become null asymptotic states.

Flat-space scattering can also be formulated in a conformal primary basis in which massless and massive external states are treated uniformly \cite{Lam:2017ofc}.  The corresponding conformal transform of an amplitude has the schematic form
\be
\begin{split}
  \widetilde{\mathcal A}
  &=
  \prod_{\ell\,\in\,{\rm massless}}
  \int_0^\infty d\omega_\ell\,\omega_\ell^{\Delta_\ell-1}
  \prod_{j\,\in\,{\rm massive}}
  \int[d\hat p_j]\,G_{\Delta_j}(\hat p_j;\vec x_j)
  \mathcal T(k_\ell,p_j)\,
  \delta^{(d+2)}\!\left(\sum_\ell k_\ell+\sum_j p_j\right).
\end{split}
\label{eq:lam-shao-conformal-basis}
\ee
Here $\mathcal T$ is the ordinary momentum-space transition amplitude, $k_\ell=\omega_\ell q_\ell$ are null momenta, $p_j=m_j\hat p_j$ are massive momenta on the unit hyperboloid, and $G_\Delta$ is the bulk-to-boundary propagator on that hyperboloid.  Eq.(\ref{eq:lam-shao-conformal-basis}), which is Eq.(2.10) of ref.~\cite{Lam:2017ofc}, is important for the flat-limit program because it shows that the celestial transform is not tied only to strictly massless four-dimensional kinematics.  It is a representation-theoretic transform of flat-space scattering data.

The more recent flat-limit construction of ref.~\cite{deGioia:2024yne} makes the connection to AdS correlators sharper.  Boundary insertions are placed near the past and future tips of the Lorentzian boundary cylinder,
\be
  t_i=\eta_i\frac{\pi}{2}+\frac{u_i}{R},
  \qquad \eta_i=\pm 1,
\label{eq:ads-boundary-strips}
\ee
where $R$ is the AdS radius and $u_i$ remains finite as $R\to\infty$.  The variables $u_i$ parametrize infinitesimal boundary strips that become retarded or advanced times in the flat limit.  A key technical step is that integrating the AdS bulk-to-boundary propagator over these strips produces a flat-space conformal primary wavefunction.  In the notation of ref.~\cite{deGioia:2024yne},
\be
\begin{split}
  \lim_{R\to\infty}\int_{-\infty}^{\infty}du\,
  (u\pm i\epsilon)^{-\lambda}
  K_\Delta(P;X)\big|_{\tau_p=\pm\pi/2+u/R}
  &=
  N_\Delta\,
  \varphi^{\pm}_{\Delta+\lambda-1}(x;\hat q).
\end{split}
\label{eq:ads-strip-to-celestial-wavefunction}
\ee
Here $K_\Delta(P;X)$ is the AdS bulk-to-boundary propagator, $P$ is a boundary point, $X$ is a bulk point, and $\varphi^{\pm}$ is the outgoing or incoming flat-space conformal primary wavefunction. This formula, Eq.(2.10) of ref.~\cite{deGioia:2024yne}, explains why the flat limit can produce celestial amplitudes rather than merely momentum-space amplitudes: the Mellin-type weight in $u$ converts the shrinking boundary strip into a conformal primary external leg.

For cubic scalar Witten diagrams this procedure yields directly
\be
  \lim_{R\to\infty}
  \prod_i\frac{1}{N_{\Delta_i}}
  \int_{-\infty}^{\infty}du_i\,(u_i+i\eta_i\epsilon)^{-\lambda_i}
  \left\langle \mathcal O_1\mathcal O_2\mathcal O_3\right\rangle_{\rm AdS}
  =\widetilde A_3\!\bigl(\Delta_1^{\rm CCFT},\Delta_2^{\rm CCFT},\Delta_3^{\rm CCFT}\bigr),
\label{eq:ads-three-point-to-celestial}
\ee
after the dimension shifts specified in ref.~\cite{deGioia:2024yne}.  Eq.(\ref{eq:ads-three-point-to-celestial}) is the conceptual prototype for the flat-limit map: a Lorentzian boundary correlator, integrated over the appropriate near-null boundary strips, becomes a celestial amplitude.  The construction is more subtle for four-point functions because bulk-point kinematics, analytic continuation and intermediate singularities must be controlled, but the same physical idea applies.

\subsection{Carrollian correlators from Witten diagrams}
\label{subsec:carrollian-from-witten}

The same limit can be organized without immediately Mellin transforming to boost eigenstates.  This gives Carrollian amplitudes: correlators at null infinity that retain the retarded-time variables $u_i$.  This is in many ways the most direct boundary description emerging from the flat limit of AdS/CFT.  Celestial amplitudes are then obtained, when the transform is well-defined, by Mellin or Fourier-Mellin transforms of the energy or retarded-time dependence.  The Witten-diagram derivation of Carrollian correlators \cite{Bagchi:2023fbj}, its extension to holographic correlators in four and three bulk dimensions \cite{Alday:2024yyj,Surubaru:2025fmg}, and recent work on a flat-space Carrollian hologram from AdS$_4$/CFT$_3$ \cite{Lipstein:2025jfj} all support this ordering of ideas: first take the null-boundary limit, then decide which transform best exposes the desired scattering observable.

This relation can be made explicit at the level of propagators and Witten diagrams \cite{Alday:2024yyj}.  In Bondi coordinates, the AdS bulk-to-boundary propagator has a smooth flat limit,
\be
  \mathcal G^{\rm AdS,\Delta}_{Bb,\epsilon}(x;X)
  \xrightarrow[\rm Bondi]{\ell\to\infty}
  \frac{-i\alpha(\Delta)}{2(2\pi)^2}
  \frac{\Gamma(\Delta)}{\left(-u-q\cdot X+i\epsilon\varepsilon\right)^\Delta}
  =\alpha(\Delta)\mathcal G^{\rm Flat,m}_{Bb}(x;X).
\label{eq:ads-bulk-boundary-flat-limit}
\ee
This is Eq.(3.28) of ref.~\cite{Alday:2024yyj}.  The boundary point is $x=(u,z,\bar z)$, $q^\mu(z,\bar z)$ is the associated null vector, $X$ is a bulk point, and $\epsilon=\pm1$ distinguishes outgoing and incoming boundary conditions.  The parameter $m$ labels retarded-time derivatives of the Carrollian primary, related to the AdS scaling dimension by $\Delta=m+1$ in the scalar example.  The equation shows that the AdS external leg itself becomes a flat-space bulk-to-boundary object at null infinity.

This conclusion is consistent with the Lorentzian AdS quantization viewpoint developed in \cite{Navarro:2025xln,Navarro:2026rna}.  There, null foliations of AdS and time-ordered bulk-to-boundary propagators are used to decompose bulk fields into positive- and negative-frequency sectors, with CFT primary operators and their shadows playing complementary roles.  In the large-radius limit, these AdS reconstruction formulae reduce either to Carrollian expansions at null infinity or, after diagonalizing boosts, to the conformal-primary wavefunctions used in celestial amplitudes.

The corresponding flat bulk-to-boundary propagator can be written as a half-line Fourier transform.  This Fourier representation is also the starting point for differential-equation approaches to Carrollian amplitudes, which make precise how retarded-time dependence encodes momentum-space constraints before one Mellin transforms to the celestial basis \cite{Ruzziconi:2024zkr}.
\be
  \mathcal G^{\rm Flat,m}_{Bb,\epsilon}(x;X)
  =\frac{-\epsilon}{2(2\pi)^2}
  \int_0^\infty d\omega\,(-i\epsilon\omega)^m e^{-\varepsilon_{\rm reg}\omega}
  e^{-i\epsilon\omega(u+q\cdot X)} .
\label{eq:flat-carrollian-bulk-boundary-fourier}
\ee
This is Eq.(2.26) of ref.~\cite{Alday:2024yyj}.  It makes the relation between Carrollian and celestial bases transparent.  Keeping $u$ gives a Carrollian correlator at null infinity; Mellin transforming the energy $\omega$ produces the boost eigenstates used in celestial amplitudes.  Thus the two descriptions are not unrelated, but they diagonalize different structures.

The Carrollian limit of ordinary CFT correlators also exhibits the expected localization on the celestial sphere.  For example, the two-point function in the contracted theory has the form
\be
  \left\langle T\{\mathcal O_\Delta(x_1)\mathcal O_\Delta(x_2)\}\right\rangle
  =\frac{\widetilde C_2(\Delta)}{\bigl(-c^2u_{12}^2+2|z_{12}|^2+i\varepsilon\bigr)^\Delta},
\label{eq:carrollian-cft-two-point-before-limit}
\ee
before taking the strict Carrollian limit $c\to0$ \cite{Alday:2024yyj}.  In that limit the correlator becomes distributional in the angular separation, matching the intuition that null infinity records radiation along fixed null directions.  This is precisely the feature that celestial amplitudes often obscure by integrating over energies: angular localization, retarded-time dependence and conformal weights are tightly linked.

\subsection{Two-point correlator on a shockwave background}
\label{subsec:eikonal-flat-limit-ads}

The shockwave two-point function provides a useful bridge between the AdS flat limit and the UV/IR issues discussed in Section~\ref{sec:uv-ir}.  In momentum space, the corresponding eikonal regime is controlled by high energy and small momentum transfer.  In celestial variables, this regime is encoded in the analytic structure at large total boost weight and small cross ratio.  Celestial two-point functions on Kerr-Schild backgrounds give a complementary flat-space viewpoint: the background can absorb momentum, so Coulomb, Schwarzschild, Kerr and Aichelburg-Sexl shockwave geometries lead to non-distributional celestial observables, with the shockwave case admitting an interpretation in terms of a conformal correlator with a background insertion \cite{Gonzo:2022tjm}.  The celestial eikonal amplitude can also be recovered from the flat-space limit of an AdS shockwave computation \cite{deGioia:2022fcn}.

A scalar crossing a shockwave with profile $h(x_\perp)$ obeys the matching condition
\be
  \phi(\epsilon,x^+,x_\perp)
  =\phi(-\epsilon,x^+-h(x_\perp),x_\perp),
  \qquad
  \Delta x^+=h(x_\perp),
\label{eq:shockwave-time-delay}
\ee
as reviewed in ref.~\cite{deGioia:2022fcn}.  In the conformal primary basis this produces a shockwave two-point function
\be
  \left\langle \mathcal O_\Delta({\bf p}_2)\mathcal O_\Delta({\bf p}_4)\right\rangle_{\rm shock}
  =\mathcal C_\Delta\int_{H_2}d^2{\bf x}_\perp\,
  \frac{\Gamma(2\Delta)}{\left(2q\cdot{\bf x}_\perp+h({\bf x}_\perp)-i\epsilon\right)^{2\Delta}} .
\label{eq:celestial-shock-two-point}
\ee
This is Eq.(4.15) of ref.~\cite{deGioia:2022fcn}.  It is valuable because it gives an explicit AdS/CFT-inspired observable whose flat limit reproduces celestial eikonal physics.  The lesson is deliberately modest: AdS limits are most tractable in special kinematic regimes, such as shockwave scattering, while a complete map for general quantum-gravitational scattering remains much less understood.  The most direct large-radius limit appears to produce Carrollian observables at null infinity, as also suggested by Carrollian partition functions obtained from AdS limits \cite{Kraus:2024gso}; celestial amplitudes arise after an additional transform that trades retarded-time or energy dependence for conformal dimension.

\subsection{Entanglement in flat holography}

Entanglement provides another diagnostic of flat holography, complementary to scattering amplitudes and boundary correlators.  In three-dimensional flat holography, the first concrete prescriptions were formulated in BMS-invariant field theories and their flat-space gravitational duals \cite{Jiang:2017ecm}.  This line of work was sharpened by covariant constructions of holographic entanglement and Poincar\'e, or global BMS, blocks in three-dimensional flat space \cite{Hijano:2017eii}, by generalized gravitational-entropy prescriptions adapted to non-Lorentz-invariant duals and applied to flat space \cite{Wen:2018mev}, and by attempts to reconstruct aspects of gravitational dynamics in Minkowski space from entanglement principles \cite{Godet:2019wje}.  Related flat-limit calculations have also studied holographic entanglement in massive three-dimensional gravity models \cite{Setare:2021auh}.  More refined probes include swing-surface prescriptions for general states and intervals in flat$_3$/BMSFT \cite{Apolo:2020bld}, modular Hamiltonians in flat holography \cite{Apolo:2020qjm}, and mixed-state measures such as entanglement negativity, entanglement-wedge cross sections and balanced partial entanglement \cite{Basu:2021axf,Basu:2021awn,Camargo:2022mme}.  These developments are reviewed from the broader Carrollian viewpoint in \cite{Bagchi:2025vri}.

From the perspective of this review, the important lesson is not that these constructions already define the celestial dual, but that entanglement probes make the codimension-one Carrollian side of flat holography more concrete.  In the Carrollian description, boundary regions live naturally on null infinity and retain retarded-time information.  This is conceptually different from the celestial basis, where a Mellin transform trades energy or retarded-time dependence for conformal dimensions.  For free Maxwell theory, vacuum entanglement across cuts of future null infinity can be organized in terms of soft or edge modes associated with asymptotic charges, and the corresponding conformally soft configurations have a natural celestial interpretation \cite{Chen:2023tvj,Chen:2024kuq}.  A complementary celestial-CFT perspective studies subregion R\'enyi entropies through twist operators and their proposed bulk cosmic-brane duals \cite{Capone:2024oim}.  Recent work on BMS-invariant free theories and Carrollian/Galilean conformal blocks has also begun to connect modular Hamiltonians and excited-state entanglement to concrete boundary dynamics \cite{Hao:2025hfa,Hao:2025naz}.

A related top-down perspective comes from Coulomb-branch geometries in which the infrared region contains a flat-space bubble; holographic entanglement and complexity probes in such backgrounds give another controlled way of asking how flat-space physics can emerge from an AdS/CFT setting \cite{Jorstad:2026jlg}.  Pseudoentropy in Carrollian conformal field theories provides a further diagnostic of how extremal-curve prescriptions behave in flat limits \cite{Fareghbal:2025dns}.  Taken together, these results suggest that entanglement may help bridge the Carrollian and celestial descriptions, but they also underline a conceptual gap: the standard celestial amplitude is an $S$-matrix observable, while entanglement is naturally associated with subregions, modular flow and state-dependent boundary algebras.  The relation between these two kinds of observables is one of the issues returned to in the Outlook.

\section{Outlook}
\label{sec:outlook}

The preceding sections describe a subject whose central structures are now quite concrete, but whose final holographic interpretation remains unsettled.  Celestial amplitudes give a Lorentz-covariant conformal basis for the flat-space S-matrix.  Soft theorems and asymptotic symmetries become Ward identities and current algebras at null infinity.  Collinear limits lead to celestial OPEs, conformally soft modes generate infinite-dimensional symmetry algebras, and twistor, Liouville and AdS-inspired constructions provide complementary ways of organizing the same scattering data.  The main open questions are therefore no longer simply whether flat-space scattering has hidden two-dimensional structures; the preceding sections show that it does.  The sharper question is whether these structures can be assembled into a complete boundary description of quantum gravity in asymptotically flat spacetime. Here we list a few open questions from various aspects.

\subsection*{Boundary observables}

The first unresolved issue is the definition of the fundamental observables.  The conventional celestial transform is anchored directly in the S-matrix and therefore preserves the connection to measurable scattering amplitudes.  At the same time, exact plane-wave amplitudes inherit the rigid support imposed by momentum conservation.  This is why many celestial amplitudes are distributions rather than ordinary functions on the celestial sphere.  The block decompositions, differential equations, shadow transforms and non-distributional constructions reviewed above should be understood as attempts to sharpen the same question: which class of celestial objects is regular enough to behave like a boundary correlator, while still retaining the physical content of scattering?

Several answers are now available in controlled settings.  Background-field amplitudes, leaf amplitudes and Liouville-type correlators produce smoother celestial objects \cite{Fan:2022vbz,Stieberger:2022zyk,Melton:2023bjw,Melton:2024jyq}.  Carrollian correlators keep retarded time explicit and are geometrically closer to null infinity.  Extrapolate dictionaries define AdS-inspired celestial observables from bulk time-ordered correlators and radial Mellin transforms \cite{Sleight:2023ojm,Iacobacci:2024nhw}.  Exactly solvable lower-dimensional examples provide another useful diagnostic: Mellin transforms of integrable two-dimensional S-matrices give nonperturbative celestial correlators whose analytic properties can be studied without relying on a weak-coupling expansion \cite{Duary:2022onm,Kapec:2022ift}.  These constructions should not be viewed as competitors to the S-matrix basis.  Rather, they suggest that flat-space holography may require a hierarchy of related observables.  A central task is to determine which observables are fundamental, which are transforms or limits of one another, and which are useful only in special sectors. Very recently, we have seen interesting connections between flat holography and more general observables. See, e.g. \cite{Gonzalez:2025eds,Ruan:2026xyd, Moult:2025njc,Elkhidir:2024izo}. It would also be interesting to relate what we discussed in the review to other approaches in flat holography; see, e.g. \cite{Ammon:2025avo} which follows more closely the GKPW \cite{Gubser:1998bc,Witten:1998qj} method while remaining intrinsic to asymptotically flat spacetimes and adapted to scattering data.

\subsection*{Quantum corrections and infrared safety}

A second major obstacle is the quantum consistency of the celestial basis.  At tree level, soft theorems and collinear limits give comparatively sharp OPEs and current algebras.  Loop corrections change the problem qualitatively.  Infrared divergences, soft anomalous dimensions, logarithms of cross ratios, dimension shifts and scheme dependence all become part of the celestial correlator.  The relation between momentum-space exponentiation and celestial anomalous dimensions is understood in important examples, but a systematic all-order framework is still missing.  Such a framework should explain how soft exponentiation, logarithmic celestial correlators and renormalization-group flow appear in a boundary language.

This issue is inseparable from infrared safety.  Exclusive Fock-space amplitudes are not the natural finite observables of massless gauge theory or gravity.  Inclusive observables, dressed states, Carrollian fluxes and eikonal or shockwave correlators retain different parts of the infrared data.  The shockwave examples reviewed in Section~\ref{sec:ads-flat-limits} show that celestial observables can be defined in controlled semiclassical regimes, but subleading eikonal corrections, radiation and inelastic channels must be incorporated before one obtains a genuinely unitary celestial object.  A further test is whether the large-radius limit of AdS loop diagrams commutes with the infrared limits that define flat-space scattering \cite{Li:2021snj,Iacobacci:2022yjo,deGioia:2024yne,Alday:2024yyj}.  Related exact two-dimensional examples show that even the operation of Mellin transforming an S-matrix need not commute with perturbative expansion, sharpening the order-of-limits problem for bottom-up tests of proposed celestial duals \cite{Tropper:2026chb}.  If these limits fail to commute, celestial, Carrollian and AdS-limit observables may still be related, but only through a nontrivial order-of-limits problem.

\subsection*{OPEs and analytic structure}

The conformal-block and analytic-correlator program raises a third question: whether celestial OPE data define a genuine boundary spectrum or merely repackage four-dimensional factorization.  In ordinary two-dimensional CFT, conformal blocks decompose correlators into exchanged operator families.  In celestial holography the situation is subtler.  The exchanged data can correspond to bulk particles, shadows, light transforms, residues of analytic continuations or continuous $SL(2,\mathbb C)$ representations.  The conformal block expansion is therefore a powerful diagnostic of representation theory, but not by itself proof that an autonomous Euclidean CFT has been found.

Liouville theory and leaf amplitudes sharpen this issue.  Semiclassical Liouville correlators, Dotsenko-Fateev integrals and PDE descriptions provide a concrete language for smooth celestial correlators in selected Yang-Mills and MHV sectors \cite{Stieberger:2022zyk,Taylor:2023bzj,Stieberger:2023fju,Giribet:2024vnk}.  The open question is whether these structures are special to protected sectors or reveal a more general celestial CFT mechanism.  One particularly important target is to understand whether finite-central-charge or finite-$b$ Liouville corrections have an interpretation as loop corrections to celestial OPE data, or instead describe a different deformation of the celestial theory.  Another is to decide whether the differential equations found in analytic celestial amplitudes are computational consequences of Mellin transforms or genuine Ward identities of a boundary theory.

\subsection*{Symmetries and protected sectors}

The celestial $w_{1+\infty}$ algebra provides one of the sharpest structural successes of the program.  The conformally soft graviton OPE, the wedge algebra, the phase-space charges and the twistor-space realization all point toward the same infinite-dimensional symmetry structure.  The remaining question is how universal this structure is.  At tree level and in chiral or self-dual sectors, the algebraic picture is remarkably coherent.  Beyond these regimes, loop corrections, infrared divergences, possible central extensions and the quantization of radiative phase space may modify the soft algebra.

The deformation problem gives a useful way to organize this uncertainty.  Flat-space $w$-algebras, Moyal deformations, cosmological-constant deformations, higher-spin interpretations and twistor realizations should reduce to the same tree-level wedge algebra in appropriate limits.  Recent work on leaf-amplitude soft algebras, supersymmetric chiral soft algebras and $\mathcal L_\Lambda w_{1+\infty}$ actions in CFT$_3$ tests these ideas in smooth observables, supersymmetric gauge theory and AdS/CFT-like settings \cite{Melton:2024jyq,Crawley:2024cak,Sheta:2025oep,Strominger:2026yrh}.  What remains unclear is whether these constructions are different presentations of one celestial symmetry algebra or distinct protected subsectors.

Twistor and ambitwistor methods are central to this question.  They naturally encode null directions, conformal covariance, scattering equations, soft currents and chiral OPEs.  They also explain why protected celestial structures often look holomorphic or chiral.  The limitation is equally important: current twistor constructions are most powerful in self-dual, MHV, chiral or tree-level regimes.  A complete holographic theory must explain how these protected structures are embedded in the full non-self-dual, loop-corrected gravitational S-matrix.

\subsection*{AdS/CFT, Carrollian data and curved backgrounds}

The relation between celestial and Carrollian holography remains one of the central conceptual problems.  Celestial holography diagonalizes boosts and uses Mellin-transformed scattering amplitudes.  Carrollian holography keeps retarded time at null infinity and is closer to the intrinsic geometry of the flat boundary.  The two descriptions are therefore not equivalent at the level of variables, even when they encode related physics. 

The flat limit of AdS/CFT sharpens the same point.  The AdS boundary is timelike, whereas null infinity is Carrollian.  Large-radius limits of Witten diagrams, bulk-point singularities and extrapolate dictionaries produce flat-space observables only after delicate Lorentzian scaling limits.  What is established is that controlled AdS limits can reproduce conformal-primary wavefunctions, null-boundary observables and special celestial correlators.  What remains conjectural is that these limits determine the full celestial S-matrix or a complete flat-space hologram.  Massive particles, spinning states and nonzero cosmological constant deformations belong naturally to this discussion: they are unavoidable from the AdS perspective, but most celestial symmetry constructions are currently sharpest for massless particles at null infinity \cite{Lam:2017ofc,Melton:2024pre,Melton:2025ecj,Melton:2025jee,Guevara:2026qzd,Guevara:2026qwa}.

\section*{Acknowledgements}
I would like to thank Stephan Stieberger for inviting me to write this report. I thank Stephan Stieberger and Tomasz Taylor for useful comments on the manuscript. I also thank all my wonderful collaborators during our journey on flat holography, in particular to Tim Adamo, Stephan Stieberger, and Tomasz Taylor.  BZ is supported by the Fundamental Research Funds for the Central Universities (010-63263123). 

\section*{Declaration of generative AI and AI-assisted technologies in the manuscript preparation process}

During the preparation of this work, the author used GPT 5.5 for spell check. The author reviewed and edited the output as needed and takes full responsibility for the content of the published article. 

\bibliographystyle{jhep}
\bibliography{cope}

\end{document}